\documentclass[namedreferences]{solarphysics}

\usepackage[hyperref,optionalrh,showbiblabels]{spr-sola-addons} 
\usepackage{graphicx}        
\usepackage{color}           
\usepackage{breakurl}        
\usepackage{rotating}
\usepackage{tabularx}

\chardef\us=`\_

\begin{document}

\begin{article}
\begin{opening}

\title{The Wind/EPACT proton event catalog (1996$-$2016)}

\author[addressref={aff1},corref,email={rmiteva@space.bas.bg}]{\inits{R.}\fnm{Rositsa}~\lnm{Miteva}}
\author[addressref=aff2,email={samwelsw@nriag.sci.eg}]{\inits{S. W.}\fnm{Susan W.}~\lnm{Samwel}}
\author[addressref=aff3,email={mvcduarte@iag.usp.br}]{\inits{M. V.}\fnm{Marcus V.}~\lnm{Costa-Duarte}}

\address[id=aff1]{Space Research and Technology Institute, Bulgarian Academy of Sciences (SRTI-BAS), 1113 Sofia, Bulgaria}
\address[id=aff2]{National Research Institute of Astronomy and Geophysics (NRIAG), 11421 Helwan, Cairo, Egypt}
\address[id=aff3]{Institute of Astronomy, Geophysics and Atmospheric Sciences, University of S\~{a}o Paulo (IAG-USP), 05508-090 S\~{a}o Paulo, Brazil}

\runningauthor{Miteva et al.}
\runningtitle{Wind/EPACT proton catalog}

\begin{abstract}
We present the finalized catalog of solar energetic proton events detected by Wind/EPACT instrument over the period 1996$-$2016. Onset times, peak times, peak proton intensity and onset-to-peak proton fluence are evaluated for the two available energy channels, at about 25 and 50 MeV. We describe the procedure utilized to identify the proton events and to relate them to their solar origin (in terms of flares and coronal mass ejections). The statistical relationships between the energetic protons and their origin (linear and partial correlation analysis) are reported and discussed in view of earlier findings. Finally, the different trends found in the first eight years of solar cycles 23 and 24 are discussed. 
\end{abstract}
\end{opening}

\section{Introduction}
     \label{S-Introduction}  

Energetic particles originating from eruptive processes at the Sun can escape the solar atmosphere and propagate into the heliosphere along the interplanetary [IP] magnetic field lines. When these field lines connect to a space-borne detector, the particles can be observed in situ. There is a limitation on the number and positioning of space probes capable to detect particles at present. The majority of the current solar dedicated space missions are orbiting L1, in addition to the twin-STEREO mission \citep{2008SSRv..136....5K} that have already surpassed the conjunction point. All of them are situated in the ecliptic plane. In summary, in situ observation can be regarded as single-point sampling of the heliosphere.

In addition to solar flares [SFs] and coronal mass ejections [CMEs] solar energetic particles [SEPs] (protons, electrons and heavy ions from keV to GeV range, \opencite{2006SSRv..123..217K,2016LRSP...13....3D}) contribute to space weather \citep{2006LRSP....3....2S}. The SEPs can cause various effects on spacecraft and terrestrial systems \citep{2007LRSP....4....1P} and the aim is to minimize the negative influences. Thus timely and reliable SEP-forecasting is of increasing importance for different space weather effects, in addition to the space travel safety of humans and satellite longevity. 

One way to approach the SEP topic is from an historical perspective based on observation of numerous events in addition to the theoretical models. Several decades of observations already exist, and multi-spacecraft data is available since about 1996 (start of solar cycle [SC] 23). Several consistent lists of SEP events (only protons, however), are compiled. Below well-known proton catalogs with a focus on lists extending to recent years are described, which does not substitute a complete review on the topic.

The longest proton series covering the period 1976$-$present is provided by the Geostationary Operational Environmental Satellite [GOES] instrument\footnote{https://ngdc.noaa.gov/stp/satellite/goes/datanotes.html} \citep{1996SPIE.2812..281O}. The GOES proton list\footnote{\url{http://umbra.nascom.nasa.gov/SEP/}} reports proton enhancements $>$10 MeV when the 10 proton flux units [pfu]\footnote{1 pfu = 1 proton/(cm$^2\,$s$\,$sr)} threshold level is surpassed. Due to this particular intensity level chosen, when a new SEP event occurs on an already enhanced background (during an ongoing SEP event) it is usually omitted from this list. Since the GOES catalog is intended to serve specific customer needs, the caveats of the list need to be taken into account while using it for scientific purposes.

The SEPServer project \citep{2013JSWSC...3A..12V} produced several proton and electron event lists during 1997$-$2010. Among them are the 55$-$80 MeV proton events recorded by the Energetic and Relativistic Nuclei and Electron [ERNE] instrument aboard Solar and Heliospheric Observatory [SOHO] \citep{1995SoPh..162..505T}. This catalog was updated on-line\footnote{\url{http://server.sepserver.eu/}, status: July 2017} to cover the period 1997$-$2015. A recent summary over the period 1996$-$2016 is presented by \cite{2017JSWSC...7A..14P}.

In addition, there is the SEPEM\footnote{The SEPEM 7.23$-$10.45 reference proton event list: \url{http://dev.sepem.oma.be/help/event_ref.html}} event list (1973$-$2013, \opencite{2015SpWea..13..406C}), based on re-calibrated GOES data. Two different studies have already utilized the database with a focus on either energy-dependent statistics (\opencite{2015SoPh..290..841D}), or, among others, the distribution with respect to NOAA Solar Radiation Storm scale (\opencite{2016JSWSC...6A..42P}).

Recently, a new list of high energy protons (1995$-$2015), $>$500 MeV from the Electron Proton Helium Instrument [EPHIN] instrument aboard SOHO \citep{1995SoPh..162..483M} was presented by \cite{2017SoPh..292...10K}.

Among the proton catalogs covering only SC23 is the 25 MeV proton database from the IMP-8 instrument \citep{2010JGRA..11508101C}, as well as other listings\footnote{Several proton lists are reported at \url{http://www.wdcb.ru/stp/index.en.html}.} as noted by \inlinecite{2016JSWSC...6A..42P}. Other particle lists using GOES and SOHO/COSTEP \citep{1995SoPh..162..483M} data were created for the purpose of testing forecasting procedures \citep{2009SpWea...7.4008L,SWE:SWE185}.

Numerous statistical studies between SEP events and their solar origin (flares and CMEs) are known. Different works use differential or integral proton fluxes, flare class or/and fluence, CME projected speed or/and angular width, investigate the longitudinal effects, probability of occurrence or propose new statistical methods. The strength of the correlation between the protons and their solar origin is used to argue in favor of or against the particle driver. Below we outline studies covering at least one solar cycle. Presenting a complete account on all previous works is not the objective of this study.

A comprehensive study is the energy dependent statistics provided by \cite{2015SoPh..290..841D} based on re-calibrated data from GOES and IMP-8 instruments over 10 energy channels (from 5 to 200 MeV) and in the entire SC23 (using SEPEM data\footnote{\url{http://dev.sepem.oma.be}}, \opencite{2015SpWea..13..406C}). The work provides results based on differential and integral proton intensities. The main result is the larger correlation between flares and high energy protons, whereas the low energy protons correlate better with CME speed. Among the other works over SC23 that provide also new particle lists are: \cite{2010JGRA..11508101C} based on IMP-8 $>25$ MeV protons; \cite{2013SoPh..282..579M} based on SOHO and Wind-protons and ACE-electron counterparts of the former proton-list; \cite{2013JSWSC...3A..12V} based on SOHO-proton and ACE-electron lists with extension by \cite{2017JSWSC...7A..14P} based on 55$-$80 MeV SOHO/ERNE protons; \cite{2016JSWSC...6A..42P} based on GOES $>$10, 30, 60 and 100 MeV SEPEM protons. In addition, a number of statistical works utilize the GOES proton event list\footnote{\url{https://umbra.nascom.nasa.gov/SEP/}}, however, listing all papers based on this on-line catalog is beyond the scope of our study. In relation to the earlier SEP catalogs and analysis, we note that the present report adds to the works presenting an event list and statistics, in this case on the Wind/EPACT $\sim$25 and $\sim$50 MeV proton events.

Until recently, all earlier statistical works use the Pearson correlation coefficients; however \cite{2015SoPh..290..819T} showed that interdependencies need to be filtered out from the correlations by utilizing the partial correlation coefficients. They argued that the flare fluence and not the flare class should be used. We continue this line of research by applying the partial correlations on a larger event sample.

Furthermore, with the progress of the current SC24, one is able to test quantitatively the solar event productivity trends. Various catalogs have been utilized for this purpose. The earliest attempts were based on three \citep{2013AdSpR..52.2102C} and reaching up to 5.8 \citep{2015ICRC..M} and seven-year \citep{2017SunGe..12...11M} comparisons, respectively. Other studies focus on larger intensity events \citep{2014EP&S...66..104G}. 

The aim of our study is to explore the finalized list of proton events from the Energetic Particles Acceleration, Composition, and Transport [EPACT] instrument \citep{1995SSRv...71..155V} aboard the Wind satellite covering over a 20-year period of time. This data was previously used {to evaluate the proton counterparts in SC23 based on another particle list by} \cite{2013SoPh..282..579M}. In order to construct a proton event sample based on exploring the data itself and independently on proton identifications using other instruments, all enhancements observed in the Wind/EPACT proton channels are re-examined. We adopted a 3-sigma criterion\footnote{As a confidence measure, one selects a threshold on the proton intensity in order to identify a new proton event, {\it e.g.} applies a factor over the standard deviation (sigma) evaluated usually during quiet-time periods.} to identify individual proton enhancements as an intermediate between the 2-sigma \citep{2003ICRC....6.3305T} and 4-sigma \citep{1999ApJ...519..864K} selections. The preliminary on-line version of the proton catalog (1996$-$2015) was presented by \cite{2016simi.conf...27M} at \url{http://www.stil.bas.bg/SEPcatalog/} and used for the investigation the SC trends of protons \citep{2017SunGe..12...11M}. We subsequently re-analyzed complex proton profiles and identified all conspicuous enhancements as new proton injections. The data analysis was extended to include 2016. This finalized proton list was explored for the purpose of comparative analyses with other catalogs over SC23 by \cite{2017JASTP...M}. In the present study, the entire proton catalog (1996$-$2016) is considered and both the standard and newly proposed statistical analysis are applied, namely between the proton peak intensity (and fluence) and the properties of the solar source of the particles, {\it e.g.}, SFs (intensity, fluence and location) and CMEs (projected speed and angular width [AW]). Taking advantage of the catalog coverage (1996$-$2016), we also complete an eight-year comparative study in SC23 and SC24 of protons and related solar activity phenomena and report the findings as function of event intensity, location and solar cycle. 

In Section 2 the data and techniques used in this work are described, separately for the protons, their solar origin, and the criteria for the association. Results are presented and discussed in Section 3 and cover the properties of the proton and proton-related flare/CME sample, the differences in the first eight years from SC23 and SC24, the occurrence probabilities for the protons events and the statistical analysis in terms of linear and partial correlations. The main findings are summarized in Section~\ref{S-Conclusion}.

\section{Data and techniques}
\label{S-Data}

In this section the proton and solar data used, as well as the analysis performed are described. 

\subsection{Proton data}
\label{S-Proton}

\begin{figure}[t!]    
\centerline{\includegraphics[width=0.99\textwidth,clip=]{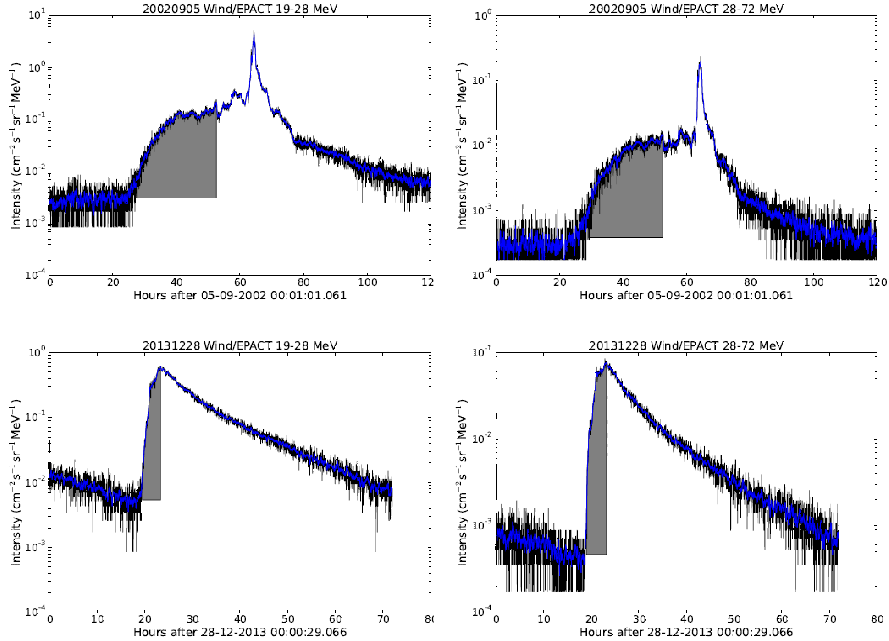}}
\caption{Example profiles of Wind/EPACT proton events for the low (left) and high (right) energy channel. The proton intensity is given in log-scale. The onset-to-peak fluence is calculated over the shaded area. The blue curve (see on-line version) denotes the smoothed data.}
   \label{F-SEPexample}               
\end{figure}

For this study the Wind/EPACT proton catalog is used, based on omni-directio\-nal Wind/EPACT data provided by the CDAWeb\footnote{\url{http://cdaweb.gsfc.nasa.gov/istp_public/}} database with 92-second time resolution. Two observers visually inspected over 20 years of data (1996 to 2016), and identified independently all proton enhancements in the two energy ranges available, 19$-$28 MeV (denoted further as low energy channel, $\sim$25 MeV) and 28$-$72 MeV (as high energy channel, $\sim$50 MeV). Example plots are given in Fig.~\ref{F-SEPexample}. The finalized Wind/EPACT proton catalog used in this study contains 429 events in the low energy channel and 397 in the high energy channel in the period 1996$-$2016.

When compiling the Wind/EPACT catalog we started by manually selecting a pre-event (quiet-time) interval, individually chosen for each case. An averaged value of the proton intensity within this interval is calculated (referred to as background level). A smoothing over five data points is applied over the data due to the large amount of noise present especially for the low intensity proton events that are affected for the most by different instrumental and sampling issues. The uncertainty value for the so-identified time markers (e.g., onset and peak times) is within $\pm$8 min\footnote{Since an averaging over 5 data points (which are 92 seconds apart from each other) is performed, the final time resolution of the smoothed data is about 7.7 minutes.} which is sufficient for the objectives of the present study. Then, a routine calculates the onset time of the proton event when the intensity reaches three standard deviations (sigmas) above the background level\footnote{Occasionally, due to large background intensity fluctuations, the 3-sigma level requirement cannot be fulfilled and no onset time is provided, thus a notation `N/A' is used.}. The peak proton intensity value ($J_{\rm p}$) is read from the smoothed data and the values are reported after subtraction of the pre-event background (namely the amplitude of the proton event is used). Furthermore, the proton intensity is integrated from the so-identified onset time to its peak, termed further onset-to-peak proton fluence ($F_{\rm p}$). Contributions from local proton acceleration in the IP space (due to the shock arrival at Earth, termed energetic storm particles, ESPs) are easy to identify and are excluded from the analysis (see the spike-like increase in Fig.~\ref{F-SEPexample}, upper plots).

\begin{figure}[t!]   
 \centerline{\includegraphics[width=0.99\textwidth,clip=]{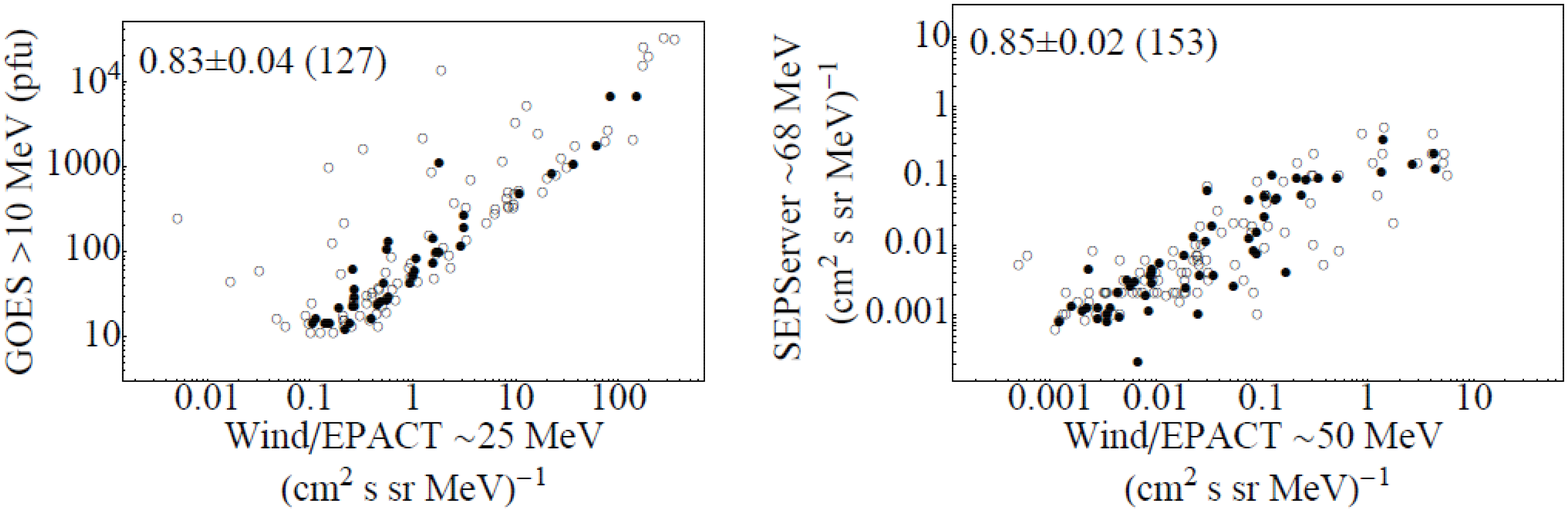}}
\caption{Scatter plot between the peak proton intensity from Wind/EPACT $\sim$25 MeV with GOES $>$10 MeV (on the left) and Wind/EPACT $\sim$50 MeV with SEPServer $\sim$68 MeV events (right). Open circles denote the proton events in SC23 (1996$-$2008) and filled circles are for SC24 (2009$-$2016). In each case, the correlation coefficients and the uncertainties, given on each plot, are calculated for the entire sample.}
   \label{F-scatter}               
\end{figure}

The trajectory of the Wind satellite during the period after its launch is constantly changing\footnote{The Wind orbit can be viewed by the orbit plotting tool provided at \url{http://cdaweb.gsfc.nasa.gov/cgi-bin/gif_walk?plot_type=wind_orbit}}, and only after mid-2004 the spacecraft follows an orbit around L1. In order to cross-check the detectability of a SEP event by Wind/EPACT during the period of observations considered in this work (1996$-$ 2016), we compared the list of proton events identified from Wind and from another spacecraft, either at geostationary orbit ({\it e.g.}, GOES), or at L1 ({\it e.g.}, SOHO). Namely, at low energies, the reported GOES proton events at $>$10 MeV proton flux and the low energy channel from Wind/EPACT $\sim$25 MeV are compared. The scatter plot is given in Figure~\ref{F-scatter}, left, and the linear correlation coefficient for the entire sample (127 events with both Wind and GOES enhancements) is 0.83$\pm$0.04. Alternatively, the high energy channel $\sim$50 MeV Wind/EPACT channel is compared with $\sim$68 MeV SOHO/ERNE proton events from SEPServer and the plot is given in Figure~\ref{F-scatter}, right (0.85$\pm$0.02 for 153 events in common). Different symbols denote data in the two solar cycles\footnote{The correlations in SC23 (1996$-$2008) are 0.80$\pm$0.05 at low energies (88 events) and 0.83$\pm$0.03 at high energies (39 events). These values are consistent with the reported scatter-plot results in \inlinecite{2017SunGe..12...11M}, based on the preliminary results of Wind/EPACT catalog. The correlations in SC24 (2009$-$2016) are 0.83$\pm$0.03 (99 events) and 0.88$\pm$0.03 (54 events), respectively for low and high energy data samples.}, namely open circles are used in SC23 (1996$-$2008) and filled circles for SC24 (2009$-$2016)\footnote{Solar cycle 24 is still ongoing, see sunspot data provided by the World Data Center SILSO, Royal Observatory of Belgium, Brussels: \url{http://www.sidc.be/silso/}.}. The high correlations of the two samples denote the good proton event correspondence observed by different SEP instruments, despite the Wind satellite motion. A larger scatter is observed for the low energy sample in SC23 (open circles). The GOES proton catalog inaccuracy\footnote{Since the GOES proton catalog automatically identifies the highest proton intensities, occasionally, energetic storm particles (due to local shock passages) are erroneously reported as SEP events.} contributes to the spread, since the scatter is visually less for the Wind high energy channel for the same period (compare the distribution of the open circles in both plots).

The proton events as identified from the Wind/EPACT instrument in two available energy channels and used in this study over the period 1996$-$2016 are listed in Table~\ref{T-datatable}. For the low energy channel the peak time and peak intensity is reported by us, whereas for the high energy channel we present the peak intensity only. The associated flares (class, onset time and location) and CMEs (time of first appearance, projected speed and angular width) are also listed in the respective columns. An ASCII version of the table is also available as supplementary material.

\setlength{\tabcolsep}{5pt}

\begin{table}[ht!]
\caption[]{List of Wind/EPACT proton events and their solar origin, flares and CMEs, in the period 1996$-$2016. Time in UT; peak proton intensity $J$ in (cm$^2$ sr s MeV)$^{-1}$; peak time is for the 25 MeV protons; speed in km$\,$s$^{-1}$; angular width (AW) in degrees; dp: data problem; nd: next day; pd: previous day; v: visual estimation; u: uncertain; -: no SEP event.}
\label{T-datatable}
\tiny
\vspace{0.01cm}
\begin{tabular}{lllll}
\hline
Event date     & \multicolumn{2}{c}{Proton event}           &      Flare SXR         &     CME         \\
yyyy-mm-dd     & peak/$J_{\rm 25\,MeV}$ & $J_{\rm 50\,MeV}$ & class/onset/location   & time/speed/AW    \\
(1)            &  (2)                   & (3)               &  (4)                   & (5)             \\
\hline
1996-10-04 & 12:15/0.3006 & 0.0042 & uncertain         & 03:15/255/28       \\
1996-11-30 & 12:32/0.0089 & -      & M1.0/20:16/S06W47 & uncertain \\ 
1996-12-24 & 15:58/0.0092 & 0.001  & C2.1/13:03/N05W95$^{\rm v}$ & 13:29/325/69 \\ 
1997-04-07 & 23:39/0.0118 & 0.0014 & C6.8/13:50/S30E19 & 14:27/878/360 \\
1997-05-12 & 12:05/0.0202 & 0.0031 & C1.3/04:42/N21W09 & 05:30/464/360 \\
1997-05-21 & 00:33$^{\rm nd}$/0.004  & 0.0006 & M1.3/20:03/N05W12 & 21:01/296/165 \\
1997-07-25 & 00:43$^{\rm nd}$/0.0083 & 0.0009 & uncertain & 21:01/611/84 \\ 
1997-09-24 & 07:28/0.0059 & 0.0017 & M5.9/02:43/S31E19 & 03:38/532/76 \\
1997-09-24 & 11:11/0.0048 & 0.0013 & uncertain & uncertain \\
1997-10-07 & 16:39/0.0108 & 0.0012 & uncertain & 13:30/1271/167 \\
1997-11-04 & 13:34/0.9518 & 0.2148 & X2.1/05:52/S14W33 & 06:10/785/360 \\
1997-11-06 & 02:19$^{\rm nd}$/18.13 & 2.933 & X9.4/11:49/S18W63 & 12:11/1556/360 \\
1997-11-13 & 02:25$^{\rm nd}$/0.0365 & 0.006 & uncertain & 22:26/546/288 \\
1997-11-14 & 16:47/0.0101 & 0.0014 & C4.6/09:05/N21E70 & 10:14/1042/86 \\
1998-01-26 & 04:21$^{\rm nd}$/0.005 & 0.0005 & M1.3/21:26$^{\rm pd}$/N22E53 & 22:20$^{\rm pd}$/596/84 \\
1998-04-05 & 08:32$^{\rm nd}$/0.008 & 0.0007 & uncertain & uncertain \\
1998-04-20 & 04:59$^{\rm nd}$/37.6 & 5.628 & M1.4/09:38/S40W90$^{\rm v}$ & 10:07/1863/243 \\
1998-04-30 & 16:31/0.0173 & 0.0024 & M6.8/16:06$^{\rm pd}$/S18E20 & 16:59$^{\rm pd}$/1374/360 \\
1998-05-02 & 17:28/1.423  & 0.3025 & X1.1/13:31/S15W15 & 14:06/938/360 \\ 
1998-05-06 & 09:19/5.162  & 0.8831 & X2.7/07:58/S11W65 & 08:29/1099/190 \\ 
1998-05-09 & 12:17/0.0964 & 0.0191 & M7.7/03:04/S15W95$^{\rm v}$ & 03:36/2331/178 \\
1998-05-09 & 22:05/0.1334 & 0.0136 & uncertain & 15:18/533/46 \\
1998-05-31 & 09:18/0.004  & 0.0008 & uncertain & 15:54$^{\rm pd}$/367/92 \\ 
1998-06-04 & 21:49/0.0123 & 0.0006 & uncertain & 02:05/1802/360 \\
1998-06-17 & 17:38/0.0336 & 0.0034 & M1.0/18:03$^{\rm pd}$/S20W95$^{\rm v}$ & 18:27$^{\rm pd}$/1484/281 \\
1998-07-24 & 23:49/0.0068 & -      & uncertain & data gap \\
1998-08-19 & 10:22/0.0056 & 0.001  & X4.9/22:10$^{\rm pd}$/N33E87 & data gap \\
1998-08-20 & 07:51/0.0127 & 0.0032 & X3.9/21:35$^{\rm pd}$/N32E75 & data gap \\
1998-08-22 & 01:34$^{\rm nd}$/0.0326 & 0.0023 & M9.0/23:57$^{\rm pd}$/N42E51 & data gap \\
1998-08-24 & 03:59$^{\rm nd}$/3.609  & 0.5184 & X1.0/21:50/N31E08 & data gap \\
1998-09-06 & 10:03/0.02   & 0.003  & C2.6/05:56/u      & data gap \\
1998-09-09 & 07:01/0.0077 & 0.0009 & M2.8/04:52/u      & data gap \\
1998-09-23 & 23:31/0.0167 & 0.0038 & M7.1/06:40/N18E09 & data gap \\
1998-09-27 & 10:20/0.0069 & -      & C2.6/08:06/N21W48 & data gap \\
1998-09-30 & 22:29/27.52  & 2.896  & M2.8/13:08/N23W81 & data gap \\
1998-10-18 & 01:57$^{\rm nd}$/0.0529 & 0.0094 & X4.9/22:10/N33E87 & data gap \\
1998-10-19 & 06:30/0.0822 & 0.0099 & uncertain & data gap \\
1998-10-21 & 10:28/0.0042 & -      & uncertain & data gap \\
1998-11-06 & 10:26/0.0405 & 0.0012 & M8.4/19:00$^{\rm pd}$/N22W18 & 20:44$^{\rm pd}$/118/360 \\
1998-11-07 & 13:31/0.0442 & 0.0048 & M2.4/11:02/N20W50 & 11:54/632/360 \\
1998-11-07 & 02:15$^{\rm nd}$/0.1007 & 0.0061 & C2.4/20:07/S20W67 & 20:54/750/96 \\
1998-11-14 & 14:21/6.314  & 1.117  & C1.7/05:15/N20W60 & data gap \\ 
1998-11-22 & 09:08/0.0609 & 0.0145 & X3.7/06:30/S27W82 & data gap  \\
1998-11-24 & 10:32/0.0315 & 0.0078 & X1.0/02:07/S25W95$^{\rm v}$ & 02:30/1798/360 \\
1999-01-20 & 03:40$^{\rm 22nd}$/0.0962 & 0.0168 & M5.2/19:06/u & data gap  \\
1999-02-16 & 17:23/0.0038 & -      & uncertain & data gap \\
1999-03-29 & 06:21/17.85  & 0.0005 & uncertain & uncertain \\
1999-04-24 & 22:56/0.4006 & 0.0308 & uncertain & 13:31/1495/360 \\
1999-05-04 & 00:30$^{\rm nd}$/0.0259 & 0.0013  & M4.4/05:36$^{\rm pd}$/N15E32 & 06:06$^{\rm pd}$/1584/360 \\
1999-05-09 & 20:18/0.0395 & 0.0047 & M7.6/17:53/N19W86 & 18:28/615/172 \\
1999-05-27 & 13:00/0.1536 & 0.0233 & uncertain & 11:06/1691/360\\
\hline
\end{tabular}
\end{table}

\begin{table}[ht!]
\addtocounter{table}{-1}
\caption[]{cont'd.}
\tiny
\vspace{0.01cm}
\begin{tabular}{lllll}
\hline
(1)            &  (2)                   & (3)               &  (4)                   & (5)             \\
\hline
1999-06-01 & 08:46$^{\rm nd}$/0.9889 & 0.1139 & C1.2/18:53/u & 19:38/1772/360 \\
1999-06-04 & 16:27/0.9046 & 0.0764 & M3.9/06:52/N17W69 & 07:27/2230/150 \\
1999-06-11 & 02:22/0.0814 & 0.0141 & uncertain & 01:27/719/101 \\ 
1999-06-27 & 15:24/0.0059 & 0.0006 & M1.0/08:34/N23W25 & 09:06/903/86 \\
1999-09-14 & 11:34/0.0033 & -      & C2.6/16:30$^{\rm pd}$/N15E06 & 17:31$^{\rm pd}$/444/184 \\
1999-12-28 & 05:29/0.0077 & 0.0008 & M4.5/00:39/N20W56 & 00:54/672/82 \\
2000-01-09 & 01:32$^{\rm nd}$/0.0265 & 0.0021 & C1.1/22:17$^{\rm pd}$/N27E53 & data gap \\
2000-01-18 & 01:47$^{\rm nd}$/0.0201 & 0.0049 & M3.9/17:07/S19E11  & 17:54/739/360 \\
2000-01-19 & 07:29$^{\rm nd}$/0.0126 & 0.002  & uncertain & 02:54/800/68 \\
2000-02-12 & 10:04/0.0451 & 0.0045 & M1.7/03:51/N26W23 & 04:31/1107/360 \\
2000-02-17 & 03:10$^{\rm nd}$/0.0196 & 0.0034 & M1.3/20:17/S29E07 & 21:30/728/360 \\
2000-02-18 & 10:47/0.4559 & 0.0789 & C1.1/09:21/S16W78 & 09:54/890/118 \\
2000-03-02 & 15:21/0.0112 & 0.0023 & X1.1/08:20/S14W66 & 08:54/776/62 \\
2000-03-03 & 05:41/0.0074 & 0.0013 & M3.8/02:08/S15W60 & 02:30/841/98 \\
2000-03-22 & 23:27/0.0101 & 0.0014 & X1.1/18:34/N14W57 & 19:31/478/154 \\
2000-03-23 & 13:18/0.0043 & 0.0005 & C7.5/07:56/N19W39 & data gap \\ 
2000-04-04 & 07:38$^{\rm nd}$/0.5441 & 0.0183 & C9.7/15:12/N16W66  & 16:33/1188/360 \\ 
2000-04-23 & 08:45$^{\rm nd}$/0.014  & 0.0033 & uncertain & 12:54/1187/360 \\
2000-04-27 & 19:15/0.0035 & -      & uncertain & 14:30/1110/138 \\
2000-05-01 & 11:46/0.0045 & 0.0005 & M1.1/10:16/N20W60$^{\rm v}$ & 10:54/1360/54 \\ 
2000-05-04 & 19:21/0.0043 & 0.0005 & M6.8/10:57/S15W90$^{\rm v}$ & 11:26/1404/170 \\
2000-05-06 & 03:08$^{\rm nd}$/0.0066 & 0.0006 & M1.5/15:19$^{\rm pd}$/S15W95  &  15:50$^{\rm pd}$/1594/360 \\
2000-05-08 & 05:51/0.0074 & -      & uncertain & 20:50$^{\rm pd}$/1781/360 \\
2000-05-15 & 01:11$^{\rm nd}$/0.0167 & 0.0013 & C7.8/15:46/S24W67 & 16:26/1212/165 \\
2000-05-18 & 02:38/0.0062 & 0.0003 & uncertain & 19:27$^{\rm pd}$/777/109 \\
2000-06-06 & 09:01$^{\rm 8th}$/0.6162 & 0.0172 & X2.3/14:58/N20E18 & 15:54/1119/360 \\
2000-06-10 & 19:40/1.608 & 0.307 & M5.2/16:40/N22W38 & 17:08/1108/360 \\
2000-06-17 & 09:34/0.0087 & 0.0009 & M3.5/02:25/N22W72 & 03:28/857/133 \\
2000-06-18 & 06:30/0.0405 & 0.0039 & X1.0/01:52/N23W85 & 02:10/629/132 \\
2000-06-23 & 17:08/0.017  & 0.0016 & M3.0/14:18/N26W72 & 14:54/847/198 \\
2000-06-25 & 22:47/0.0327 & 0.0015 & M1.9/07:17/N16W55 & 07:54/1617/165 \\
2000-07-11 & 09:17$^{\rm nd}$/0.0088 & 0.0011 & M5.7/21:05$^{\rm pd}$/N18E49 & 21:50$^{\rm pd}$/1352/289 \\
2000-07-14 & 15:24$^{\rm nd}$/174.3 & 13.85 & X5.7/10:03/N22W07 & 10:54/1674/360 \\
2000-07-22 & 15:26/0.3039 & 0.0415 & M3.7/11:17/N14W56 & 11:54/1230/229 \\
2000-07-22 & 20:56/0.0962 & 0.0056 & uncertain & uncertain \\
2000-07-28 & 04:39/0.1457 & 0.0154 & uncertain & 19:54$^{\rm pd}$/905/360 \\
2000-07-28 & 06:40/0.4932 & 0.0248 & uncertain & uncertain \\
2000-07-28 & 09:43/0.4393 & 0.024  & C4.2/00:00/N10W95 & 00:54/447/57 \\
2000-07-28 & 12:25/0.3881 & 0.0163 & uncertain & uncertain \\
2000-07-28 & 22:45/0.0261 & 0.0011 & uncertain & 18:30/864/163 \\
2000-08-11 & 15:17/0.0892 & 0.0013 & C3.2/07:13/N26E66 & 07:31/1071/70 \\
2000-08-11 & 16:43/0.1905 & 0.0022 & uncertain & uncertain \\
2000-08-12 & 12:15/0.022  & 0.0022 & M1.1/09:45/S15W95$^{\rm v}$ & 10:35/662/168 \\
2000-08-12 & 13:36/0.0454 & 0.0027 & uncertain & uncertain \\
2000-08-12 & 18:35/0.0096 & 0.0007 & C3.2/13:48/N11W59 & 14:54/876/161 \\
2000-08-13 & 04:53/0.0076 & -      & uncertain & 15:54$^{\rm pd}$/499/117 \\
2000-08-13 & 07:37$^{\rm nd}$/0.0096 & -  & C1.7/05:10/S14E39 & 06:06/883/154 \\
2000-09-07 & 04:20$^{\rm nd}$/0.005  & 0.0004 & C7.2/20:32/N06W47 & 21:30/422/169 \\
2000-09-09 & 15:57/0.0086 & 0.0008 & M1.6/08:28/N07W67 & 08:57/554/180 \\
2000-09-12 & 03:53$^{\rm nd}$/3.257 & 0.1634 & M1.0/11:31/S17W09 & 11:54/1550/360 \\
2000-09-19 & 19:41/0.0114 & 0.0011 & M5.1/08:06/N14W46 & 08:50/766/76 \\
2000-10-10 & 09:34$^{\rm nd}$/0.0075 & 0.0008 & C6.7/23:19$^{\rm pd}$/N01W14 & 00:26/506/360  \\
2000-10-16 & 15:50/0.3778 & 0.0536 & M2.5/06:40/N05W95$^{\rm v}$ & 07:27/1336/360 \\
2000-10-25 & 01:57$^{\rm nd}$/0.2115 & 0.0189 & C4.0/08:45/N10W80$^{\rm v}$ & 08:26/770/360 \\
2000-10-29 & 23:26/0.0114 & 0.0006 & uncertain & data gap \\
2000-10-31 & 15:45/0.0333 & 0.0015 & uncertain & 18:07$^{\rm pd}$/785/106 \\ 
2000-11-08 & 00:07$^{\rm 10th}$/173.2 & 14.44 & M7.4/22:42/N10W77 & 23:06/1738/170 \\
\hline
\end{tabular}
\end{table}

\begin{table}[ht!]
\addtocounter{table}{-1}
\caption[]{cont'd.}
\tiny
\vspace{0.01cm}
\begin{tabular}{lllll}
\hline
(1)            &  (2)                   & (3)               &  (4)                   & (5)             \\
\hline
2000-11-24 & 12:38/0.1501 & 0.0303 & X2.0/04:55/N20W05 & 05:30/1289/360 \\
2000-11-24 & 20:52/1.643  & 0.206  & X2.3/14:51/N22W07 & 15:30/1245/360  \\
2000-11-26 & 20:53/31.41  & 1.752  & M8.2/00:59$^{\rm pd}$/N07E50 & 01:31$^{\rm pd}$/2519/360  \\
2000-12-28 & 22:09/0.0128 & 0.0021 & uncertain & 03:30/524/199  \\
2000-12-29 & 14:19/0.0055 & 0.0007 & uncertain & 12:06/930/360  \\
2001-01-05 & 00:49$^{\rm nd}$/0.0285 & 0.0032 & uncertain & 17:06/828/360 \\
2001-01-21 & 07:26/0.0052 & 0.0014 & uncertain & uncertain \\
2001-01-21 & 12:48/0.0038 & 0.0009 & uncertain & uncertain \\
2001-01-21 & 18:37$^{\rm nd}$/0.0399 & 0.0042 & C3.0/19:17/S08E36 & 19:54/664/213  \\
2001-01-28 & 06:11$^{\rm nd}$/0.9456 & 0.1071 & M1.5/15:40/S04W59 & 15:54/916/360  \\
2001-02-11 & 09:41/0.0115 & 0.0011 & C6.5/00:57/N24W57 & 01:32/1183/360  \\
2001-02-26 & 13:54/0.0094 & 0.0006 & uncertain & 05:30/851/152  \\
2001-03-26 & 08:42$^{\rm nd}$/0.0163 & - & C9.0/16:25$^{\rm pd}$/N16E25 & 17:06$^{\rm pd}$/677/360 \\
2001-03-29 & 02:40$^{\rm nd}$/0.6266 & 0.0799 & X1.7/09:57/N20W19 & 10:26/942/360  \\
2001-04-02 & 12:26/0.2886 & 0.0224 & X1.4/10:04/N17W60 & uncertain \\ 
2001-04-02 & 16:38/0.0542 & 0.0082 & X1.1/10:58/N20W60$^{\rm v}$ & 11:26/992/80  \\
2001-04-03 & 06:15/7.389  & 4.012  & X20/21:32$^{\rm pd}$/N20W75$^{\rm v}$ & 22:06$^{\rm pd}$/2505/244  \\
2001-04-09 & 22:12/0.0892 & 0.026  & M7.9/15:20/S21W04 & 15:54/1192/360  \\
2001-04-10 & 09:45$^{\rm nd}$/2.499 & 0.2863 & X2.3/05:06/S23W09 & 05:30/2411/360  \\
2001-04-12 & 20:55/0.6537 & 0.1586 & X2.0/09:39/S19W43 & 10:31/1184/360  \\
2001-04-15 & 19:01/31.3   & 5.124  & X14.4/13:19/S20W85 & 14:07/1199/167 \\
2001-04-18 & 11:05/8.5    & 1.377  & C2.2/02:11/S15W90$^{\rm v}$ & 02:30/2465/360  \\
2001-04-27 & 01:57$^{\rm nd}$/0.0318 & - & M7.8/11:26$^{\rm pd}$/N17W31 & 12:30$^{\rm pd}$/1006/360 \\
2001-05-07 & 00:38$^{\rm nd}$/0.3542 & 0.0248  & C3.9/11:36/N26W35 & 12:06/1223/205  \\
2001-05-20 & 14:55/0.1787 & 0.038  & M6.4/06:00/S20W95$^{\rm v}$   & 06:26/546/179  \\
2001-06-04 & 20:56/0.0361 & 0.0043 & C3.2/16:11/N24W59 & 16:30/464/89 \\
2001-06-11 & 01:05/0.0039 & -      & uncertain & uncertain \\
2001-06-12 & 09:02/0.0043 & -      & uncertain & uncertain \\
2001-06-15 & 18:35/0.6736 & 0.1097 & uncertain & 15:56/1701/360 \\
2001-06-19 & 06:27/0.0091 & 0.0025 & C4.7/21:50$^{\rm pd}$/u  & 22:47$^{\rm pd}$/1301/59 \\
2001-06-19 & 11:55/0.0058 & 0.0012 & uncertain & 03:54/817/58 \\   
2001-08-09 & 03:48$^{\rm nd}$/0.04 & 0.0011 & uncertain & 10:30/479/175  \\
2001-08-10 & 11:17/0.2071 & 0.0089 & C1.4/20:35$^{\rm pd}$/u &  21:30$^{\rm pd}$/909/100 \\
2001-08-14 & 23:34/0.0107 & 0.0008 & C2.3/11:30/N25W25$^{\rm v}$ & 12:00$^{\rm v}$/618/360  \\
2001-08-16 & 04:04/8.382  & 2.71   & uncertain & 23:54$^{\rm pd}$/1575/360 \\
2001-09-15 & 15:23/0.1664 & 0.0185 & M1.5/11:04/S21W49 & 11:54/478/130  \\
2001-09-18 & 09:13/0.0034 & 0.0007 & uncertain & uncertain \\
2001-09-18 & 21:30/0.0054 & 0.0005 & uncertain & 18:33/376/253  \\
2001-09-19 & 18:19/0.0045 & 0.0005 & C7.2/06:59/S20W95$^{\rm v}$ & 07:32/416/210  \\
2001-09-24 & 15:24$^{\rm nd}$/1.88 & 0.2002 & X2.6/09:32/S16E23 & 10:31/2401/360 \\
2001-10-01 & 22:18/16.34  & 1.745  & M9.1/04:41/S30W95$^{\rm v}$ & 05:30/1405/360  \\
2001-10-06 & 07:36$^{\rm nd}$/0.118 & 0.009 & uncertain  & 10:30$^{\rm pd}$/1537/360 \\
2001-10-09 & 09:53/0.0133 & 0.0013 & C4.3/08:09$^{\rm pd}$/N25E95 & 08:30$^{\rm pd}$/845/138 \\
2001-10-09 & 00:58$^{\rm nd}$/0.0351 & 0.0028 & M1.4/10:46/S28E08 & 11:30/973/360  \\
2001-10-19 & 05:57/0.1249 & 0.0238 & X1.6/00:47/N16W18 & 01:27/558/360 \\ 
2001-10-19 & 20:17/0.1236 & - & X1.6/16:13/N15W29 & 16:50/901/360 \\   
2001-10-22 & 22:36/0.3525 & 0.067 & M6.7/14:27/S21E18 & 15:06/1336/360 \\
2001-11-04 & 20:24$^{\rm nd}$/277.3  & 29.42  &  X1.0/16:03/N06W18  & 16:35/1810/360  \\
2001-11-17 & 07:41$^{\rm nd}$/0.0412 & 0.0029 &  M2.8/04:49/S13E42  & 05:30/1379/360  \\
2001-11-22 & 21:28$^{\rm nd}$/196.7  & 16.65  &  M3.8/20:18/S25W67  & 20:31/1443/360  \\
2001-12-16 & 11:22/0.0077 & 0.0009 & M1.5/01:14/N15W23 & 02:06/343/66 \\
2001-12-26 & 10:48/24.47  & 4.107  & M7.1/04:32/N08W54 & 05:30/1446/212  \\
2001-12-30 & 14:14$^{\rm nd}$/1.972 & 0.1166 & uncertain & uncertain \\
2002-01-10 & 01:53$^{\rm nd}$/1.667 & 0.0893 & uncertain & uncertain \\  
2002-01-15 & 23:02/0.2095 & 0.0245 & M4.4/05:29$^{\rm pd}$/S20W80$^{\rm v}$  & 05:35$^{\rm pd}$/1492/360   \\
2002-01-27 & 17:02/0.2095 & 0.0254 & uncertain &  12:30/1136/360  \\ 
2002-02-20 & 07:27/0.2128 & 0.0264 & M5.1/05:52/N12W72 & 06:30/952/360   \\
\hline
\end{tabular}
\end{table}

\begin{table}[ht!]
\addtocounter{table}{-1}
\caption[]{cont'd.}
\tiny
\vspace{0.01cm}
\begin{tabular}{lllll}
\hline
(1)            &  (2)                   & (3)               &  (4)                   & (5)             \\
\hline  
2002-03-16 & 03:38/0.0054 & 0.0008 & M2.2/22:09$^{\rm pd}$/S08W03  & 23:06$^{\rm pd}$/957/360   \\
2002-03-16 & 14:02/0.0129 & 0.0009 & uncertain & uncertain \\
2002-03-17 & 08:54/0.0574 & 0.0010 & uncertain & uncertain \\
2002-03-18 & 11:27/0.0488 & 0.007  & M1.0/02:16/S15W35$^{\rm v}$ & 02:54/989/360   \\
2002-03-18 & 23:37/0.1983 & 0.019  & uncertain & uncertain \\
2002-03-19 & 06:55/0.5505 & 0.0263 & uncertain & uncertain \\
2002-03-22 & 21:27/0.0467 & 0.0007 & M1.6/10:12/S20W90$^{\rm v}$ & 11:06/1750/360  \\
2002-04-11 & 22:35/0.0172 & 0.0014 & C9.2/16:16/S15W33 & 16:50/540/70   \\
2002-04-14 & 00:17$^{\rm nd}$/0.007 & 0.0007 & C9.6/07:28/N19W57 & 07:50/757/76   \\
2002-04-17 & 14:33/0.1024 & 0.0094 & M2.6/07:46/S14W34  & 08:26/1240/360   \\
2002-04-19 & 16:24/0.0161 & 0.0003 & uncertain & uncertain \\
2002-04-21 & 22:52/79.07  & 10.84  &  X1.5/00:43/S14W84  & 01:27/2393/360   \\
2002-05-01 & 04:15/0.0239 & -      &  uncertain &  23:26$^{\rm pd}$/1103/199 \\
2002-05-20 & 19:28/0.0049 & 0.0006 &  X2.1/15:21/S21E65  & 15:50/553/69   \\
2002-05-22 & 04:15$^{\rm nd}$/1.517 & 0.0825 & C5.0/03:18/S20W85$^{\rm v}$  & 03:50/1557/360  \\ 
2002-07-07 & 22:33/0.4989 & 0.0553 &  M1.0/11:15/S20W90$^{\rm v}$ & 11:30/1423/228  \\
2002-07-10 & 06:33/0.0154 & 0.0011 & uncertain & 19:32$^{\rm pd}$/1079/360 \\
2002-07-16 & 01:18$^{\rm nd}$/0.0051 & 0.0006 & X3.0/19:59$^{\rm pd}$/N19W01 & 20:30$^{\rm pd}$/1151/360  \\
2002-07-22 & 06:59$^{\rm nd}$/0.3956 & 0.0327 & uncertain & uncertain \\
2002-08-04 & 01:46/0.0048 & -      & X1.0/18:59$^{\rm pd}$/S16W76  & 19:32$^{\rm pd}$/1150/138   \\
2002-08-05 & 12:17/0.0053 & -      & C4.8/04:21/S10W43 & 07:31/689/43 \\
2002-08-14 & 03:25/0.0206 & 0.0022 & M2.3/01:47/N09W54 & 02:30/1309/133 \\
2002-08-14 & 08:28/0.2654 & 0.0260 & uncertain & uncertain \\
2002-08-14 & 12:18/0.0702 & 0.0021 & uncertain & uncertain \\
2002-08-14 & 15:47/0.1452 & 0.0009 & uncertain & uncertain \\
2002-08-16 & 11:32/0.0076 & 0.0014 & M2.4/05:46/N07W83 & 06:06/1378/162 \\
2002-08-16 & 09:35$^{\rm nd}$/0.0349 & 0.0013 & M5.2/11:32/S14E20   & 12:30/1585/360 \\
2002-08-18 & 02:02$^{\rm nd}$/0.0585 & 0.0074 & M2.2/21:12/S12W19   & 21:54/682/140 \\
2002-08-19 & 11:47/0.031  & 0.0021 & uncertain & uncertain \\
2002-08-20 & 10:08/0.0297 & 0.0045 & M3.4/08:22/S10W38 & 08:55/1099/122 \\
2002-08-22 & 08:11/0.4628 & 0.1076 & M5.4/01:47/S07W62 & 02:06/998/360  \\
2002-08-24 & 08:02/9.4    & 1.421  & X3.1/00:49/S02W81 & 01:27/1913/360  \\
2002-09-06 & 04:35/0.2118 & 0.0145 & C5.2/16:18$^{\rm pd}$/N09E28 & 16:54$^{\rm pd}$/1748/360  \\ 
2002-09-27 & 03:14/0.0034 & 0.0003 & C5.0/01:18/u  & 01:54/1502/59 \\
2002-10-30 & 20:58$^{\rm nd}$/0.0207 & 0.0029 & uncertain & uncertain \\
2002-11-09 & 02:41$^{\rm nd}$/8.09 & 0.3812 & M4.6/13:08/S12W29 & 13:32/1838/360   \\
2002-12-19 & 01:26$^{\rm nd}$/0.1002 & 0.0102 & M2.7/21:34/N15W09 & 22:06/1092/360   \\
2002-12-22 & 19:57/0.0096 & 0.0011 & M1.1/02:14/N23W42 & 03:30/1071/272 \\
2003-03-17 & 22:07/0.0114 & 0.0005 & X1.5/18:50/S14W39 & 19:54/1020/96  \\
2003-03-18 & 18:48/0.008  & 0.0011 & X1.5/11:51/S15W46 & 12:30/1601/209  \\
2003-04-07 & 22:38/0.0077 & 0.0004 & B7.3/09:25/S11W80 & 09:50/719/69  \\
2003-04-21 & 21:37/0.0082 & 0.0016 & M2.8/12:54/N18E02 & 13:36/784/163 \\
2003-04-23 & 08:45/0.0146 & 0.0033 & M5.1/00:39/N22W25 & 01:27/916/248  \\
2003-04-24 & 14:16/0.015  & 0.002  & M3.3/12:45/N21W39 & 13:27/609/242 \\
2003-05-28 & 19:39/0.1636 & 0.0193 & X3.6/00:17/S07W20 & 00:50/1366/360  \\
2003-05-30 & 22:32/0.0132 & 0.0005 & C8.6/06:39/S05W55 & 06:26/836/69 \\ 
2003-05-31 & 05:35/0.5543 & 0.1077 & M9.3/02:13/S07W65 & 02:30/1835/360  \\
2003-06-02 & 06:52/0.0048 & 0.0006 & M6.5/00:07/S06W90 & 00:30/1656/172 \\
2003-06-18 & 03:27$^{\rm nd}$/0.2634 & 0.0063 & M6.8/22:27$^{\rm pd}$/S08E58   & 23:18$^{\rm pd}$/1813/360  \\
2003-08-19 & 12:18/0.0051 & 0.0004 & M2.0/07:38/S12W63 & 08:31/412/35 \\
2003-10-22 & 17:32/0.0113 & 0.067  & uncertain & 08:31/719/267 \\
2003-10-23 & 14:01/0.0149 & -      & M9.9/19:47/S18E78 & 20:06/1085/134 \\
2003-10-26 & 04:00$^{\rm nd}$/9.707 & 0.9038 & X1.2/17:21/N02W38 & 17:54/1537/171 \\
2003-10-28 & 10:41$^{\rm nd}$/353.2 & 32.04  & X17.2/09:51/S16E08  & 11:30/2459/360  \\
2003-11-02 & 14:10/0.325  & 0.0271  & uncertain  & 09:30/2036/360  \\
2003-11-02 & 00:24$^{\rm nd}$/50.99 & 7.662 & X8.3/17:03/S14W56  & 17:30/2598/360  \\
2003-11-04 & 06:57$^{\rm nd}$/9.54  & 1.075 & X28/19:29/S19W83  & 19:54/2657/360  \\
\hline
\end{tabular}
\end{table}

\begin{table}[ht!]
\addtocounter{table}{-1}
\caption[]{cont'd.}
\tiny
\vspace{0.01cm}
\begin{tabular}{lllll}
\hline
(1)            &  (2)                   & (3)               &  (4)                   & (5)             \\
\hline
2003-11-10 & 15:05/0.0175 & 0.0017 & uncertain & uncertain \\
2003-11-11 & 05:32/0.0088 & 0.0009 & uncertain & 02:30/1359/360 \\
2003-11-18 & 22:59/0.0111 & 0.0008 & M3.9/08:12/S02E18 & 08:50/1660/360  \\
2003-11-20 & 11:35/0.0694 & 0.0077 & M9.6/07:35/N01W08 & 08:06/669/360  \\
2003-11-21 & 23:49/0.2540 & 0.0114 & M5.8/23:42$^{\rm pd}$/N02W17 & 00:26/494/52 \\
2003-12-02 & 18:03/2.225  & 0.0627 & C7.2/09:40/S19W89 & 10:50/1393/150   \\
2004-02-04 & 16:14/0.0035 & 0.0004 & C9.9/11:12/S05W48 & 11:54/764/33  \\
2004-04-11 & 16:20/0.4638 & 0.0308 & C9.6/03:54/S16W46 & 04:30/1645/314  \\
2004-07-13 & 04:46/0.0186 & 0.0027 & M6.7/00:09/N14W45 & 00:54/607/88  \\
2004-07-22 & 18:29/0.0263 & 0.0023 & C5.8/11:02/N05E11 & 11:54/574/45 \\
2004-07-23 & 02:12/0.0135 & 0.0025 & M1.6/22:40$^{\rm pd}$/N05E04 & 23:54$^{\rm pd}$/448/46 \\
2004-07-23 & 05:01/0.019  & 0.0037 & uncertain & uncertain \\
2004-07-23 & 09:01/0.0349 & 0.005  & uncertain & uncertain \\
2004-07-23 & 18:24/0.0707 & 0.0087 & C2.1/06:41/N03W03 & 07:31/459/138 \\
2004-07-25 & 20:05/1.245  & 0.0667 & M1.1/14:19/N08W33 & 14:54/1333/360  \\
2004-07-30 & 03:15$^{\rm nd}$/0.0213 & 0.0029 & C2.1/11:42$^{\rm pd}$/S07W88 & 12:06$^{\rm pd}$/1180/360  \\
2004-08-18 & 22:59/0.0042 & 0.0005 & X1.8/17:29/S14W90 & 17:54/602/120  \\
2004-09-12 & 19:24/0.0296 & 0.0021 & uncertain & uncertain \\
2004-09-13 & 23:33/6.227  & 0.2109 & uncertain & uncertain \\
2004-09-19 & 23:16/1.339  & 0.1682 & M1.9/16:46/N03W58 & data gap \\
2004-10-30 & 15:37/0.03   & 0.0033 & M4.1/06:08/N13W22 & 06:54/422/360 \\
2004-10-30 & 20:30/0.0276 & 0.0026 & M5.9/16:18/N13W28 & 16:54/690/360 \\
2004-11-01 & 07:49/2.303  & 0.5299 & M1.1/03:04/N12W49 & 03:54/459/192  \\
2004-11-07 & 21:23/10.77  & 0.6909 & X2.0/15:42/N09W17 & 16:54/1759/360 \\
2004-11-10 & 09:29/0.0804 & 0.0066 & X2.5/01:59/N09W49 & 02:26/3387/360  \\
2004-12-03 & 23:26/0.0655 & 0.0034 & M1.5/23:44$^{\rm pd}$/N08W02 & 00:26/1216/360  \\
2005-01-15 & 11:55/0.1771 & 0.0244 & M8.6/05:54/N11E06 & 06:30/2049/360  \\
2005-01-16 & 17:50/12.71  & 1.234  & X2.6/22:25$^{\rm pd}$/N14W08 & 23:07$^{\rm pd}$/2861/360  \\
2005-01-17 & 04:29$^{\rm nd}$/197.4 & 18.06 & X3.8/06:59/N15W25 & 09:54/2547/360 \\
2005-01-20 & 08:53/58.58  & 8.249  & X7.1/06:36/N14W61 & 06:54/882/360 \\
2005-05-06 & 12:09/0.0314 & 0.0009 & C7.8/20:09$^{\rm pd}$/S04W67 & 20:30$^{\rm pd}$/1180/360  \\
2005-05-06 & 17:59/0.0158 & 0.0005 & M1.3/11:11/S04W76 & 11:54/1144/129 \\
2005-05-07 & 10:51/0.01   & 0.0004 & C8.5/16:03$^{\rm pd}$/S09E28 & 17:29$^{\rm pd}$/1128/360         \\
2005-05-11 & 22:38/0.0117 & 0.0011 & M1.1/19:22/S10W47 & 20:13/550/360 \\
2005-05-13 & 20:49$^{\rm nd}$/9.876 & 0.1406 & M8.0/16:13/N12E11 & 17:12/1689/360 \\
2005-06-16 & 03:07$^{\rm nd}$/1.109 & 0.2925 & M4.0/20:01/N08W90 & data gap \\
2005-07-10 & 04:50/0.0447 & 0.0055 & M2.8/21:47$^{\rm pd}$/N12W28 & 22:30$^{\rm pd}$/1540/360  \\
2005-07-10 & 07:05/0.0225 & 0.0046 & C1.6/05:09/N14W33 & 06:30/265/41 \\
2005-07-10 & 10:41/0.0565 & 0.0046 & uncertain & 08:54/835/182 \\
2005-07-12 & 20:57/0.0041 & 0.0004 & M1.5/15:47/N09W67 & 16:54/523/360 \\
2005-07-13 & 03:11$^{\rm nd}$/0.2097 & 0.0165 & M5.0/14:01/N11W90 & 14:30/1423/360  \\
2005-07-14 & 04:01$^{\rm nd}$/3.319  & 0.3018 & M1.2/21:49$^{\rm pd}$/N08W90 & 22:30$^{\rm pd}$/539/34  \\ 
2005-07-17 & 19:22/0.7705 & 0.1043 & uncertain & 11:30/1527/360  \\
2005-07-26 & 17:56$^{\rm nd}$/0.2123 & 0.0272 & uncertain & 04:54/1458/360   \\
2005-07-27 & 13:05$^{\rm nd}$/0.7035 & 0.0515 & M3.7/04:33/N11E90  & 04:54/1787/360 \\  
2005-08-22 & 06:24/0.1792 & 0.0124 & M2.6/00:44/S11W45 & 01:32/1194/360 \\
2005-08-22 & 08:10$^{\rm nd}$/8.385  & 0.5301 &  M5.6/16:46/S13W65  & 17:30/2378/360  \\
2005-08-29 & 00:40$^{\rm nd}$/0.0328 & 0.0066 & uncertain & 10:54/1600/360  \\
2005-09-01 & 03:16$^{\rm nd}$/0.0444 & 0.0092 & uncertain & 22:30$^{\rm pd}$/1808/360  \\
2005-09-07 & 04:19$^{\rm 11th}$/74.61 & 3.684 & X17/17:17/S11E77 & data gap  \\
2005-09-14 & 06:47$^{\rm nd}$/5.515  & 0.255  & X1.5/19:19$^{\rm pd}$/S09E10 & 20:00$^{\rm pd}$/1866/360 \\
2006-07-06 & 15:12/0.0543 & 0.0104 & M2.5/08:13/S09W34  & 08:54/911/360  \\
2006-12-05 & 19:12/0.0812 & 0.0125 & X9.0/10:18/S07E68 & data gap \\
2006-12-06 & 18:05$^{\rm nd}$/138.8 & 7.311   & M6.0/08:02/S04E63 & data gap \\
2006-12-13 & 09:18/20.17  & 5.288 & X3.4/02:14/S06W23 & 02:54/1774/360 \\
2006-12-15 & 02:25/0.519  & 0.1289  & X1.5/21:07$^{\rm pd}$/S06W46 & 22:30$^{\rm pd}$/1042/360 \\
2006-12-15 & 04:56/0.4638 & 0.066  & uncertain & uncertain \\
\hline
\end{tabular}
\end{table}

\begin{table}[ht!]
\addtocounter{table}{-1}
\caption[]{cont'd.}
\tiny
\vspace{0.01cm}
\begin{tabular}{lllll}
\hline
(1)            &  (2)                   & (3)               &  (4)                   & (5)             \\
\hline
2006-12-15 & 08:18/0.0894 & 0.0142 & uncertain & uncertain \\
2007-2009  & \multicolumn{2}{c}{no events} &  - & - \\
2010-06-12 & 08:39/0.0123 & 0.002  & M2.0/00:30/N23W43  & 01:32/486/119  \\
2010-08-03 & 18:25/0.0478 & 0.0014 & uncertain & 11:12/221/21 \\ 
2010-08-03 & 01:45$^{\rm nd}$/0.0027 & 0.0002 & uncertain & 21:17/265/14 \\
2010-08-07 & 01:43$^{\rm nd}$/0.0111 & 0.0013 & M1.0/17:55/N11E34 & 18:36/871/360 \\
2010-08-08 & 11:22/0.0074 & 0.0008 & uncertain & uncertain \\
2010-08-08 & 19:25/0.0031 & 0.0007 & uncertain & uncertain \\
2010-08-14 & 13:05/0.1581 & 0.0185 & C4.4/09:38/N17W52 & 10:12/1205/360 \\ 
2010-08-18 & 12:18/0.0486 & 0.0034 & C4.5/04:45/N18W88 & 05:48/1471/184 \\ 
2010-09-09 & 04:25/0.0071 & 0.0007 & C3.3/23:05$^{\rm pd}$/N21W87 & 23:27$^{\rm pd}$/818/147 \\
2011-01-28 & 05:13/0.0511 & 0.0088 & M1.3/00:44/N16W88 & 01:26/606/119 \\
2011-01-28 & 13:36/0.0326 & 0.0034 & C1.5/10:05/N17W91 & 10:36/499/94 \\
2011-02-15 & 10:24/0.0382 & 0.0037 & X2.2/01:44/S20W10 & 02:24/669/360 \\
2011-03-07 & 10:03$^{\rm nd}$/1.012 & 0.0843 & M3.7/19:43/N30W48 & 20:00/2125/360 \\ 
2011-03-21 & 23:56/0.2427 & 0.03 & uncertain &  02:24/1341/360 \\
2011-05-11 & 07:11/0.0071 & 0.0007 & B8.1/02:23/N17W85 & 02:48/745/225 \\
2011-06-04 & 18:44/0.0047 & 0.0007 & uncertain & 06:48/1407/360 \\
2011-06-04 & 00:07$^{\rm nd}$/0.0153 & 0.0024 & uncertain & 22:05/2425/360 \\
2011-06-05 & 08:33/0.0115 & 0.0023 & uncertain & uncertain \\
2011-06-05 & 04:23$^{\rm nd}$/0.0724 & 0.0147  & uncertain & uncertain \\
2011-06-07 & 17:31/1.579 & 0.339 & M2.5/06:16/S21W54 & 06:49/1255/360 \\
2011-06-11 & 18:38/0.046  & 0.0048 & uncertain & uncertain \\
2011-08-02 & 09:40/0.0399 & 0.0063 & M1.4/05:19/N14W15 & 06:36/712/268 \\
2011-08-03 & 17:38/0.0085 & 0.0008 & M6.0/13:17/N16W30 & 14:00/610/360 \\
2011-08-04 & 11:43/1.79   & 0.2665 & M9.3/03:41/N19W36 & 04:12/1315/360 \\
2011-08-08 & 19:35/0.0527 & 0.0053 & M3.5/18:00/N16W61 & 18:12/1343/237 \\
2011-08-09 & 11:46/0.5535 & 0.127  & X6.9/07:48/N17W69 & 08:12/1610/360 \\
2011-09-06 & 09:31/0.0413 & 0.0092 & M5.3/01:35/N14W07 & 02:24/782/360 \\
2011-09-07 & 05:39/0.1368 & 0.0343 & X2.1/22:12$^{\rm pd}$/N14W18 & 23:05$^{\rm pd}$/575/360 \\
2011-09-22 & 13:27$^{\rm 24th}$/0.2728 & 0.0255 & X1.4/10:29/N09E89 & 10:49/1905/360 \\
2011-10-22 & 04:47$^{\rm nd}$/0.0255   & 0.0009 & M1.3/10:00/N25W77 & 10:24/1005/360 \\
2011-11-04 & 11:17/0.0492 & 0.0108 & X1.9/20:16$^{\rm pd}$/N22E63 & 23:30$^{\rm pd}$/991/360 \\
2011-11-26 & 23:31/1.093  & 0.0544 & C1.2/06:09/N11W47 & 07:12/933/360 \\
2011-12-25 & 00:38$^{\rm nd}$/0.0479 & 0.0068 & M4.0/18:11/S22W26 & 18:48/366/125 \\
2012-01-02 & 20:21/0.015  & 0.0017 & C2.4/14:31/N07W89  & 15:13/1138/360 \\
2012-01-20 & 22:26/0.0173 & 0.0006 & M3.2/13:44$^{\rm pd}$/N32E22  & 14:36$^{\rm pd}$/1120/360 \\
2012-01-22 & 07:16/0.0229 & 0.0005 & uncertain & uncertain \\
2012-01-23 & 22:03/153.2  & 9.455  &  M8.7/03:38/N28W21 & 04:00/2175/360 \\
2012-01-27 & 13:34$^{\rm nd}$/22.71 & 2.683 & X1.7/17:37/N27W71  & 18:28/2508/360 \\
2012-02-25 & 00:04$^{\rm nd}$/0.0122 & - & uncertain & 03:46$^{\rm pd}$/800/189 \\
2012-03-04 & 21:37$^{\rm nd}$/0.0384 & 0.0035 & M2.0/10:29/N19E61 & 11:00/1306/360 \\
2012-03-07 & 00:04$^{\rm nd}$/82.93  & 9.591 & X1.3/01:05/N22E12  & 01:30/1825/360 \\
2012-03-13 & 19:47/10.91 & 1.423 & M7.9/17:12/N17W66 & 17:36/1884/360 \\
2012-03-30 & 11:46/0.0099 & 0.0009 & B6.2/23:19$^{\rm pd}$/N23W43 & 23:36$^{\rm pd}$/753/36 \\
2012-04-06 & 05:55/0.0056 & 0.0005 & C1.5/20:49$^{\rm pd}$/N18W29 & 21:25$^{\rm pd}$/828/360 \\
2012-05-17 & 04:41/3.116  & 1.357  & M5.1/01:25/N11W76 & 01:48/1582/360 \\
2012-05-26 & 09:57$^{\rm nd}$/0.1051 & 0.0035 & uncertain  & 20:58/1966/360 \\
2012-06-15 & 09:51/0.0064 & - & M1.9/12:52$^{\rm pd}$/S19E06 & 14:12$^{\rm pd}$/987/360 \\
2012-06-16 & 19:44/0.1419 & 0.0055 & uncertain & uncertain \\
2012-07-07 & 07:15/0.4885 & 0.0666 & X1.1/23:01$^{\rm pd}$/S13W59 & 23:24$^{\rm pd}$/18:28/360 \\
2012-07-08 & 00:28$^{\rm nd}$/0.2148 & 0.0611 & M6.9/16:23/S17W74 & 16:54/1495/157 \\
2012-07-12 & 21:28/1.683 & 0.1327 & X1.4/15:37/S15W01 & 16:48/885/360 \\
2012-07-17 & 23:01/1.581 & 0.0902 & M1.7/12:03/S28W65 & 13:48/958/176 \\
2012-07-19 & 21:08/1.542 & 0.2402 & M7.7/04:17/S13W88 & 05:24/1631/360 \\
2012-07-23 & 16:03$^{\rm 25th}$/0.2191 & 0.0758 & uncertain & 02:36/2003/360 \\
2012-08-02 & 18:03/0.0114 & 0.001  & uncertain & uncertain \\
\hline
\end{tabular}
\end{table}

\begin{table}[t!]
\addtocounter{table}{-1}
\caption[]{cont'd.}
\tiny
\vspace{0.1cm}
\begin{tabular}{lllll}
\hline
(1)            &  (2)                   & (3)               &  (4)                   & (5)             \\
\hline 
2012-09-01 & 20:21/0.2617 & 0.0073 & C8.4/19:45$^{\rm pd}$/S19E42 & 20:00$^{\rm pd}$/1442/360 \\
2012-09-08 & 17:34/0.016  & 0.0028 & data gap  & 10:00/734/360 \\
2012-09-22 & 21:43/0.0063 & 0.001  & uncertain & uncertain \\
2012-09-24 & 12:32/0.0075 & 0.0005 & C1.7/15:02$^{\rm pd}$/S07E88 & 14:48$^{\rm pd}$/939/258 \\
2012-09-28 & 03:24/0.2716 & 0.0393 & C3.7/23:36$^{\rm pd}$/N06W34 & 00:12/947/360 \\
2012-11-08 & 17:51$^{\rm nd}$/0.0321 & 0.0053 & uncertain & 11:00/972/360 \\
2012-12-15 & 02:05/0.0084 & - & uncertain & 02:00$^{\rm pd}$/763/149 \\
2012-12-16 & 05:44/0.005  & - & uncertain & uncertain \\
2013-01-17 & 05:46/0.0182 & 0.001  & C2.2/18:21$^{\rm pd}$/S32W89$^u$ & 19:00$^{\rm pd}$/648/250  \\
2013-02-27 & 10:16/0.009  & 0.0009 & uncertain & 09:12$^{\rm pd}$/987/360 \\
2013-03-06 & 00:17$^{\rm 8th}$/0.0076 & 0.0014 & uncertain & 03:48$^{\rm pd}$/1316/360  \\
2013-03-15 & 23:41/0.009  & 0.0004 & M1.1/05:46/N11E12 & 07:12/1063/360 \\
2013-03-16 & 16:03/0.1159 & 0.0085 & uncertain & uncertain \\
2013-04-11 & 12:17/2.932  & 0.5116 & uncertain & 07:24/919/360 \\
2013-04-21 & 13:39/0.0572 & 0.0052 & C1.4/21:05$^{\rm pd}$/N10W65 & 22:12$^{\rm pd}$/594/360 \\
2013-04-25 & 09:05/0.0185 & 0.0057 & uncertain & uncertain \\
2013-05-13 & 13:44$^{\rm nd}$/0.0217  & 0.0023 & X2.8/15:48/N11E85 & 16:08/1850/360 \\
2013-05-15 & 03:28$^{\rm nd}$/0.5234  & 0.0352 & X1.2/01:25/N12E64 & 01:48/1366/360 \\
2013-05-22 & 03:13$^{\rm nd}$/61.84   & 4.404  & M5.0/13:08/N14W87 & 12:26/1466/360 \\
2013-06-21 & 10:15$^{\rm nd}$/0.1223  & 0.0073 & M2.9/02:30/S16E73 & 03:12/1900/207 \\
2013-06-23 & 00:21$^{\rm nd}$/0.1502  & 0.007  & C1.5/17:31/N05W27 & 16:24/477/227  \\
2013-06-28 & 13:16/0.0086 & 0.0005 & C4.4/01:36/S18W19  & 02:00/1037/360 \\
2013-07-22 & 14:57/0.0061 & 0.0006 & uncertain & 06:24/1004/360  \\
2013-08-17 & 03:49$^{\rm nd}$/0.0049 & 0.0003 & M3.3/18:16/S07W30 & 19:12/1202/360 \\
2013-08-21 & 16:45/0.0256 & 0.0028 & uncertain & 08:12$^{\rm pd}$/784/360 \\
2013-09-30 & 17:38/3.208  & 0.172  & C1.2/21:43$^{\rm pd}$/N10W43 & 22:12$^{\rm pd}$/1179/360 \\
2013-10-13 & 11:06/0.0056 & 0.0021 & uncertain & uncertain \\
2013-10-15 & 01:14/0.0098 & 0.0013 & uncertain & data gap \\
2013-10-23 & 03:16/0.0119 & 0.002  & M4.2/21:15$^{\rm pd}$/N04W01 & 21:48$^{\rm pd}$/459/360 \\
2013-10-25 & 13:30$^{\rm nd}$/0.0128 & 0.0022 & X2.1/14:51/S06E69 & 15:12/1081/360 \\
2013-10-28 & 14:47/0.0701 & 0.0079 & M5.1/04:32/N08W71 & 04:48/1201/156 \\
2013-11-02 & 19:23/0.0489 & 0.0142 & C8.2/04:40/S23W04 & 04:48/828/360 \\
2013-11-07 & 03:09/0.0578 & 0.0058 & M1.8/23:44$^{\rm pd}$/S11W88 & 00:00/1033/360 \\
2013-11-08 & 11:10$^{\rm nd}$/0.0192 & 0.0014 & C2.1/10:26$^{\rm pd}$/S11E25 & 10:36$^{\rm pd}$/1405/360 \\
2013-11-19 & 16:02/0.0974 & 0.0136 & X1.0/10:14/S70W14 & 10:36/740/360 \\
2013-12-12 & 12:35/0.0043 & 0.0003 & C4.6/03:11/S23W46 & 03:36/1002/276 \\
2013-12-14 & 22:33/0.0149 & 0.0008 & uncertain & 21:24$^{\rm pd}$/518/360\\
2013-12-26 & 21:24/0.0483 & 0.0048 & uncertain & 03:34/1336/360 \\
2013-12-28 & 23:20/0.5896 & 0.0758 & C9.3/17:53/S18E07 & 17:36/1118/360 \\
2014-01-05 & 10:51/0.0067 & - & M4.0/18:47$^{\rm pd}$/S11E33 & 21:23$^{\rm pd}$/977/360 \\
2014-01-06 & 14:50/0.9507 & 0.2141 & uncertain & 08:00/1402/360 \\
2014-01-07 & 13:00$^{\rm nd}$/36.67 & 4.284 & X1.2/18:04/S15W11 & 18:24/1830/360 \\
2014-02-14 & 20:04/0.005  & 0.0007 & uncertain & 08:48/1165/360 \\
2014-02-18 & 11:49/0.0071 & -      & uncertain & 01:36/779/360 \\
2014-02-19 & 15:59/0.0199 & -      & uncertain & 04:48/612/360 \\
2014-02-19 & 23:57/0.0304 & 0.0009 & uncertain & 16:00/571/267 \\
2014-02-20 & 12:30/0.2633 & 0.0309 & M3.0/07:26/S15W73 & 08:00/948/360 \\
2014-02-25 & 18:31/0.566  & 0.1061 & X4.9/00:39/S12E82 & 01:26/2147/360 \\
2014-03-25 & 08:27/0.0083 & -      & uncertain  & 05:36/651/233 \\
2014-03-29 & 04:47/0.0036 & -      & M2.6/23:44$^{\rm pd}$/N10W22 & 23:48$^{\rm pd}$/514/138\\
2014-03-29 & 20:49/0.0552 & 0.0091 & X1.0/17:35/N11W32 & 18:12/528/360 \\
2014-04-05 & 04:35$^{\rm nd}$/0.0144 & 0.001  & uncertain & 06:24/798/203 \\
2014-04-18 & 03:07$^{\rm nd}$/1.046  & 0.0906 & M7.3/12:31/S20W34 & 13:26/1203/360 \\
2014-05-07 & 21:47/0.0113 & 0.0016 & M1.2/16:07/N15E50 & 16:24/923/360 \\
2014-05-07 & 00:06$^{\rm nd}$/0.0153 & 0.0016 & uncertain & uncertain \\
2014-05-08 & 03:18/0.0051 & 0.0005 & uncertain & uncertain \\
\hline
\end{tabular}
\end{table}

\begin{table}[t!]
\addtocounter{table}{-1}
\caption[]{cont'd.}
\tiny
\vspace{0.1cm}
\begin{tabular}{lllll}
\hline
(1)            &  (2)                   & (3)               &  (4)                   & (5)             \\
\hline 
2014-05-09 & 08:46/0.0055 & 0.0012 & uncertain & 02:48/1099/360 \\
2014-06-13 & 04:40/0.008  & 0.0011 & M3.1/21:34$^{\rm pd}$/S20W55 & 22:12$^{\rm pd}$/684/186 \\
2014-08-25 & 22:17/0.0261 & 0.0028 & M2.0/14:46/N05W36 & 15:36/555/360 \\
2014-09-02 & 02:53$^{\rm 5th}$/0.0818 & 0.0225 & uncertain & 11:12$^{\rm pd}$/1901/360 \\
2014-09-10 & 05:11$^{\rm nd}$/0.5808 & 0.1399 & X1.6/17:21/N14E02 & 18:00/1267/360 \\
2014-09-22 & 12:01/0.0314 & 0.0043 & uncertain & 06:12/618/252 \\
2014-09-24 & 18:45$^{\rm nd}$/0.0048 & 0.0016 & uncertain & 21:30/1350/360 \\
2014-10-10 & 20:39/0.005  & -   & C3.0/15:42/S20W51 & 16:12/782/210 \\
2014-12-13 & 10:16$^{\rm nd}$/0.0268 & 0.0067 & data gap & 14:24/2222/360 \\
2014-12-17 & 04:36$^{\rm nd}$/0.0107 & 0.0016 & M8.7/04:25/S20E09 & 05:00/587/360 \\
2014-12-21 & 19:29/0.0258 & 0.0015 & uncertain & uncertain \\
2014-12-28 & 16:29/0.0079 & - & uncertain & uncertain \\
2015-02-21 & 11:27$^{\rm nd}$/0.0136 & 0.0011 & uncertain  & 09:24/1120/360 \\
2015-03-15 & 07:12/0.0325 & 0.0029 & C9.1/01:15/S22W25 & 01:48/719/360 \\
2015-03-24 & 21:24/0.0105 & 0.0011 & uncertain & 08:24/1794/360 \\
2015-03-27 & 19:07/0.005  & -      & uncertain & uncertain \\
2015-04-22 & 05:24$^{\rm nd}$/0.0155 & 0.0012 & M2.2/11:49$^{\rm pd}$/N10W84 & 13:26$^{\rm pd}$/1079/83 \\
2015-04-23 & 02:26$^{\rm nd}$/0.0046 & 0.0008 & uncertain & uncertain \\
2015-05-12 & 06:43/0.1547 & 0.0193 & C2.6/02:15/S21W83 & 02:48/772/250 \\
2015-06-18 & 15:26/0.3978 & 0.0252 & M1.2/00:33/S16W81 & 01:26/1714/195 \\
2015-06-21 & 12:13/0.021  & 0.0009 & M2.6/02:06/N13E10$^{\rm v}$ & 02:36/1366/360 \\  
2015-06-21 & 00:21$^{\rm nd}$/0.9349 & 0.0203 & uncertain & uncertain \\
2015-06-22 & 09:51/1.817  & 0.0387 & uncertain & uncertain \\
2015-06-24 & 17:42/0.0346 & 0.0026 & uncertain & uncertain \\
2015-06-25 & 18:09$^{\rm nd}$/0.2726 & 0.0142 & M7.9/08:02/N09W42 & 08:36/1627/360 \\
2015-07-01 & 19:06/0.045  & 0.0034 & uncertain & 14:36/1435/360 \\
2015-09-20 & 20:25/0.0402 & 0.0045 & M2.1/17:32/S20W24 & 18:12/1239/360 \\
2015-09-30 & 21:04/0.0052 & -      & uncertain & 09:36/586/102 \\
2015-10-29 & 08:59/0.449  & 0.1052 & uncertain & 02:36/530/202 \\ 
2015-11-04 & 19:02/0.005  & 0.0006 & M3.7/13:31/N08W02 & 14:48/578/291 \\
2015-11-09 & 23:56/0.0675 & 0.009  & M3.9/12:49/S11E41 & 13:25/1041/273 \\ 
2015-12-28 & 16:34/0.0115 & 0.001  & M1.8/11:20/S23W11 & 12:12/1212/360 \\ 
2015-12-29 & 00:43$^{\rm nd}$/0.0146 & 0.0006 & uncertain & uncertain \\
2016-01-02 & 00:53/0.1929 & 0.0174 & M2.3/23:10$^{\rm pd}$/S25W82 & 23:24/1730/360 \\
2016-03-16 & 10:36/0.016  & 0.0024 & C2.2/06:34/N12W88 & 07:00/592/154 \\
2016-04-18 & 12:24/0.0143 & 0.0007 & M6.7/00:14/N12W62 & 00:48/1084/162\\
2016-05-15 & 21:31/0.0126 & -      & C3.2/15:19/N21W06 & 15:12/1118/176\\
2016-07-23 & 06:51/0.0055 & 0.0004 & M7.6/05:00/N05W73 & 05:24/835/117 \\
\hline
\end{tabular}
\end{table}

The on-line version of the Wind/EPACT proton catalog \url{http://www.newserver.stil.bas.bg/SEPcatalog} will be updated for subsequent years if data continues to be provided. An yearly update is planned to be released early in the subsequent year, since the omni-data is not provided in real-time. In Table~\ref{T-datatable} and also in the web-site, the onset time of the proton event is given as the event date (yyyy-mm-dd) and time (hh:mm in UT); all other time markers are given in reference to this date. The online catalog is planned to contain more information, namely onset and peak times and peak proton amplitude (and fluence) for the low energy channel as well as the peak proton intensity (and fluence) for the high energy channel. Overview plots are given for each case and can be downloaded. Finally, the properties of the solar origin (flares and CMEs) of the proton events are also listed. 

\subsection{Flare and CME data} 

The SF date, onset/peak time, class and helio-location of the active region [AR] are taken from the GOES soft X-ray [SXR] flare listings\footnote{\url{ftp://ftp.ngdc.noaa.gov/STP/space-weather/solar-data/solar-features/solar-flares/x-rays/goes/}}. The missing flare locations were adopted from \url{SolarMonitor.org} or Solar Geophysical Data reports\footnote{\url{https://www.ngdc.noaa.gov/stp/solar/sgd.html}}. For several cases, we visually identified the helio-location according to the position of hard X-ray brightenings (with about $\pm$10 degrees uncertainty). Only for seven SFs no location could be found. 

For the CMEs we report the time of first appearance over the C2-oculting disk of SOHO/LASCO \citep{1995SoPh..162..357B} that together with the projected on-sky speed and AW are taken from the CDAW LASCO CME catalog\footnote{\url{https://cdaw.gsfc.nasa.gov/CME_list/}} \citep{2004JGRA..10907105Y,2009EM&P..104..295G}. 

In order to allow for comparison with earlier reports, the above flare and CME catalogs are utilized. These databases are common sources that have been used in previous studies when identifying the SEP origin. 

A recent study by \cite{2015SoPh..290.1741R} investigated the influence of the differently reported CME properties, by different catalogs, on the correlations coefficients with proton intensities. Reports on individual CME speed can differ significantly, however, the correlation analysis shows mostly consistent results. We are not aware whether and by what amount the occasional differences when reporting SXR flare class by different sources\footnote{The SXR flare class based on GOES data is reported by: \url{ftp://ftp.ngdc.noaa.gov/STP/space-weather/solar-data/solar-features/solar-flares/x-rays/goes/}; \url{http://legacy-www.swpc.noaa.gov/ftpmenu/indices/events.html}; \url{www.solarmonitor.org}; \url{ftp://ftp.ngdc.noaa.gov/STP/SOLAR_DATA/SGD_PDFversion/}.} will influence the correlation coefficients with the proton intensity.

\subsection{Solar origin association} 

The solar origin of the observed in situ protons is usually sought in terms of SFs and CMEs. This is because of the two main particle accelerator mechanisms acting in the solar atmosphere, namely the magnetic reconnection process (the driver of SFs) and shock acceleration (formed when the CME propagates into the IP space). Remote sensing of various emission signatures and theoretical modeling are supporting either scenario. However, the individual contribution of each mechanism to the in situ observed particle fluxes is not quantified at present  \citep{2001SSRv...95..215K,2007SSRv..129..359K,2015SoPh..290..819T,2017JPhCS.798a2034B}. That has led to year-long debates on the particle origin, with preferences either exclusively towards the CMEs ({\it e.g.}, \opencite{1982JGR....87.3439K,2008AnGeo..26.3033G,2016ApJ...832..128C}), or arguing in favor of the flares as a contributing agent ({\it e.g.}, \opencite{1986ApJ...301..448C,2008A&A...486..589K,2010SoPh..263..185K,2015SoPh..290.2827G}) and the references therein. Based on the proton rise time, composition, abundances, and related phenomena, several classifications schemes have been proposed, and periodically revised \citep{1999SSRv...90..413R,2009CEAB...33..253C,2013SSRv..175...53R}. Note that the focus of the majority of the particle studies in the literature is on protons, probably due to the availability of the GOES proton data and list. For electron event analysis fewer statistical studies and event lists with partial coverage are available, see {\it e.g.}, \cite{1999ApJ...519..864K,2002ApJ...579..841H,2007ApJ...663L.109K,2013JSWSC...3A..12V}.
 
In the present study we did not consider impulsive vs. gradual separation of the SEP events as proposed by \cite{1999SSRv...90..413R} that nowadays is considered as oversimplified. Moreover, no attempt was made to identify the proton injection time at the Sun ({\it e.g.}, \opencite{2013JSWSC...3A..12V,2015A&A...580A..80K}) via the velocity dispersion analysis [VDA], since the two energy ranges provided by Wind/EPACT are not optimal for this procedure. An alternative way is to perform time shifting analysis [TSA]. The travel time of protons propagating along a nominal Parker Spiral of length 1.2 AU ranges from $\sim$30 min ($\sim$50 MeV) up to $\sim$45 min ($\sim$25 MeV). Due to IP density and magnetic field fluctuations, the proton travel time is expected to last longer than the scatter-free travel time. In addition, both procedures (VDA and TSA) entail numerous approximations and simplifications, {\it e.g.}, simultaneous in time and identical place of injection of low and high energy particles, scatter-free propagation. 

For our analysis we decided to follow a set of guidelines. The solar origin identification proposed here is done using timing, intensity and positioning constraints. Namely, the most prominent SF (strongest in flare class) and CME (fastest and widest) within one day prior to the identified proton onset time is initially selected. Additionally, the individual proton profiles are taken into account. If considered necessary, the initial solar origin association is reexamined. For example, an earlier solar origin (and often at eastern helio-longitudes) could be preferred when the proton event rises slowly (of the order of hours). Another indicator for the SEP origin is the proxy for the particle escape from the solar corona in terms of the strongest type III radio burst (observed by the Wind/WAVES instrument). In general, we aim to identify both SF and CME as the particle origin, where possible. Nevertheless, for 62 proton events (about 14\% of the event list here) no solar origin (neither SF, nor CME) could be allocated due to multiple SF/CME choices or a high degree of uncertainty.

Despite the care taken during the solar source identification procedure, subjectivity in the analysis cannot be eliminated. Since no temporal  constraints during the solar origin identification process are implied by us, one quarter of the associated solar flares occur more than 5 hours before the proton onset, 14\% of the flares start more than 10 hours earlier and 5\% more than 20 hours before the respective proton onset (with no clear longitudinal dependence for the latter two samples). Thus the latter two cases are subject to larger uncertainty but are finally kept in our analysis. The mean/median difference between the flare and proton onsets is 4.6/2.3 hours, respectively.

Special attention in our work is given on the longitudinal dependence of the protons and their solar origin (flare$-$CME pair). The location of the solar flare is adopted by the reported AR longitude and latitude. For the same SEP event, the flare AR is considered also as the origin of the CME (since the location of the flare-AR and the direction of propagation of the CME are in the same solar disk quadrant). However, for 75 CMEs no SF counterpart could be found ({\it e.g.}, behind the limb SFs, multiple candidates/uncertain cases.). In order to recover the longitudinal information for these cases, the CME projected direction of propagation is used. The latter is given by the measurement position angle [MPA]\footnote{MPA corresponds to the position of the fastest moving segment of the CME leading edge in counterclockwise direction and is provided for all CMEs, including halo.}, reported by the LASCO CME catalog. Namely, when the CME is ejected in eastern direction, MPA ranges from 0 to 180 degrees, whereas the western direction of propagation is for angles between 181 and 360. Based on these values, the CME sample is also divided in helio-longitude. Thus, when the proton origin location is considered we use the combined information obtained by both solar eruptive phenomena, SF and CME.

\section{Results and Discussion}
\label{S-Results} 

In this Section the main outcomes from the Wind/EPACT catalog (as presented in Sect.~\ref{S-Proton}) are presented, organized in several topics. Firstly, the overall behavior of the proton samples at low and high energies is shown, followed by their solar origin: flares and CMEs. In each subsection the intensity (from the weakest to strongest events) and the longitudinal trends (differences due to eastern to western positioning of the solar source) are described. Finally, the trends in the first eight years from each SC in terms of percentage changes and probabilities of occurrences for the protons in specific SF and CME ranges are investigated.

\subsection{Properties of Wind/EPACT proton events}

\subsubsection{Proton intensity}

For the needs of the analysis, the protons in the low energy channel are binned into several representative intensity ranges, divided roughly by an order of magnitude, namely: 
\begin{itemize}
\item minor (mi): $J_{\rm p} < 0.01$
\item weak (w): $0.01 \leq J_{\rm p} < 0.1$
\item intermediate (i): $0.1 \leq J_{\rm p} < 1$
\item major (ma): $1 \leq J_{\rm p} < 10$
\item extreme (e): $J_{\rm p} \geq 10$
\end{itemize}
where $J_{\rm p}$ is in (cm$^2$ s sr MeV)$^{-1}$. For the high energy channel the same terminology is used, however the values were divided by 10 since the proton intensities for the high energy channel are lower. The exact numbers in each sub-category are summarized in Table~\ref{T-SEP_properties}. There, the mean/median values of the entire sample of protons are listed, separately for the proton sample with origin located at the west and east (in different columns). In addition, we present the mean/median values and their number for both proton energy channels and for the proton intensity sub-categories as introduced above (in different rows of the Table). The high energy protons have a lower mean/median intensity by roughly one order of magnitude in all sub-categories, compared to the low energy sample. The proton intensity of the entire sample (`All' in Table~\ref{T-SEP_properties}) is similar to the value obtained for the most abundant sub-category ({\it i.e.}, for the weak proton events).

\begin{table}[t]
\small
\caption{The mean/median values of Wind/EPACT proton intensity ($J_{\rm p}$ in (cm${^2}\,$s$\,$sr$\,$MeV)$^{-1}$), in low and high energy, over the period 1996$-$2016 in different sub-categories. The number of events used for each calculation is given in brackets.}
\begin{tabular}{llll}
\hline
Proton type             & All events         & Western-origin     & Eastern-origin      \\
\hline
\multicolumn{4}{c}{$J_{\rm 25\,MeV}$} \\
All                     & 0.0774/0.0361 (429)& 0.0946/0.0444 (283)& 0.0668/0.0264 (84)\\
extreme                 & 51.21/43.79 (28)   & 46.25/37.13 (22)   & 99.04/82.93 (5)   \\
major                   & 2.891/2.303 (41)   & 2.941/2.499 (37)   & 4.045/3.606 (5)  \\
intermediate            & 0.2939/0.2716 (87) & 0.2902/0.2675 (76) & 0.3087/0.2681 (16) \\
weak                    & 0.0266/0.0262 (170)& 0.0273/0.0298 (102)& 0.0237/0.0205 (42) \\
minor                   & 0.0059/0.0058 (103)& 0.0061/0.0059 (67) & 0.0062/0.0065 (16) \\
\hline
\multicolumn{4}{c}{$J_{\rm 50\,MeV}$} \\
All                     & 0.0093/0.0045 (397)& 0.0123/0.0064 (266)& 0.0073/0.003 (76) \\
extreme                 & 4.538/4.284 (33)   & 4.222/4.195 (28)   & 6.796/7.311 (5)    \\
major                   & 0.2205/0.2084 (40) & 0.2245/0.21 (38)   & 0.224/0.2002 (7)  \\
intermediate            & 0.0285/0.0254 (81) & 0.0295/0.0257 (56) & 0.0265/0.0249 (12) \\
weak                    & 0.0027/0.0027 (161)& 0.0029/0.0029 (97) & 0.0024/0.0023 (40) \\
minor                   & 0.0006/0.0006 (82) & 0.0006/0.0006 (64) & 0.0005/0.0005 (12)  \\
\hline
\end{tabular}
\label{T-SEP_properties}
\end{table}

\begin{figure}[t!]     
\centerline{\includegraphics[width=0.99\textwidth,clip=]{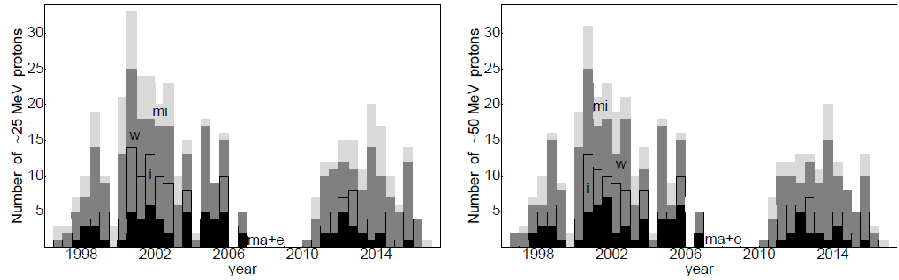}}
\caption{Number of Wind/EPACT low (left panel) and high (right panel) energy proton events over time, binned in a 6-month period. The height of the color bar denotes the number of events in given intensity interval: light-gray (for minor intensity events, mi), gray (weak, w), contour (intermediate, i) and black (major and extreme, ma+e). The total number of events is given by the sum of all colored sections. See text for abbreviations and further explanations.}
   \label{F-SEPhisto-year}               
\end{figure}

The distribution of the events over the years (1996$-$2016) is presented in Fig.~\ref{F-SEPhisto-year} using 6-month binning. The SC trend can be readily recognized. Since the difference in the number of low and high energy proton events is about 10$\%$, the distributions are visually very similar. Each bin is filled with proton events of different intensity (shown with different colors/contours) and the entire distribution is the sum of all colored sections. From the stacked histograms one can see that the weak (`w') in intensity protons are the most abundant samples at both energies, see also Table~\ref{T-SEP_properties}.  

\begin{figure}[t!]    
\centerline{\includegraphics[width=0.99\textwidth,clip=]{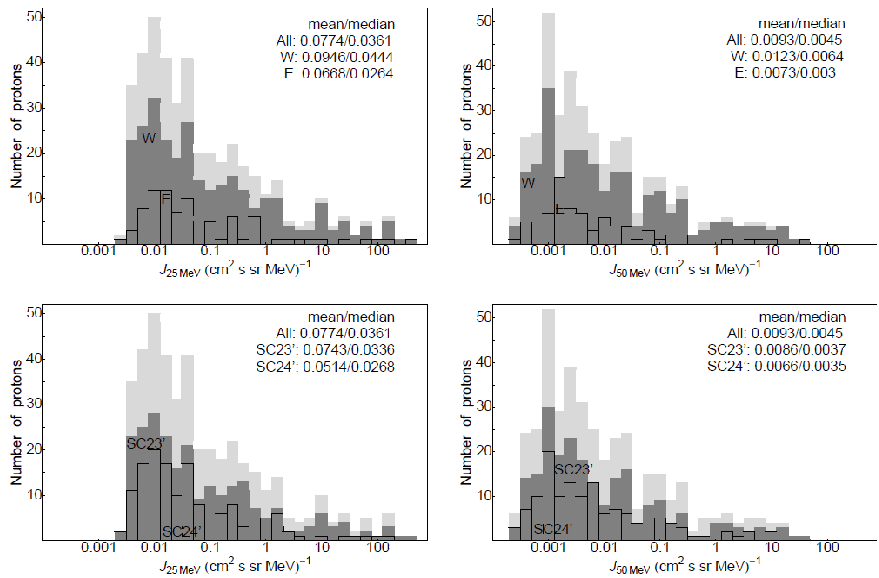}}
\caption{Number of Wind/EPACT proton events over $J_{\rm p}$ given separately for the $\sim$25 MeV (left plot) and $\sim$50 MeV (right) energy channel. The distributions are given for the entire proton sample (light gray) overlaid with the histograms for the protons with identified western/in SC23$^{\prime}$ (dark gray) and eastern origin/in SC24$^{\prime}$ (contours), respectively.}
   \label{F-SEPhisto-Jp}               
\end{figure}

The following color-code is adopted for the subsequent plots. The entire distribution is shown in light-gray color and it serves as the background histogram, for comparative purposes. Any other samples of interest are overlapped and given in different colors or contours. The low energy protons ($J_{\rm 25\,MeV}$) are given on the left, whereas the high energy protons ($J_{\rm 50\,MeV}$) are organized in the right column of plots. The mean and median values for each proton sample are shown on each plot. The division of the proton sample into W/E longitude is according to the AR of the proton-producing SF and CME measurements of the MPA. 

In Fig.~\ref{F-SEPhisto-Jp} are presented proton distributions with a focus on events with origin in eastern [E] vs. western [W] longitudes, upper panel, or the protons originating in the first eight years of each solar cycle, noted here as SC23$^\prime$ (09/1996$-$08/2004) vs. SC24$^\prime$ (01/2009$-$12/2016) in the lower panel of plots. The W-origin proton sample contains more events with highest ({\it e.g.}, extreme and major) intensity ({\it i.e.}, $\sim$21\%, 59/283) compared to the E sample ({\it i.e.}, $\sim$12\%, 10/84, respectively), evidenced also by their mean/median values, see Table~\ref{T-SEP_properties} and Fig.~\ref{F-SEPhisto-Jp} (upper plots). This is true also for the high energy protons. When the entire proton distributions in SC23$^\prime$ and SC24$^\prime$ (Fig.~\ref{F-SEPhisto-Jp}, lower plots) are compared, however, there is no clear difference in the proton intensity, since the distributions mostly overlap. 

\begin{figure}[t!]
\centerline{\includegraphics[width=0.99\textwidth,clip=]{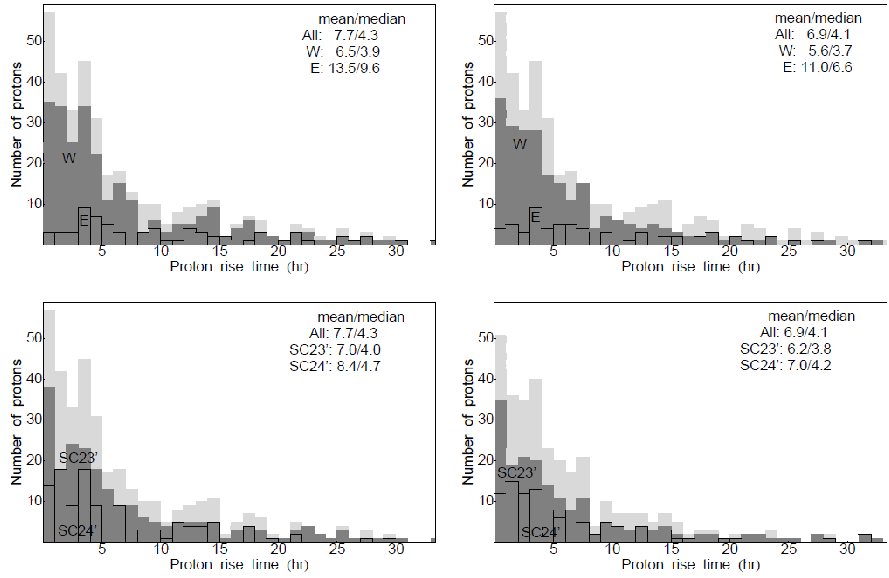}}
\caption{Number of Wind/EPACT proton events over proton rise time in the low (on the left) and high (right) energy channels. For plotting purpose the sample in first 35 hours is shown, however the entire distributions are used for the calculations of mean/median values. Color code as in Fig.~\ref{F-SEPhisto-Jp}.}
   \label{F-SEPhisto-rise}               
\end{figure}

The mean/median values of the different proton sub-categories as function of W/E location and SC are given in more details in Table~\ref{T-SEPevents}. The entire sample of proton events (see `All/All events') has similar mean/median values in each SC\footnote{The statistical difference between two samples is evaluated based on the Kolmogorov-Smirnov test selecting 95\% confidence level.}. There is a difference for the larger intensity protons (denoted by `e' and `ma'), namely the mean/median values in SC23$^\prime$ are larger than the respective values in SC24$^\prime$, which is more evident for $J_{\rm 25\,MeV}$. When the longitude is considered, there is a difference only for the extreme (`e') proton category (at low and high energy), where the E-sample has a larger intensity compared to the W-sample (note however the very low sampling of the E events). The trend is noticed in both SCs. 

\begin{sidewaystable}
\tiny
\caption{Mean/median values of proton intensity and fluence for different energy channels, categories and time periods. The number of events used in each calculation is given in brackets. The percentage change is the difference between the first eight years in SC24 (SC24$^\prime$) and the same period in SC23 (SC23$^\prime$), divided by the value in SC23$^\prime$. Negative/positive value denote a decrease/increase in SC24$^\prime$, respectively. The uncertainty is calculated as Poisson error. Abbreviations: e: extreme, ma: major, i: intermediate, w: weak, mi: minor, W: western origin, E: eastern origin.}
\begin{tabularx}{\textwidth}{llllllll}
\hline
SEP         & \multicolumn{3}{c}{SC23$^\prime$: 09/1996$-$08/2004} & \multicolumn{3}{c}{SC24$^\prime$: 01/2009$-$12/2016} &  Percentage change \\
type        & All events & W & E & All events & W & E & All events/W/E \\
\hline
\multicolumn{8}{c}{$J_{\rm 25\,MeV}$} \\
All         & 0.0743/0.0336 (239)& 0.0946/0.0405 (161)& 0.0439/0.0228 (42) & 0.0514/0.0268 (149) & 0.0655/0.0353 (94) & 0.0506/0.0217 (33) & $-$38$\pm$7/$-$42$\pm$8/$-$21$\pm$18\\
e           & 58.34/37.6 (15)    & 58.35/43.79 (12)   & 105.3 (2)          & 43.88/47.62 (6)     & 38.63/36.67 (5)    & 82.93 (1)          & $-$60$\pm$19/$-$58$\pm$22/$-$50$\pm$61\\
ma          & 3.544/3.257 (21)   & 3.803/3.609 (19)   & 2.605 (2)          & 1.729/1.631 (12)    & 1.721/1.581 (11)   & 2.932 (1)          & $-$43$\pm$21/$-$42$\pm$22/$-$50$\pm$61\\
i           & 0.2935/0.2945 (52) & 0.2929/0.2945 (44) & 0.3019/0.2634 (5)  & 0.2819/0.2633 (27)  & 0.2762/0.2633 (25) & 0.3173/0.3779 (8)  & $-$48$\pm$12/$-$43$\pm$14/+60$\pm$91\\
w           & 0.0258/0.0249 (88) & 0.0256/0.0239 (49) & 0.0228/0.0228 (24) & 0.0256/0.0256 (65)  & 0.0271/0.0285 (40) & 0.0244/0.0204 (16) & $-$26$\pm$12/$-$18$\pm$17/$-$33$\pm$22\\
mi          & 0.0059/0.0058 (63) & 0.0061/0.0064 (44) & 0.0061/0.0056 (9)  & 0.0061/0.0061 (39)  & 0.0061/0.0055 (22) & 0.0064/0.0071 (7)  & $-$38$\pm$13/$-$50$\pm$13/$-$22$\pm$39\\
\hline
\multicolumn{8}{c}{$J_{\rm 50\,MeV}$} \\
All         & 0.0086/0.0037 (221)& 0.0126/0.0072 (150)& 0.0046/0.0023 (39) & 0.0066/0.0035 (135)& 0.0085/0.0048 (88) & 0.0063/0.0041 (28) & $-$39$\pm$7/$-$41$\pm$8/$-$28$\pm$18  \\
e           & 4.63/4.059 (20)    & 4.389/4.059 (18)   & 7.492 (2)          & 3.664/4.284 (7)    & 3.121/3.39 (6)     & 9.591 (1)          & $-$65$\pm$15/$-$67$\pm$16/$-$50$\pm$61\\
ma          & 0.2222/0.2031 (18) & 0.2207/0.206 (15)  & 0.3222 (2)         & 0.1886/0.172 (11)  & 0.1857/0.1919 (8)  & 0.1966/0.1399 (3)  & $-$39$\pm$23/$-$47$\pm$23/+50$\pm$137\\
i           & 0.0283/0.0251 (46) & 0.029/0.0254 (35)  & 0.0256/0.0172 (3)  & 0.0318/0.0304 (26) & 0.035/0.0326 (18)  & 0.0278/0.0255 (5)  & $-$43$\pm$14/$-$49$\pm$15/+67$\pm$122\\
w           & 0.0025/0.0023 (89) & 0.0027/0.0026 (48) & 0.0023/0.0023 (27) & 0.0029/0.0028 (61) & 0.003/0.0029 (39)  & 0.003/0.0023 (13)  & $-$31$\pm$11/$-$19$\pm$18/$-$52$\pm$16\\
mi          & 0.0006/0.0006 (48) & 0.0006/0.0006 (34) & 0.0006/0.0006 (5)  & 0.0006/0.0007 (30) & 0.0006/0.0007 (17) & 0.0005/0.0005 (6)  & $-$38$\pm$14/$-$50$\pm$15/+20$\pm$73\\
\hline
\multicolumn{8}{c}{$F_{\rm 25\,MeV}$} \\
All          & 888/435 (201)    & 936/447 (138)    & 1364/520 (34) & 868/586 (126)    & 1042/594 (80)    & 1191/836 (31) & $-$37$\pm$7/$-$42$\pm$8/$-$9$\pm$23 \\
e            & 1.5/0.66 $|\times 10^6$ (11) & 1.4/0.66 $|\times 10^6$ (9) & 2.27$\times 10^6$ (2) & 0.85/1.1 $|\times 10^6$ (6) & 0.68/1 $|\times 10^6$ (5) & 2.67$\times 10^6$ (1) & $-$45$\pm$28/$-$44$\pm$31/$-$50$\pm$61 \\
ma           & 38973/73886 (20) & 37004/76625 (16) & 54492 (2)     & 23878/24939 (12) & 25741/24939 (10) & 9716 (1)      & $-$40$\pm$22/$-$38$\pm$25/$-$50$\pm$61 \\
i            & 2659/3152 (52)   & 2675/2619 (39)   & 6461/8090 (5) & 4117/4601 (26)   & 2917/2910 (16)   & 8091/6707 (8) & $-$50$\pm$12/$-$59$\pm$12/+60$\pm$91  \\
w            & 254/268 (83)     & 195/199 (46)     & 438/395 (22)  & 308/358 (64)     & 277/329 (40)     & 578/470 (16)  & $-$23$\pm$13/$-$13$\pm$19/$-$27$\pm$24 \\
mi           & 38/56 (35)       & 33/47 (28)       & 259/256 (3)   & 40/47 (18)       & 47/43 (9)        & 79/172 (5)    & $-$49$\pm$15/$-$68$\pm$12/+67$\pm$122 \\
\hline
\multicolumn{8}{c}{$F_{\rm 50\,MeV}$} \\
All          & 116/58 (175)     & 165/82 (121)     & 62/48 (34)  & 116/88 (110)    & 107/55 (76)     & 179/150 (24)  & $-$37$\pm$8/$-$37$\pm$9/$-$29$\pm$19 \\
e            & 44550/27266 (18) & 61742/33595 (16) & 3273 (2)    & 30867/64195 (7) & 25097/57392 (6) & 106851 (1)    & $-$61$\pm$17/$-$63$\pm$18/$-$50$\pm$61 \\
ma           & 2865/2008 (18)   & 2562/1827 (15)   & 4639 (2)    & 1814/2190 (11)  & 1719/2239 (8)   & 2093/2190 (3) & $-$39$\pm$23/$-$47$\pm$23/+50$\pm$137 \\
i            & 261/333 (45)     & 250/284 (34)     & 717/651 (3) & 397/326 (25)    & 362/326 (17)    & 694/620 (5)   & $-$44$\pm$14/$-$50$\pm$15/+67$\pm$122 \\
w            & 20/21 (79)       & 18/17 (45)       & 33/38 (24)  & 34/33 (59)      & 27/26 (39)      & 58/87 (13)    & $-$25$\pm$13/$-$13$\pm$19/$-$46$\pm$19 \\
mi           & 2/4 (15)         & 2/6 (11)         & 3/4 (3)     & 4/4 (8)         & 3/3 (6)         & 10 (2)        & $-$47$\pm$23/$-$45$\pm$28/$-$33$\pm$61 \\
\hline
\end{tabularx}
\label{T-SEPevents}
\end{sidewaystable}

In Fig.~\ref{F-SEPhisto-rise} the distributions of proton events as a function of the proton rise times (onset-to-peak, given in hours) is shown. As expected the high energy protons (plots on the right) have lower mean/median values of the rise time compared to low-energy (left-sided plots), however the difference is small since both energy channels are relatively close in energy. The trend confirms the well-known observation that high energy particles arrive first and thus tend to reach peak values sooner than the low energy particles. Considering the longitudinal trends, we obtain that the rise of the Wind/EPACT proton events with identified E-origin takes nearly twice as long, compared to the W-events. The result confirms the well-known longitude dependence of the increase of the proton intensity observed in situ. No significant differences can be found between the SC23$^\prime$ to SC24$^\prime$ proton samples in terms of rise time.

\subsubsection{Proton fluence}

\begin{table}[t!]
\small
\caption{The mean/median values of the Wind/EPACT proton fluence ($F_{\rm p}$ in (cm${^2}\,$sr$\,$MeV)$^{-1}$), at low and high energy, over the period 1996$-$2016 in different sub-categories. The number of events used for each calculation is given in brackets.}
\begin{tabular}{llll}
\hline
Proton type             & All events         & Western-origin     & Eastern-origin      \\
\hline
\multicolumn{4}{c}{$F_{\rm 25\,MeV}$} \\
All                     & 1053/557 (361)     & 1116/578 (243)     & 1942/836 (71) \\
extreme                 & 984255/818209 (24) & 758783/619100 (19) & 2.6/2.7  $|\times 10^6$ (5) \\
major                   & 35380/26200 (39)   & 31490/25030 (31)   & 71309/83119 (5) \\
intermediate            & 3075/3795 (85)     & 2709/2619 (59)     & 7428/7258 (15) \\
weak                    & 272/322 (160)      & 231/262 (97)       & 492/427 (38) \\
minor                   & 38/50 (53)         & 36/43 (37)         & 123/181 (8) \\
\hline
\multicolumn{4}{c}{$F_{\rm 50\,MeV}$} \\
All                     & 139/85 (318)       & 159/91 (222)       & 147/87 (64) \\
extreme                 & 45852/51309 (31)   & 49040/48020 (26)   & 32329/106851 (5) \\
major                   & 2346/2175 (39)     & 2085/1827 (31)     & 3468/3344 (7) \\
intermediate            & 295/300 (76)       & 272/290 (54)       & 722/648 (10) \\
weak                    & 25/27 (148)        & 22/24 (93)         & 40/49 (37) \\
minor                   & 3/4 (24)           & 3/4 (18)           & 5/5 (5) \\
\hline
\end{tabular}
\label{T-Fluence_properties}
\end{table}

The proton intensity integrated from the onset to peak time, {\it e.g.}, proton fluence, $F_{\rm p}$ in (cm$^2$ sr MeV)$^{-1}$, is also evaluated for 361 low energy protons and for 318 high energy protons. For several cases, despite the evaluation of onset and peak time, no fluence could be computed due to occasional data gaps during the proton rise time. The values for each proton event are planned to appear only in the on-line catalog. In Table~\ref{T-Fluence_properties} are summarized the mean/median values of the proton fluence for the low ($F_{\rm 25\,MeV}$) and high ($F_{\rm 50\,MeV}$) energy channel for the values of the entire sample (`All events'), and also for the proton intensity sub-samples with origin in west and east. We obtain that the values for the E-proton fluence are larger than the values for the W-sample for all proton intensity categories. The W-sample is similar to the mean/median values of the entire population, since the W-events constitute about $70\%$ of the proton sample.

\begin{figure}[t!]
\centerline{\includegraphics[width=0.99\textwidth,clip=]{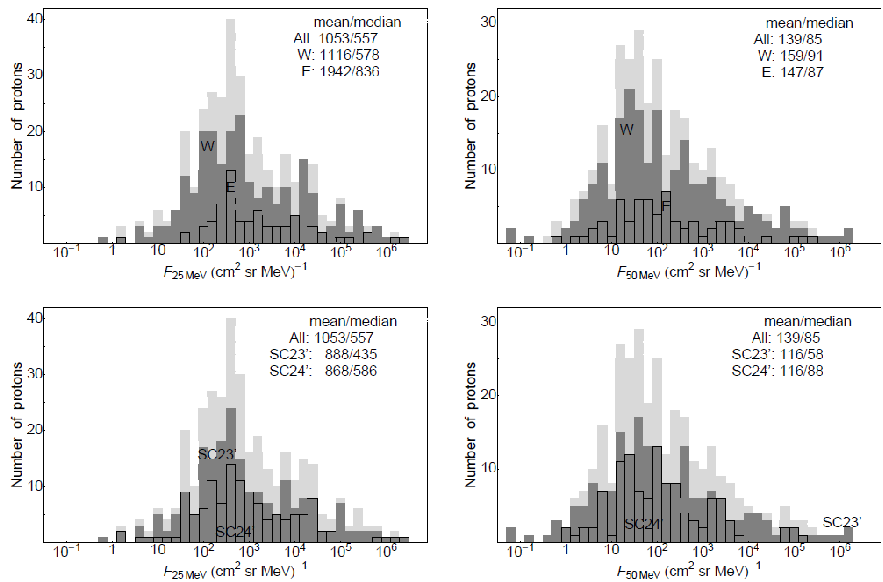}}
\caption{Number of Wind/EPACT proton events over proton fluence in the low (on the left) and high (right) energy channels. Color code as in Fig.~\ref{F-SEPhisto-Jp}.}
   \label{F-SEPhisto-fluence}               
\end{figure}

The distributions of $F_{\rm 25\,MeV}$ and $F_{\rm 50\,MeV}$ with respect to the solar origin longitude and the SC occurrence are presented in Fig.~\ref{F-SEPhisto-fluence}. In either case, a more symmetrical distribution is obtained, compared to the proton intensity. 

In the upper panels of Fig.~\ref{F-SEPhisto-fluence} are shown the E-to-W trends. The larger onset-to-peak fluence $F_{\rm 25\,MeV}$ with origin in the east compared to the respective values of the W-fluence is noticeable as the E-distribution peak is shifted towards larger fluence values. Since the fluence is intensity integration over a duration of the rise time, one could explain the observed results as follows. The eastern $J_{\rm 25\,MeV}$ is only slightly lower than the western one (in terms of mean/median values for the entire sample), however the proton rise time nearly doubled for the E-sample. Thus, the combination of the two parameters (peak intensity and rise time) leads to a larger fluence for the E-proton sample. This is not the case for the high energy channel, however, where the mean/median fluence values are similar (Table~\ref{T-Fluence_properties}). Although the proton rise time is also doubled for the E-sample, the eastern $J_{\rm 50\,MeV}$ is much lower (about half), which reflects the lack of significant difference for $F_{\rm 50\,MeV}$ between the E and W samples.

With respect to the SC (Fig.~\ref{F-SEPhisto-fluence}, lower plots), no clear difference in fluences in SC24$^\prime$ compared to SC23$^\prime$ is obtained, for both energy channels, for the entire proton sample ({\it i.e.}, no preference with location). The exact mean/median values for the proton fluence in the different proton sub-categories are summarized in Table~\ref{T-SEPevents} (last two sections, entitled $F_{\rm 25\,MeV}$ and $F_{\rm 50\,MeV}$). In specific proton categories there are occasional differences, when E and W samples in either SC are compared, however the trends cannot be generalized easily. This outcome could be partly due to the lower sampling of the E-events and lack of statistical significance.

\subsection{Properties of the SEP solar origin}

In this Subsection the properties of the SEP-associated solar origin, SFs and CMEs are given. The properties of the samples of all observed SFs and CMEs (1996$-$2015) were already examined in \inlinecite{2017SunGe..12...11M}. The presented here samples of SEP-producing SFs and CMEs are only a small sub-sample of the entire SF and CME populations, respectively ($\sim$1$-$2\%). We could associate a SF to 292 proton events and a CME to 342 events.

The SFs are naturally classified by their peak intensity in SXR GOES 1$-$8 \AA$\,$channel (known as flare class\footnote{The X-class of SFs is 10$^{-4}$ W$\,$m$^2$ followed by M, C, B, A flare classes, which are 10 times less in flux than the previous. The number following the letter is denoting a multiplier of the SXR flux.}), whereas the CME sample is divided by us into several categories in speed (in km$\,$s$^{-1}$) and AW (in degrees), as given below:
\begin{itemize}
\item slow (s): $V_{\rm CME} < 500$
\item intermediate (i): $500 \leq V_{\rm CME} < 1000$
\item fast (f): $1000 \leq V_{\rm CME} < 2000$
\item extreme (e): $V_{\rm CME} \geq 2000$
\item narrow (n): AW$<$100
\item non-halo (n-h): 100$\leq$AW$<$360
\item halo (h): AW=360
\end{itemize}

The mean/median values and the number of SFs and CMEs related to the entire population of proton events and for W and E-location are summarized in Table~\ref{T-Origin_properties}. Slightly larger values in SXR flare class are obtained for the E compared to W-samples of SFs (for entire sample and also for the sub-categories of X, M and C-class). For the CME speed, however, no clear longitudinal trends are obtained, only the narrow CMEs tend to be slightly faster when originating from the west. 

Considering the first eight years of each SC, the mean/median values for a given SF and CME intensity range are given in Table~\ref{T-SEPorigin}. Here several general dependencies are outlined. Slightly larger mean/median values are noticed in X- and C-class flares in relation to All proton events in SC23$^\prime$ compared to SC24$^\prime$ (M-class is an exception). This trend is the same for the W-flares with respect to SC. The E-flares however show slightly higher values for All and C-class samples in SC24$^\prime$. 

In general, no clear CME speed difference is observed when comparing sub-samples in the two SCs with the exception of the extreme CME group that tends to be faster in SC23$^\prime$. In view of the AW, the overall trend is that the halo CMEs (All, W, E events) are also slightly faster in SC23$^\prime$, compared to SC24$^\prime$.

\begin{table}[t!]
\small
\caption{The mean/median values of the SF class and CME linear speed (in km$\,$s$^{-1}$) in the period 1996$-$2016 in different sub-categories. The number of events used for each calculation is given in brackets.}
\begin{tabular}{llll}
\hline
Event type              & All events         & Western-origin     & Eastern-origin \\
\hline
All SFs                 & M2.8/M3.0 (292)    & M2.7/M2.7 (216)    & M3.8/M4.0 (69) \\
X-class                 & X2.4/X2.0 (67)     & X2.3/X2.0 (49)     & X2.9/X2.2 (18) \\
M-class                 & M3.2/M3.3 (148)    & M3.1/M3.0 (111)    & M3.7/M3.9 (35) \\
C-class                 & C3.4/C3.4 (77)     & C3.3/C3.2 (56)     & C4.0/C4.4 (16) \\
\hline
All CMEs                & 1140/1070 (342)    & 1130/1040 (269)    & 1170/1070 (73) \\
extreme                 & 2410/2410 (24)     & 2430/2410 (19)     & 2320/2400 (5)  \\
fast                    & 1410/1370 (158)    & 1390/1360 (123)    & 1450/1370 (35) \\
intermediate            & 750/760 (129)      & 740/750 (101)      & 780/790 (28)   \\
slow                    & 400/440 (31)       & 410/450 (26)       & 340/370 (5)  \\
halo                    & 1330/1240 (196)    & 1320/1200 (147)    & 1360/1340 (49)  \\
non-halo                & 960/850 (100)      & 970/830 (86)       & 940/870 (14)    \\
narrow                  & 680/640 (44)       & 720/730 (34)       & 570/540 (10)   \\
\hline
\end{tabular}
\label{T-Origin_properties}
\end{table}

\begin{sidewaystable}
\scriptsize
\caption{Mean/median values SF class and CME speed for solar events related to protons. Different sub-categories and time periods as for Table~\ref{T-Origin_properties}, and the respective percentage change as for Table~\ref{T-SEPevents}. Abbreviations: X/M/C: X/M/C flare class; e: extreme in speed CMEs; f: fast; i: intermediate; s: slow; h: halo; n-h: non-halo; n: narrow; W: western origin, E: eastern origin.}
\begin{tabularx}{\textwidth}{llllllll}
\hline
Solar       & \multicolumn{3}{c}{SC23$^\prime$: 09/1996$-$08/2004} & \multicolumn{3}{c}{SC24$^\prime$: 01/2009$-$12/2016} &  Percentage change \\
type        & All events & W & E & All events & W & E & All events/W/E \\
\hline
\multicolumn{8}{c}{{\bf SEP-related flares}} \\
All      & M2.9/M2.9 (170) & M3.1/M3.0 (125) & M3.0/M4.1 (38) & M2.2/M2.6 (90) & M1.8/M2.3 (67) & M3.7/M3.2 (23) & $-$47$\pm$7/$-$46$\pm$8/$-$39$\pm$16   \\
X        & X2.5/X2.0 (40)  & X2.4/X1.8 (33)  & X3.3/X2.6 (7)  & X1.8/X1.6 (17) & X1.7/X1.4 (9)  & X1.9/X1.7 (8)  & $-$58$\pm$12/$-$73$\pm$10/$+$14$\pm$59  \\
M        & M3.2/M3.6 (82)  & M3.0/M2.9 (60)  & M4.2/M5.0 (20) & M3.2/M3.2 (47) & M3.5/M3.6 (36) & M2.4/M2.6 (11) & $-$43$\pm$10/$-$40$\pm$13/$-$45$\pm$21  \\
C        & C3.8/C4.2 (48)  & C4.0/C4.3 (32)  & C3.7/C4.3 (11) & C2.6/C2.5 (26) & C2.4/C2.5 (22) & C4.1/C4.2 (4)  & $-$46$\pm$13/$-$31$\pm$19/$-$64$\pm$21  \\
\hline
\multicolumn{8}{c}{{\bf SEP-related CMEs}} \\
All      & 1110/1030 (188)& 1110/1080 (153)& 1110/910 (35)  & 1110/1020 (122) & 1100/1000 (90) & 1160/1090 (32) & $-$36$\pm$8/$-$42$\pm$8/$-$9$\pm$22   \\
e        & 2420/2440 (12) & 2400/2410 (9)  & 2460 (3)       & 2230/2180 (7)   & 2240/2200 (6)  & 2150 (1)       & $-$42$\pm$28/$-$33$\pm$35/$-$67$\pm$38   \\
f        & 1410/1380 (84) & 1410/1380 (71) & 1400/1370 (13) & 1390/1340 (57)  & 1370/1330 (40) & 1440/1370 (17) & $-$33$\pm$12/$-$45$\pm$11/+31$\pm$48   \\
i        & 750/770 (71)   & 760/770 (55)   & 750/760 (16)   & 750/750 (51)    & 730/730 (39)   & 830/900 (12)   & $-$30$\pm$13/$-$31$\pm$15/$-$25$\pm$29   \\
s        & 400/440 (21)   & 410/450 (18)   & 400/380 (3)    & 400/460 (7)     & 460/480 (5)    & 240 (2)        & $-$67$\pm$15/$-$72$\pm$14/$-$33$\pm$61   \\
h        & 1360/1330 (85) & 1330/1240 (68) & 1470/1380 (17) & 1240/1170 (86)  & 1240/1170 (58) & 1230/1190 (26) & +1$\pm$15/$-$15$\pm$15/+53$\pm$48   \\
n-h      & 1020/890 (66)  & 1050/990 (56) & 850/850 (10)   & 860/760 (31)     & 810/710 (27)   & 1170/990 (4)   & $-$53$\pm$10/$-$52$\pm$11/$-$60$\pm$24   \\
n        & 720/690 (37)   & 730/750 (29)   & 650/560 (8)    & 560/500 (5)     & 780/750 (3)    & 240 (2)        & $-$86$\pm$6/$-$90$\pm$6/$-$75$\pm$20   \\  
\hline
\end{tabularx}
\label{T-SEPorigin}
\end{sidewaystable}

\begin{figure}[t!]    
\centerline{\hspace*{0.015\textwidth}
               \includegraphics[width=0.99\textwidth,clip=]{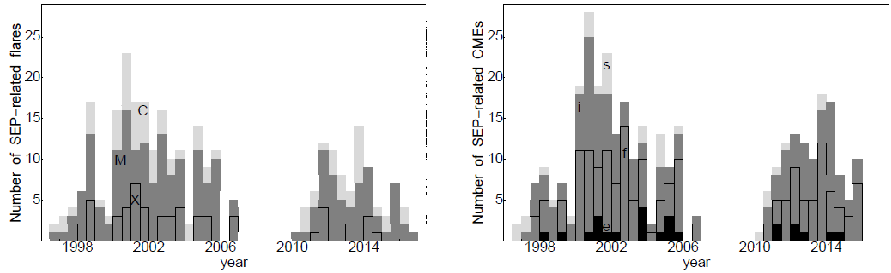}}
\caption{Temporal distributions of flare (left) and CME (right) events related to Wind/EPACT protons. Left: light-gray for C-class flares, dark-gray for M-class and contours $-$ for X-class. Right: Light-gray for slow (s) in speed CMEs, dark-gray $-$ for intermediate (i), contours $-$ for fast (f) and black $-$ for extreme (e) CMEs. See text for further details.}
   \label{F-SEPorigin_year}               
\end{figure}

\begin{figure}[t!]    
\centerline{\hspace*{0.015\textwidth}
               \includegraphics[width=0.99\textwidth,clip=]{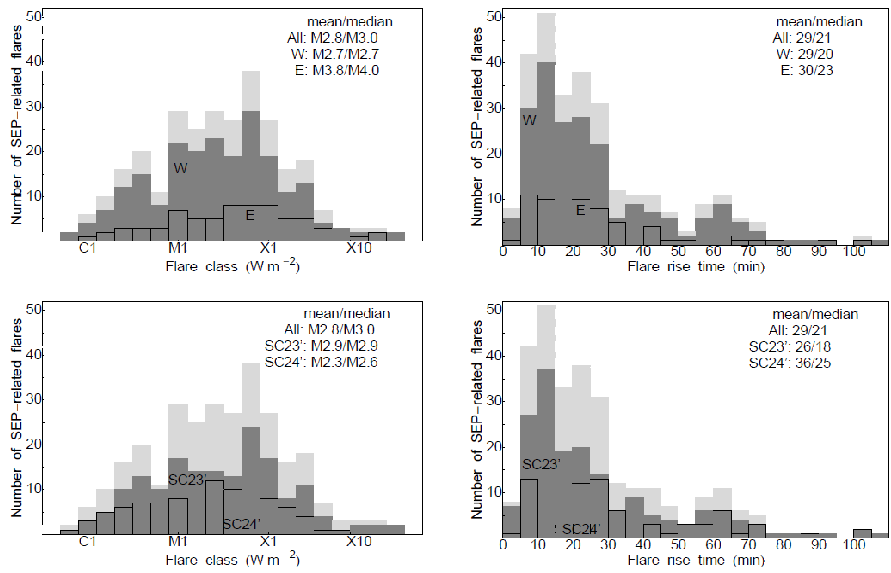}}
\caption{Distributions of Wind/EPACT proton related flare class and rise time. Color code as in Fig.~\ref{F-SEPhisto-Jp}.}
   \label{F-SEPorigin-class}               
\end{figure}

The temporal distributions of SEP-producing flares (on the left) and CMEs (right) are given in Fig.~\ref{F-SEPorigin_year}, where the flare class/CME speed is given in color-code. The majority of the SEP-productive flares are of M-class ($\sim$51\%, 148/292, see also Fig.~\ref{F-SEPorigin-class}), with fewer C ($\sim$26\%, 77/292) and X-class flares ($\sim$23\%, 67/292) related to in situ proton events. For the CMEs, the fast CMEs ($\sim$46\%, 158/342) form the most numerous group giving rise to a proton event. For both SEP-solar origins, the effect of strength vs. appearance is able to explain the results. Namely, weak flares (C-class) and slow CMEs (of the order of the solar wind speed) are not expected to be efficient particle accelerators, thus their number as SEP-producing origin is not very high. In contrast the X-flares and very fast (and halo) CMEs are expected to efficiently energize particles, however they do not occur often\footnote{In the period 1996$-$2016, among all SFs we found $\lesssim$1\% to be of X-class and $\sim$9\% of M-class. Among the CMEs, 2\% are fast and 0.2\% are extreme in speed.}. The exact number of events in each sub-category is given in the respective Table~\ref{T-Origin_properties} and \ref{T-SEPorigin}.

\begin{figure}[ht!]
\centerline{\hspace*{0.015\textwidth}
               \includegraphics[width=0.99\textwidth,clip=]{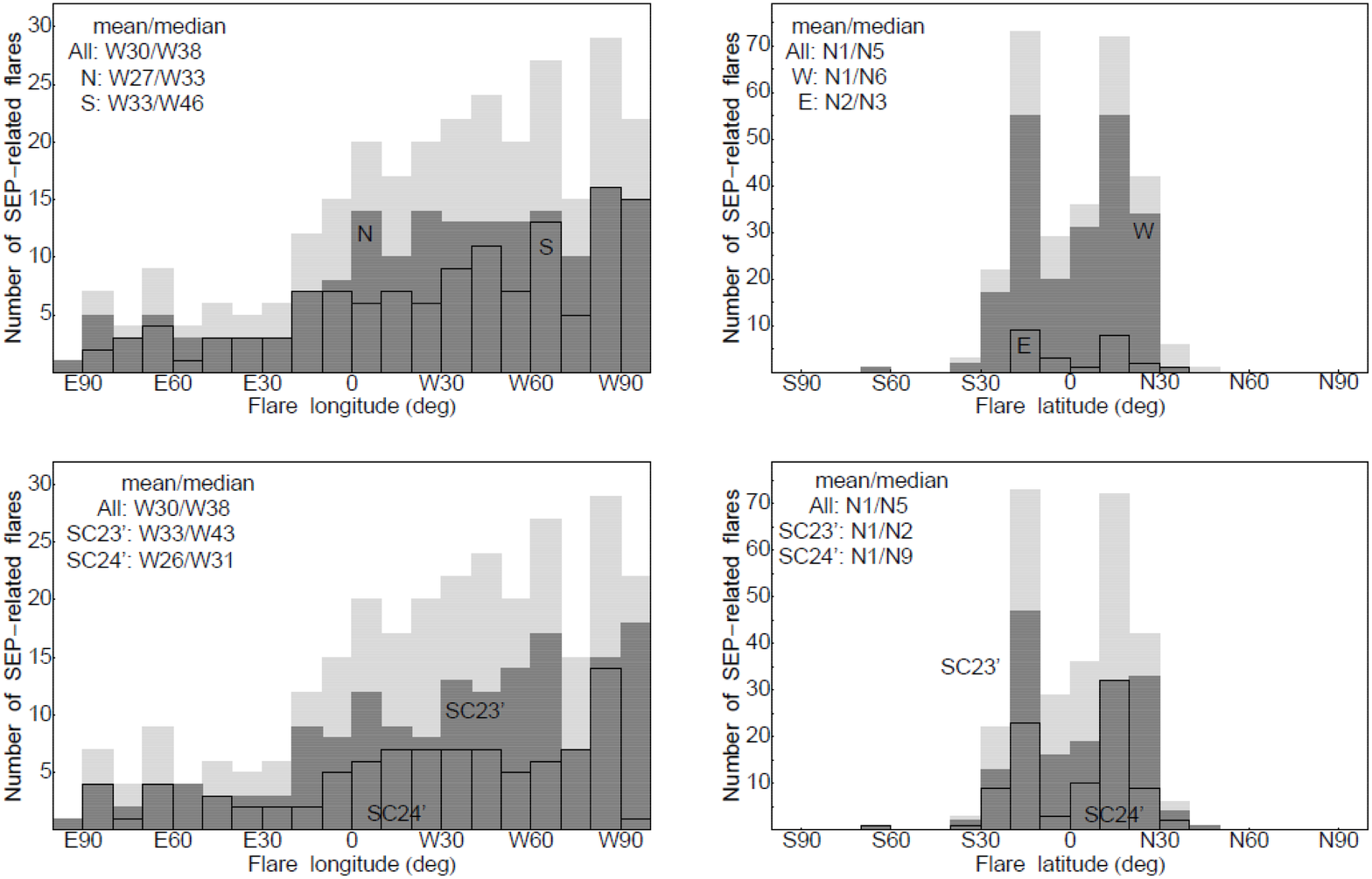}}
\caption{Distributions of Wind/EPACT proton related flare location. Color code as in Fig.~\ref{F-SEPhisto-Jp}.}
   \label{F-SEPOrigin-location}               
\end{figure}

The detailed distribution of several properties of the proton-producing flares are shown in Fig.~\ref{F-SEPorigin-class} (namely flare class and rise time) and \ref{F-SEPOrigin-location} (flare longitude and latitude). Again for comparison, the distribution of the entire flare sample is presented in light-gray color, on the background. We found that the histograms for low and high energy Wind/EPACT protons to be similar, thus only the flares related to the low energy proton sample are further presented. 

The distribution of E-flares is slightly shifted to larger flare class values (Fig.~\ref{F-SEPorigin-class}, left), also confirmed by the mean/median values: M3.8/M4.0 for E-flares, compared to M2.7 for W-flares, respectively. In terms of flare rise times (Fig.~\ref{F-SEPorigin-class}, right), the distribution is fairly flat for E-flare sample with similar mean/median values with respect to longitude. When the flare sample is split into events occurring during SC23 and SC24 (and independent on location, Fig.~\ref{F-SEPorigin-class}, lower plots), the differences in flare class are not clearly identifiable, however slightly longer rise times are obtained in SC24$^\prime$ ({\it e.g.}, 36/25 min for the mean/median flare rise time compared to 26/18 min, respectively).

The distribution of the longitude and latitude of the SEP-producing flares is shown in Fig.~\ref{F-SEPOrigin-location}. With respect to the longitude (plots on the left), the flare sample is divided into northern [N] and southern [S] latitude and into flare samples occurring in SC23$^\prime$ and SC24$^\prime$. The flares originating at the northern solar hemisphere have a longitudinal distribution with mean/median values at W27/W33, whereas the southern flare sample is shifted westward to W33/W46. With respect to SC, the events in the SC23$^\prime$ have a distribution that has mean/median values at W33/W43, whereas for the ongoing SC, the value is shifted eastward and flattens at around W30. 

The latitudinal distribution (Fig.~\ref{F-SEPOrigin-location}, right plots) is shown for E-to-W and SC behavior. Overall, the distribution is double peaked, centered at about S20 and N20 helio-latitudes, respectively. The entire sample (irrespective of flare intensity, shown by light-gray color on the plots) has mean/median values that are in the northern hemisphere (N1/N5), with nearly the same values (N1/N6) for the western sample, shown in dark color, and slightly lower for the eastern sample (N2/N3) shown in contours. When calculated over the first eight years of each SC but now irrespective on the longitude, the mean/median values range from N1/N2 (for SC23$^\prime$ in dark color) to N1/N9 (for SC24$^\prime$ in contour). Since earlier studies have reported S-to-N trend with the SC \citep{2013AdSpR..52.2102C}, we investigated why for the entire Wind/EPACT proton-related flare sample no such clear southern trend in SC23$^\prime$ is seen. Our analysis showed that when only proton events of larger intensity are considered, the S-to-N trend becomes more evident: namely when the large (or/and extreme) proton sample is selected, the related flare sample in SC23$^\prime$ has median value at southern longitudes, namely at S3 (S4), respectively.

\begin{figure}[t!]    
\centerline{\hspace*{0.015\textwidth}
               \includegraphics[width=0.99\textwidth,clip=]{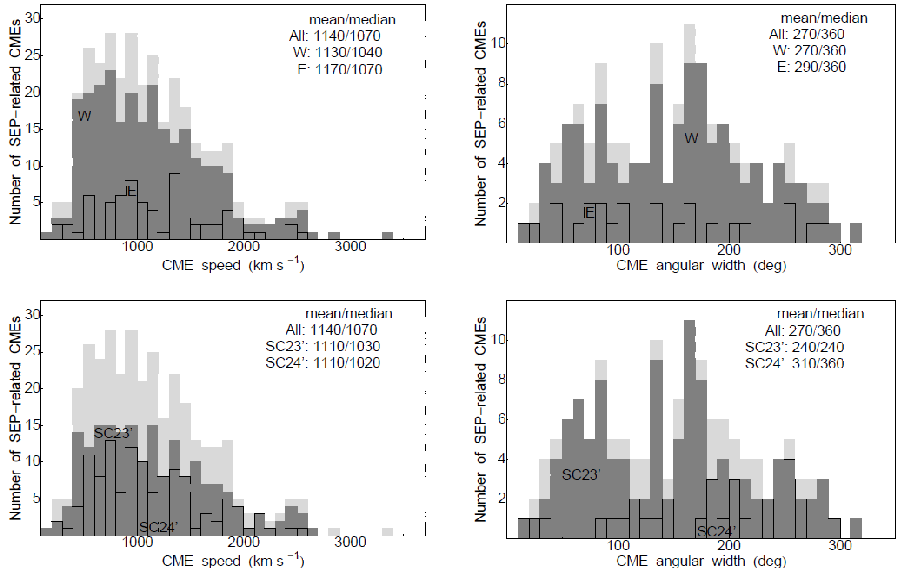}}
\caption{Distributions of Wind/EPACT proton related CME speed and angular width. For visibility purpose, the bin of halo CMEs is omitted, however the mean/median values are calculated using the entire sample. Color code as in Fig.~\ref{F-SEPhisto-Jp}.}
   \label{F-SEPorigin-CME}               
\end{figure}

Finally, we present the distributions of proton-producing CMEs, namely the number of CMEs vs. their speed and AW, shown in Fig.~\ref{F-SEPorigin-CME}. For each case, the samples for W/E origin CME-events and SC23$^\prime$/SC24$^\prime$ are shown. There is no difference for the entire sample in mean/median speed neither when longitude sub-samples are considered, nor with respect to SC. For the AW, to optimize the presentation on the plots, the CMEs without the halo events are given, however the calculation of the mean/median values as shown on the plots considers the entire sample. There is no clear difference in AW with respect to longitude (upper plot). However, with respect to the SC, the AW for the CME sample in SC23$^\prime$ peaks at 240 degree, whereas in the ongoing SC, the CMEs are exclusively wider (the median values denotes halo CMEs), and the distributions are visually different. For the entire sample, $\sim$58\% (197/342) are halo CMEs, whereas the fraction of halo-CMEs in SC23$^\prime$ is $\sim$45\% (85/188) compared to $\sim$72\% (87/121) in SC24$^\prime$. This finding confirms earlier results for a larger fraction of halo-CMEs in the ongoing SC \citep{2012AIPC.1500...14G}. Note that the number of halo-CME events is nearly the same, however the entire CME sample in the ongoing SC is less which increases the halo-CME fraction. 

In addition, we evaluated the linear correlations between the flare class ($I_{\rm X}$)/ fluence ($F_{\rm X}$) and the CME speed ($V_{\rm C}$)/AW. When considering the entire proton-related solar origin sample, larger correlation coefficients are obtained for $r_{F_{\rm X}V_{\rm C}}=0.47\pm 0.06$ compared to $r_{I_{\rm X}V_{\rm C}}=0.37\pm 0.06$. The respective correlations with the AW are slightly lower compared to $V_{\rm C}$, namely $r_{F_{\rm X}{\rm AW}}=0.43\pm 0.05$ and significantly lower for $r_{I_{\rm X}{\rm AW}}=0.29\pm 0.05$. The longitudinal dependence for the entire sample shows a slight increase of all coefficients only for the eastern located solar origin. When we compare the correlations of the proton origin sample in SC23$^\prime$ and SC24$^\prime$ (irrespective on longitude trends), a slight increase in the number of the correlations during the ongoing SC are obtained. Nevertheless, most of the differences of the correlation coefficients are within the uncertainty ranges.

\subsection{Percentage change in SC24 compared to SC23}

Our results confirm that the number of protons in the ongoing SC24 is less than in the previous SC23. In order to present quantitatively the change observed in SC24 compared to SC23 the percentage change is used. It is defined as the ratio of the difference between the event number in SC24 and SC23 to the event number in SC23. The negative/positive values denote the decreasing/increasing trend in SC24 compared to the same period in SC23. In Tables \ref{T-SEPevents} and \ref{T-SEPorigin} are listed all values of the percentage change, calculated for the entire samples (protons, flares and CMEs) and separately for the W and E sub-samples.

There is a decrease for low and high energy protons (as a whole) by about 40\% in the ongoing SC, that is roughly the same for the events of intensity $J_{\rm 25\,MeV}<10$, within the uncertainties (see Table~\ref{T-SEPevents}, last column). The drop is largest (60$-$65\%) for the extreme proton events and smallest ($\sim$25\%) for the weak proton events. The results are reminiscent to the obtained values for the western origin protons (where the drop of weak events is less, $\sim$20\%). For the eastern events, due to the small number of events, the trends are not easy to interpret due to the large uncertainties. Occasionally, an increase of protons in the SC24 is obtained for specific proton sub-categories, however the error bars are too large to be conclusive.

For completeness we calculated the probability of occurrence based on the number of events that led to the evaluation of the proton fluence. As expected, the results are similar to those based on the proton intensity, since the event samples are nearly the same.

In a similar way, the percentage change for SEP-producing flares and CMEs is calculated (see Table~\ref{T-SEPorigin}, last column), that also are fewer in numbers during the first eight years of SC24, compared to the same period in SC23. The entire flare sample decreases nearly in half ($47 \pm 7$ \% less) whereas the entire CME sample drops by about $36 \pm 8$ \%. The X-class flares decrease by about 60\%, whereas on the west this sample decreases by about 70\% (the eastern X-class flares are too few). The C/M-class flares drop from about 45\% for the entire sample to about 30$-$40\% when occurring on the west, respectively. 

For the CMEs, we obtain no change in the SC for the halo CMEs with some positive trend for the eastern halos (however, with very large uncertainty). The non-halo CMEs show a larger drop for the eastern CMEs (about 60\%) compared to all/western (about 50\%) and even larger drops are obtained for the narrow CMEs (90\% for the western sample). Results for the different CME speed categories are summarized in Table~\ref{T-SEPorigin}. Note that larger uncertainties are obtained for the CME subsamples with origin on the east.

\subsection{Occurrence probabilities}

We calculated the occurrence probabilities (in \%) and the results are summarized in Table~\ref{T-SEPprob}. The probabilities are calculated as the ratio of SEP-producing solar samples ({\it e.g.}, number of flares and CMEs adopted from the Wind/EPACT catalog, respectively) to the entire solar event sample (adopted from flare and CME catalogs). Subsequently, the solar event sample is divided according to specific criteria ({\it i.e.}, speed and AW for the CMEs and for the flares there are sub-categories for C, M, X-class).

When considering the entire sample of flares and CMEs, we obtain only about 1\% probability to produce a particle event. This is due to the large number of solar eruptive events observed remotely, compared to the SEPs observed in situ near Earth. The probability rises very slowly with increase of flare class and CME speed. 

The largest values for the occurrence probabilities are reached for the largest flares (X-class in SXR intensity) and CMEs (extreme in speed, namely $\geq$2000 km$\,$s$^{-1}$, Table~\ref{T-SEPprob}, last column). Namely, X-class flares have about 39\% probability to be followed by a proton event observed in situ (only 7\% for M-class), and the extreme CMEs have about 50\%. The probability for the halo-CME sample (irrespective on speed) to produce protons ahead of Earth is slightly below 30\%, whereas for the non-halo CME sample it is only a few percent. 

In general no differences in occurrence are obtained for protons with respect to the SC (comparing columns~2 and 3 from Table~\ref{T-SEPprob}), due to the simultaneous decline in number of the proton-producing and the entire samples of both SFs and CMEs, in the respective time periods. The only exception for a statistically significant difference is obtained for the fast CME sample occurring in SC24, when the occurrence probability shows an increasing trend ($31 \pm 3$) compared to the value in SC23 ($21 \pm 2$). For the flares there is no SC dependence within the uncertainty ranges. 

\begin{table}[t!]
\small
\caption{Occurrence probabilities and its binomial error (in \%) of all Wind/EPACT protons for different ranges of flare class and CME speed and angular width.}
\begin{tabular}{lccc}
\hline
Solar            & SC23$^\prime$:     & SC24$^\prime$:    & Entire period: \\
event type       & 09/1996$-$08/2004  & 01/2009$-$12/2016 & 01/1996$-$12/2016\\
\hline
All SFs          & $1$           & $1$           & $1$       \\
C-class          & $0.4$         & $0.3$         & $0.4$     \\
M-class          & $8 \pm 1$     & $7 \pm 1$     & $7 \pm 1$  \\
X-class          & $40 \pm 5$    & $38 \pm 7$    & $39 \pm 4$ \\
\hline
All CMEs         & $2$           & $0.9$         & $1$ \\
slow             & $0.4$         & $0.7$         & $0.1$  \\
intermediate     & $2.5$         & $2.5$         & $2$  \\
fast             & $21 \pm 2$    & $31 \pm 3$    & $25 \pm 2$  \\
extreme          & $48 \pm 10$   & $64 \pm 15$   & $50 \pm 7$  \\
narrow           & $0.5$         & $0.04$        & $0.2$  \\
non-halo         & $6 \pm 1$     & $2$           & $4$  \\
halo             & $29 \pm 3$    & $28 \pm 2$    & $28 \pm 2$  \\
\hline
\end{tabular}
\label{T-SEPprob}
\end{table}

\subsection{Statistical analysis}

In this subsection we complete the statistical analysis between the properties of the protons from one side and the SFs and CMEs, from another. Namely, the linear and partial correlation coefficients (log$_{10}-$log$_{10}$ values) are calculated between the relevant parameters of the protons ($J_{\rm p}$ and $F_{\rm p}$) and the parameters of the particle origin (SF class, onset-to-peak flare fluence, CME speed and AW). All results are summarized in the Appendix (Tables \ref{T-lowE-cc_Jp}$-$\ref{T-highE-cc_Fp}) for the low and high proton energy and for the proton intensity and fluence, respectively.

\subsubsection{Linear correlation analysis} 

A well-explored method in SEP studies used for deducing the solar origin influence is by calculating Pearson correlations (log or linear). In the present work, several samples are considered. The entire proton sample with their associated origin (denoted with `All events' in the respective tables) over the entire period of interest is the most numerous sample. In addition this sample is divided according to the helio-location of the proton related SFs and CMEs into W and E samples. The correlation coefficients of these three samples are shown in the last column in Tables~\ref{T-lowE-cc_Jp}$-$\ref{T-highE-cc_Fp} (for the entire period). For the purpose of our work, we selected the proton sample over SC23$^{\prime}$ (col.~2) and SC24$^{\prime}$ (col.~3, respectively) and in addition specified sub-samples in W vs. E. All calculations are done for the low and high energy protons. 

The bootstrapping procedure of \cite{2003psa..book.....W} is used for evaluating all uncertainty of the linear correlation coefficients, in the same way as discussed in \cite{2013SoPh..282..579M}. Namely, from the original sample we randomly select events (allowing the same event to be chosen more than once) and compile 1000 alternative samples with their respective linear correlations. The final uncertainty on all linear correlation coefficients reported throughout this work is the standard deviation based on the respective 1000 correlation coefficients.

Overall, for the entire sample and the entire period, the largest correlations are those between the proton peak intensity and the CME speed, followed by those with the SXR flare class/fluence, and the lowest are with AW. The W-subsample has the same correlations within the uncertainties as for the entire sample. In contrast, the E-subsamples show increasing values for the correlations, namely between $J_{\rm p}-I_{\rm X}$, $J_{\rm p}-V_{\rm C}$ and $F_{\rm p}-I_{\rm X}$, $F_{\rm p}-V_{\rm C}$, compared to the ones using the entire sample, with the exception of E-value for $J_{\rm p}-F_{\rm X}$ when a decreasing trend is obtained. The latter trends are more pronounced for the eastern sub-samples during SC23$^{\prime}$, whereas in SC24$^{\prime}$ there are no correlations obtained with the $F_{\rm X}$ and very low with AW.

The entire proton sample with its enriched statistics ($\sim$250$-$340 data points and uncertainty of the order of 0.05) has correlations around 0.4 with the flare (similar to \opencite{2003GeoRL..30lSEP3G}) and 0.5 with the CME speed and 0.3 with the AW. These values are lower than the 0.6 value obtained with both flares/CMEs in SC23 by \inlinecite{2010JGRA..11508101C,2013SoPh..282..579M} and with the CMEs (partial analysis in SC23) by \inlinecite{2003GeoRL..30lSEP3G}. Even larger values (0.7) between protons and CME speed are reported by \inlinecite{2001JGR...10620947K} and protons with flare class (0.5) in an earlier study by \inlinecite{1982JGR....87.3439K}. Energy trends in the correlations have been shown by \inlinecite{2015SoPh..290..841D}, and for the 25-to-50 MeV protons they report correlations of about 0.6 with the flare class and 0.54-to-0.46 with the CME speed. Only the latter correlation value is confirmed also by our current analysis, although log-values are used. 

Similarities are also found with the recent study by \inlinecite{2016JSWSC...6A..42P} that reports linear correlations of 0.51 between 30 MeV protons and CME speed and 0.43 when considering 60 MeV protons. The correlations with the solar flare class however are about 0.47 for both energy channels (note that \inlinecite{2016JSWSC...6A..42P} uses the same data as \opencite{2015SoPh..290..841D}). For the 30 MeV proton fluence and the flare class \inlinecite{2016JSWSC...6A..42P} reports the slightly lower values of 0.42, whereas with the flare fluence the correlation rises to 0.49. These results are at the upper limit of the respective results obtained by us.

\subsubsection{Partial correlation analysis} 

An earlier study by \cite{2015SoPh..290..819T} utilized for the first time the partial correlation in SEP studies and showed that SF onset-to-peak fluence ($F_{\rm X}$) is to be considered in correlation analysis instead of the usually utilized flare class ($I_{\rm X}$).

In this study we consider the first order partial correlation coefficients. Name\-ly, when the partial correlations between proton ($J_{\rm p}$ or $F_{\rm p}$) and flare ($I_{\rm X}$ or $F_{\rm X}$) parameters are calculated, from the correlations is excluded the influence either of $V_{\rm C}$ or AW. Alternatively, when the correlation with the CME speed and AW is performed, the influence of $I_{\rm X}$ or $F_{\rm X}$ is excluded. The generalized expression reads as $r_{xy.z}=(r_{xy}-r_{xz}r_{yz})/((1-r^2_{xz})(1-r^2_{yz}))^{0.5}$, where the influence of $z$ (control variable) is excluded from the correlation between $x$ and $y$. In our study the variable $x$ is $J_{\rm p}$ or $F_{\rm p}$, whereas $y$ and $z$ are $I_{\rm X}$, $F_{\rm X}$, $V_{\rm C}$, or AW, respectively. 

We calculated the partial correlation coefficients for the entire proton sample and also for the W and E sub-samples. In addition, the entire period of observation, SC23$^\prime$ and SC24$^\prime$, is considered. The uncertainty range for each partial correlation is calculated by the propagation of uncertainty method using the formula given above. All results are summarized in the Appendix (Tables~\ref{T-lowE-cc_Jp}$-$\ref{T-highE-cc_Fp}). The correlation analysis using more control variables (namely second, third and higher order partial correlations) is, however, beyond the scope of the present study.

In general, lower values for the partial coefficients over the entire sample are obtained compared to the linear coefficients. When excluding the influence of $V_{\rm C}$, it is found that $r_{J_{\rm p}I_{\rm X}.V_{\rm C}}\sim 0.2$ with a higher value for $r_{J_{\rm p}F_{\rm X}.V_{\rm C}}\sim 0.3$. Thus we confirm the earlier results by \cite{2015SoPh..290..819T} based on a smaller event sample. Alternatively, the partial correlation of $r_{J_{\rm p}V_{\rm C}.I_{\rm X}}$ is $\sim$0.25, but increases to $\sim$0.35 when the control variable is $F_{\rm X}$ (the differences are not statistically significant, however). Similar results, within the uncertainty ranges, are obtained when performing the correlation with the AW. A slight increase is noticed only for the latter values in the E-sample, however it is not statistically significant. The results are valid both for the low and high energy protons (see Tables~\ref{T-lowE-cc_Jp} and \ref{T-highE-cc_Jp} in the Appendix).

The results for the respective partial correlations in SC23$^{\prime}$ are statistically similar to those for the entire sample. In SC24$^{\prime}$ a slight increase in the correlations is obtained, however within very large uncertainty ranges due to small event samples. Namely, the sub-samples with W and E origin in SC23 and SC24 contain very small number events and thus no statistical significant difference is obtained (due to the large error bars), especially when the individual SC is considered. Nevertheless, there is a tendency of increased correlations between the protons and $V_{\rm C}$/AW to above 0.4 (E-sample in SC23$^{\prime}$ and W-sample in SC24$^{\prime}$), when the effect of $F_{\rm X}$ is excluded. The same value is obtained for the partial correlation with the $F_{\rm X}$ when using AW as a control parameter. The Eastern sample in SC24 presents a striking exception with its very low partial correlation coefficients (occasionally, the uncertainty could not be evaluated and is regarded as uncertain by us). The above trends are similar for the low and high energy proton samples, and also when we utilize particle fluence (see Tables~\ref{T-lowE-cc_Fp} and \ref{T-highE-cc_Fp} in the Appendix). 

Despite the large uncertainty ranges, the partial correlation analysis shows a link between the protons and $F_{\rm X}$ even after excluding the underlying influence of the CME speed and especially AW (since there is a moderate liner correlation found between the CME and flare parameters, see Section~3.2). Other flare parameters, {\it e.g.}, based on hard X-ray, radio or newly suggested by theoretical models could be potentially explored, compared to the usually utilized flare class, $I_{\rm X}$, which turns out to be influenced by the CME.

Finally, we have to point out that during correlation studies a rather crude approximation scheme is adopted, namely by selecting a single, `representative' value of protons, flares and CMEs ({\it e.g.}, peak values in intensity or speed) in order to describe a particle acceleration and propagation process that is expected to be time-, energy- and location-dependent.  

\section{Conclusion}
\label{S-Conclusion}

In this study we present the overall characteristics of the proton events from the finalized Wind/EPACT catalog as well as the properties of the flares/CMEs associated to these protons. All comparisons between the first eight years of solar cycles 23 and 24 are presented for the first time here. Below the main outcomes of our work are summarized, focusing on general results from the correlation analysis and the obtained longitudinal and solar cycle trends.

\subsection{Wind/EPACT proton events}

\begin{itemize}
 \item[{\it Energy trend}] The 25 and 50 MeV energy channel samples contain similar number of events, 429 and 397, respectively. The respective mean/median values of the associated solar origin samples (in terms of flare class and CME speed) are nearly the same for the two energy channels.
 \item[{\it Intensity trend}] The peak intensity of the higher energy channel ($J_{\rm 50\,MeV}$) is about one order of magnitude lower compared to the lower energy channel ($J_{\rm 25\,MeV}$). Nearly half of the entire proton population consists of minor and weak in intensity events ({\it i.e.}, $J_{\rm 25\,MeV} < 0.01$ and $J_{\rm 50\,MeV} < 0.001$, respectively).
 \item[{\it Solar origin location trend}] From the entire proton sample, 32\% (136/429) could not be associated to solar flares, 20\% (87/429) could not be associated to CMEs and 14\% (61/429) could not be related to either of the solar origin. Considering the combined information from the flare-AR and CME-MPA, 66\% (283/429) protons of western and 20\% (84/429) of eastern directivity are obtained. The results are similar when using the high energy proton sample. This finding is consistent with earlier SEP studies.
 \item[{\it Solar cycle trend}] The entire sample of (high and low energy) protons shows a decrease of about 40\% in occurrence in the ongoing SC24, based on 8 years of comparison. Similar percentage is obtained for minor and intermediate in intensity protons. The largest in intensity protons are much less (to about 60$-$65\%). This trend is similar (within uncertainties) for the western origin protons, whereas for the eastern protons there are occasional differences in some of the $J_{\rm p}$-categories, however not statistically significant. In comparison,  \cite{2015ICRC..M} reported 23\% of decrease of 10 MeV protons based on 5.8 year solar cycle-comparison, which is similar to $\sim$22\% drop of GOES protons reported by \cite{2012AIPC.1500...14G} based on 4.5 years, whereas \cite{2013AdSpR..52.2102C} (their Table~3) presents only $\sim$9\% drop of the SEPs (sum of major, minor and weak but also associated to type II radio bursts), during the first 3 years.
 \item[{\it Occurrence probabilities of protons}] With the increase of the flare class and CME speed, larger probability of occurrence for $\sim$25 and $\sim$50 MeV protons is obtained. The trends are consistent with the results presented by \cite{2015SoPh..290..841D}, using data over the entire SC23 and different sub-division in flare class and CME speed. With respect to the SC, however, no difference is noticed in the probability of proton occurrence for the majority of the flare class or CME speed range, within the uncertainty.
\end{itemize}

\subsection{SEP origin: flares}
\begin{itemize}
 \item[{\it Intensity trend}] M-class flares are the most abundant flare class producing proton events (51\% of the entire proton sample), followed by C-class (26\%) and X-class (23\%).
 \item[{\it Longitudinal trend}] Proton-related flares originating at eastern location have sli\-ghtly larger flare class compared to the western flares. There is no longitudinal difference with respect to the flare rise time.
 \item[{\it Latitudinal trend}] The flares at the northern hemisphere are positioned slightly to the disk center compared to the southern flares. 
 \item[{\it Solar cycle trend}] Slightly larger flare class is observed in SC23, compared to SC24 for all flares (confirming earlier tendencies reported by \opencite{2013AdSpR..52.2102C}). Flare rise time is slightly longer in SC24$^\prime$. There is a slight tendency in SC23$^\prime$ for flares to occur to the west compared to those in SC24$^\prime$. The latitudinal trend with SC, as discussed by earlier works, e.g., \cite{2013AdSpR..52.2102C} and references therein, appears only for the large intensity proton sample (southern shift for SC23$^\prime$ and vice versa).
\end{itemize}

\subsection{SEP origin: CMEs}
\begin{itemize}
 \item[{\it Projected speed and width trends}] The fast and intermediate CMEs are the most numerous sub-category (46\% and 39\%, respectively). The halo CMEs (irrespective of speed) constitute 58\% from the proton-producing CME sample.
 \item[{\it Longitudinal trend}] No clear velocity and angular width trends is observed for the entire distribution of proton-producing CMEs propagating to the east or to the west.
 \item[{\it Solar cycle trend}] There are more halo CMEs in SC24$^\prime$ (as shown also by \opencite{2012AIPC.1500...14G,2013AdSpR..52.2102C}, both based on the rise phases of both solar cycles). Slightly faster speeds are obtained for the extreme and halo CMEs.
\end{itemize}

\subsection{Correlation analysis}
In this work the linear and partial correlation coefficients are calculated. We used $log_{10}J_{\rm p}$ ($log_{10}F_{\rm p}$) from one side and $log_{10}I_{\rm X}$ ($log_{10}F_{\rm X}$) or $log_{10}V_{\rm C}$ ($log_{10}{\rm AW}$), from another. Several general trends in the correlation analysis are described below. The exceptions and statistical significance of the results are given in the respective tables in the Appendix.
\begin{itemize}
 \item[{\it Linear correlations}] The correlations of the $F_{\rm p}$ are slightly higher than the respective ones of the $J_{\rm p}$ confirming earlier results (\opencite{2015SoPh..290..819T,2016JSWSC...6A..42P}). The correlation coefficients with the CME speed are in general higher compared with SF class/fluence, but mostly not statistically significant. There is an increase in the correlations with CME speed for the E-sample in SC23$^\prime$.
 \item[{\it Partial correlations}] The partial correlations are lower than the linear as overall values, as shown for the first time by \cite{2015SoPh..290..819T} based on smaller event sample. The results show that the relationship with CME speed is of the order of 0.25/0.35 after removing the contribution of the flare class/flu\-ence, respectively, whereas the relationship with SF class is of the order of 0.2 after removing the CME influence. When using the flare fluence, instead of class, correlations of about 0.3 are obtained. The respective correlations with AW are 0.2/0.3, depending on the control parameter used, flare class/fluence, respectively. Slight increase for the correlations over the E-sample is noticed, whereas the results for the W-sample are similar to those for the entire sample. Increase is observed in SC24$^\prime$, for some of the partial correlations.
\end{itemize}

\subsection{Future prospects}

The cataloging of solar events is expected to cover the entire solar cycle 24. Subsequently, a comparison between the entire in length solar cycles can be performed that will allow to deduce the final occurrence trends of solar eruptive events. The small event size of SEP events is one of the reasons for the large uncertainties on the correlations between the particles and the respective solar origin thus larger particle samples are desirable. The utilized here partial correlations are based on the exclusion of one out of many solar parameters. The use of multiple control variables can be the research topic of future detailed analysis. In addition to the linear and partial correlations, other statistical methods could be employed to disentangle the inter-correlations between the particles and the other solar phenomena. For example, principal component analysis, cluster analysis, etc., can be potentially implemented. In addition to the academic applications, large data samples (SEP, flare, CME data) are employed in different models for building, testing and validating the performance of forecasting methods. Empirical models are based on such long series of data and their performance is trained on an actually observed set of events. Forecasting and theoretical modeling need to comply with observations, which requires a detailed analysis and compilation of event lists.

\begin{acks}
We acknowledge the open data policy from the CDAWeb data base, GOES flare listings and CDAW LASCO CME catalog. The CME catalog is generated and maintained at the CDAW Data Center by NASA and the Catholic University of America in cooperation with the Naval Research Laboratory. SOHO is a project of international cooperation between ESA and NASA. MVCD thanks FAPESP project 2016/05254-9. 
\end{acks}
\appendix   

\begin{sidewaystable}
\tiny
\caption{Table of log$-$log correlation coefficients between low energy proton intensity $J_{\rm 25\,MeV}$ and flare/CME properties. The number of events used for specific calculation is given in brackets. u - uncertain.}
\begin{tabular}{clll}
\hline
Correlation             & SC23$^\prime$: 09/1996$-$08/2004           & SC24$^\prime$: 01/2009$-$12/2016           & Entire period: 01/1996$-$12/2016 \\
coefficient             & All events/W-origin sample/E-origin sample & All events/W-origin sample/E-origin sample & All events/W-origin sample/E-origin sample \\
\hline
\multicolumn{4}{c}{{\bf Linear}} \\
$r_{J_{\rm p}I_{\rm X}}$ & $0.36^{\pm 0.08}$ (171)/$0.37^{\pm 0.08}$ (129)/$0.35^{\pm 0.20}$ (38) & $0.35^{\pm 0.08}$ (90)/$0.35^{\pm 0.10}$ (68)/$0.44^{\pm 0.16}$ (23)  & $0.41^{\pm 0.05}$ (293)/$0.40^{\pm 0.06}$ (221)/$0.44^{\pm 0.10}$ (68) \\
$r_{J_{\rm p}F_{\rm X}}$ & $0.34^{\pm 0.08}$ (169)/$0.36^{\pm 0.10}$ (129)/$0.31^{\pm 0.17}$ (37) & $0.33^{\pm 0.08}$ (89)/$0.44^{\pm 0.09}$ (67)/$0.02^{\pm 0.24}$ (22)  & $0.38^{\pm 0.05}$ (290)/$0.42^{\pm 0.07}$ (220)/$0.30^{\pm 0.10}$ (67) \\
$r_{J_{\rm p}V_{\rm C}}$ & 0.52$^{\pm 0.05}$ (188)/$0.49^{\pm 0.06}$ (153)/$0.70^{\pm 0.08}$ (35) & $0.45^{\pm 0.06}$ (120)/$0.47^{\pm 0.08}$ (88)/$0.44^{\pm 0.11}$ (32) & $0.49^{\pm 0.04}$ (340)/$0.47^{\pm 0.04}$ (267)/$0.61^{\pm 0.06}$ (73) \\
$r_{J_{\rm p}{\rm AW}}$  & $0.40^{\pm 0.05}$ (188)/$0.40^{\pm 0.06}$ (153)/$0.38^{\pm 0.09}$ (35) & $0.23^{\pm 0.06}$ (120)/$0.25^{\pm 0.06}$ (88)/$0.21^{\pm 0.14}$ (32) & $0.31^{\pm 0.04}$ (340)/$0.31^{\pm 0.05}$ (267)/$0.32^{\pm 0.07}$ (73) \\
\hline
\multicolumn{4}{c}{{\bf Partial}} \\
$r_{J_{\rm p}I_{\rm X}.V_{\rm C}}$ & $0.19^{\pm 0.10}$ (153)/$0.12^{\pm 0.20}$ (40)/$0.22^{\pm 0.34}$ (31) & $0.23^{\pm 0.17}$ (58)/$0.24^{\pm 0.27}$ (36)/$0.20^{\pm 0.21}$ (22) & $0.22^{\pm 0.07}$ (267)/$0.23^{\pm 0.09}$ (180)/$0.20^{\pm 0.12}$ (87) \\
$r_{J_{\rm p}I_{\rm X}.{\rm AW}}$  & $0.18^{\pm 0.09}$ (153)/$0.17^{\pm 0.21}$ (40)/$0.26^{\pm 0.25}$ (31) & $0.25^{\pm 0.12}$ (58)/$0.26^{\pm 0.19}$ (36)/$0.22^{\pm u}$ (22) & $0.21^{\pm 0.06}$ (267)/$0.22^{\pm 0.07}$ (180)/$0.28^{\pm 0.09}$ (87) \\
$r_{J_{\rm p}F_{\rm X}.V_{\rm C}}$ & $0.25^{\pm 0.11}$ (152)/$0.21^{\pm 0.20}$ (40)/$0.28^{\pm 0.32}$ (30) & $0.26^{\pm 0.20}$ (57)/$0.29^{\pm 0.26}$ (35)/$0.07^{\pm 0.33}$ (22) & $0.29^{\pm 0.08}$ (265)/$0.29^{\pm 0.10}$ (179)/$0.28^{\pm 0.12}$ (86) \\
$r_{J_{\rm p}F_{\rm X}.{\rm AW}}$  & $0.28^{\pm 0.10}$ (152)/$0.23^{\pm 0.20}$ (40)/$0.44^{\pm 0.26}$ (30) & $0.46^{\pm 0.16}$ (57)/$0.45^{\pm 0.21}$ (35)/$0.14^{\pm u}$ (22) & $0.33^{\pm 0.07}$ (265)/$0.33^{\pm 0.09}$ (179)/$0.41^{\pm 0.10}$ (86) \\
$r_{J_{\rm p}V_{\rm C}.I_{\rm X}}$ & $0.23^{\pm 0.06}$ (153)/$0.12^{\pm 0.13}$ (40)/$0.31^{\pm 0.12}$ (31) & $0.30^{\pm 0.07}$ (58)/$0.32^{\pm 0.10}$ (36)/$0.25^{\pm 0.13}$ (22) & $0.25^{\pm 0.05}$ (267)/$0.25^{\pm 0.06}$ (180)/$0.26^{\pm 0.07}$ (87) \\
$r_{J_{\rm p}V_{\rm C}.F_{\rm X}}$ & $0.30^{\pm 0.07}$ (152)/$0.23^{\pm 0.13}$ (40)/$0.43^{\pm 0.12}$ (30) & $0.41^{\pm 0.08}$ (57)/$0.45^{\pm 0.10}$ (35)/$0.14^{\pm 0.11}$ (22) & $0.34^{\pm 0.05}$ (265)/$0.33^{\pm 0.07}$ (179)/$0.38^{\pm 0.07}$ (86) \\
$r_{J_{\rm p}{\rm AW}.I_{\rm X}}$  & $0.19^{\pm 0.06}$ (153)/$0.18^{\pm 0.13}$ (40)/$0.23^{\pm 0.13}$ (31) & $0.20^{\pm 0.10}$ (58)/$0.23^{\pm 0.12}$ (36)/$0.14^{\pm u}$ (22) & $0.19^{\pm 0.05}$ (267)/$0.21^{\pm 0.06}$ (180)/$0.25^{\pm 0.07}$ (87) \\
$r_{J_{\rm p}{\rm AW}.F_{\rm X}}$  & $0.30^{\pm 0.06}$ (152)/$0.27^{\pm 0.13}$ (40)/$0.44^{\pm 0.15}$ (30) & $0.46^{\pm 0.11}$ (57)/$0.45^{\pm 0.13}$ (35)/$0.14^{\pm u}$ (22) & $0.31^{\pm 0.06}$ (265)/$0.32^{\pm 0.07}$ (179)/$0.39^{\pm 0.08}$ (86) \\
\hline
\end{tabular}
\label{T-lowE-cc_Jp}
\end{sidewaystable}

\begin{sidewaystable}
\tiny
\caption{Table of log$-$log correlation coefficients between the high energy proton intensity $J_{\rm 50\,MeV}$ and flare/CME properties. The number of events used for specific calculation is given in brackets. u - uncertain.}
\begin{tabular}{clll}
\hline
Correlation             & SC23$^\prime$: 09/1996$-$08/2004           & SC24$^\prime$: 01/2009$-$12/2016           & Entire period: 01/1996$-$12/2016 \\
coefficient             & All events/W-origin sample/E-origin sample & All events/W-origin sample/E-origin sample & All events/W-origin sample/E-origin sample \\
\hline
\multicolumn{4}{c}{{\bf Linear}} \\
$r_{J_{\rm p}I_{\rm X}}$ & $0.40^{\pm 0.08}$ (162)/$0.40^{\pm 0.08}$ (123)/$0.43^{\pm 0.17}$ (35) & $0.41^{\pm 0.08}$ (85)/$0.39^{\pm 0.09}$ (65)/$0.53^{\pm 0.14}$ (20) & $0.44^{\pm 0.05}$ (278)/$0.43^{\pm 0.06}$ (212)/$0.51^{\pm 0.09}$ (63)\\
$r_{J_{\rm p}F_{\rm X}}$ & $0.35^{\pm 0.08}$ (161)/$0.35^{\pm 0.11}$ (123)/$0.36^{\pm 0.15}$ (35) & $0.35^{\pm 0.08}$ (84)/$0.44^{\pm 0.08}$ (64)/$0.11^{\pm 0.25}$ (20) & $0.39^{\pm 0.06}$ (277)/$0.42^{\pm 0.07}$ (211)/$0.34^{\pm 0.10}$ (63) \\
$r_{J_{\rm p}V_{\rm C}}$ & $0.52^{\pm 0.05}$ (177)/$0.51^{\pm 0.05}$ (145)/$0.62^{\pm 0.10}$ (32) & $0.43^{\pm 0.07}$ (110)/$0.43^{\pm 0.09}$ (83)/$0.46^{\pm 0.11}$ (27)& $0.48^{\pm 0.04}$ (319)/$0.47^{\pm 0.04}$ (254)/$0.57^{\pm 0.06}$ (65) \\
$r_{J_{\rm p}{\rm AW}}$  & $0.41^{\pm 0.05}$ (177)/$0.42^{\pm 0.05}$ (145)/$0.38^{\pm 0.08}$ (32) & $0.23^{\pm 0.06}$ (110)/$0.20^{\pm 0.07}$ (83)/$0.31^{\pm u}$ (27)  &  $0.32^{\pm 0.04}$ (319)/$0.31^{\pm 0.05}$ (254)/$0.36^{\pm 0.06}$ (65) \\
\hline
\multicolumn{4}{c}{{\bf Partial}} \\
$r_{J_{\rm p}I_{\rm X}.V_{\rm C}}$ & $0.19^{\pm 0.10}$ (146)/$0.09^{\pm 0.20}$ (39)/$0.25^{\pm 0.31}$ (28) & $0.26^{\pm 0.15}$ (53)/$0.26^{\pm 0.24}$ (33)/$0.24^{\pm 0.18}$ (20) & $0.22^{\pm 0.07}$ (256)/$0.23^{\pm 0.08}$ (173)/$0.22^{\pm 0.12}$ (83)\\
$r_{J_{\rm p}I_{\rm X}.{\rm AW}}$  & $0.18^{\pm 0.09}$ (146)/$0.17^{\pm 0.21}$ (39)/$0.30^{\pm 0.25}$ (28) & $0.26^{\pm 0.12}$ (53)/$0.30^{\pm 0.18}$ (33)/$0.21^{\pm u}$ (20) & $0.21^{\pm 0.06}$ (256)/$0.22^{\pm 0.07}$ (173)/$0.29^{\pm 0.09}$ (83) \\
$r_{J_{\rm p}F_{\rm X}.V_{\rm C}}$ & $0.25^{\pm 0.12}$ (146)/$0.19^{\pm 0.20}$ (39)/$0.34^{\pm 0.28}$ (28) & $0.31^{\pm 0.18}$ (53)/$0.35^{\pm 0.24}$ (33)/$0.12^{\pm 0.31}$ (20) & $0.30^{\pm 0.08}$ (255)/$0.30^{\pm 0.10}$ (173)/$0.31^{\pm 0.11}$ (82) \\
$r_{J_{\rm p}F_{\rm X}.{\rm AW}}$  & $0.29^{\pm 0.11}$ (146)/$0.23^{\pm 0.21}$ (39)/$0.50^{\pm 0.25}$ (28) & $0.48^{\pm 0.17}$ (53)/$0.51^{\pm 0.20}$ (33)/$0.11^{\pm u}$ (20) & $0.34^{\pm 0.07}$ (255)/$0.35^{\pm 0.09}$ (173)/$0.41^{\pm 0.10}$ (82) \\
$r_{J_{\rm p}V_{\rm C}.I_{\rm X}}$ & $0.21^{\pm 0.06}$ (146)/$0.10^{\pm 0.12}$ (39)/$0.28^{\pm 0.18}$ (28) & $0.27^{\pm 0.10}$ (53)/$0.29^{\pm 0.13}$ (33)/$0.22^{\pm 0.18}$ (20) & $0.24^{\pm 0.05}$ (256)/$0.24^{\pm 0.07}$ (173)/$0.24^{\pm 0.08}$ (83) \\
$r_{J_{\rm p}V_{\rm C}.F_{\rm X}}$ & $0.30^{\pm 0.06}$ (146)/$0.22^{\pm 0.11}$ (39)/$0.42^{\pm 0.18}$ (28) & $0.41^{\pm 0.10}$ (53)/$0.44^{\pm 0.13}$ (33)/$0.18^{\pm 0.13}$ (20) & $0.34^{\pm 0.05}$ (255)/$0.33^{\pm 0.07}$ (173)/$0.37^{\pm 0.09}$ (82) \\
$r_{J_{\rm p}{\rm AW}.I_{\rm X}}$  & $0.18^{\pm 0.06}$ (146)/$0.18^{\pm 0.13}$ (39)/$0.24^{\pm 0.13}$ (28) & $0.19^{\pm 0.10}$ (53)/$0.23^{\pm 0.13}$ (33)/$0.13^{\pm u}$ (20) & $0.18^{\pm 0.06}$ (256)/$0.20^{\pm 0.07}$ (173)/$0.23^{\pm 0.08}$ (83) \\
$r_{J_{\rm p}{\rm AW}.F_{\rm X}}$  & $0.30^{\pm 0.07}$ (146)/$0.28^{\pm 0.13}$ (39)/$0.46^{\pm 0.14}$ (28) & $0.45^{\pm 0.10}$ (53)/$0.47^{\pm 0.14}$ (33)/$0.11^{\pm u}$ & $0.31^{\pm 0.06}$ (255)/$0.32^{\pm 0.07}$ (173)/$0.37^{\pm 0.09}$ (82) \\
\hline
\end{tabular}
\label{T-highE-cc_Jp}
\end{sidewaystable}

\begin{sidewaystable}
\tiny
\caption{Table of log$-$log correlation coefficients between the low energy proton fluence $F_{\rm 25\,MeV}$ and flare/CME properties. The number of events used for specific calculation is given in brackets. u - uncertain.}
\begin{tabular}{clll}
\hline
Correlation             & SC23$^\prime$: 09/1996$-$08/2004           & SC24$^\prime$: 01/2009$-$12/2016           & Entire period: 01/1996$-$12/2016 \\
coefficient             & All events/W-origin sample/E-origin sample & All events/W-origin sample/E-origin sample & All events/W-origin sample/E-origin sample \\
\hline
\multicolumn{4}{c}{{\bf Linear}} \\
$r_{F_{\rm p}I_{\rm X}}$ & $0.34^{\pm 0.08}$ (147)/$0.29^{\pm 0.09}$ (111)/$0.55^{\pm 0.15}$ (32) & $0.39^{\pm 0.09}$ (82)/$0.40^{\pm 0.10}$ (61)/$0.39^{\pm 0.13}$ (21)  & $0.41^{\pm 0.05}$ (256)/$0.36^{\pm 0.06}$ (193)/$0.54^{\pm 0.10}$ (59) \\
$r_{F_{\rm p}F_{\rm X}}$ & $0.38^{\pm 0.08}$ (145)/$0.35^{\pm 0.10}$ (111)/$0.49^{\pm 0.11}$ (31) & $0.34^{\pm 0.08}$ (82)/$0.44^{\pm 0.09}$ (61)/$-0.03^{\pm 0.19}$ (21) & $0.41^{\pm 0.05}$ (254)/$0.42^{\pm 0.07}$ (193)/$0.37^{\pm 0.09}$ (58) \\
$r_{F_{\rm p}V_{\rm C}}$ & $0.52^{\pm 0.05}$ (160)/$0.49^{\pm 0.06}$ (132)/$0.73^{\pm 0.09}$ (28) & $0.51^{\pm 0.07}$ (105)/$0.51^{\pm 0.09}$ (75)/$0.50^{\pm 0.09}$ (30) & $0.52^{\pm 0.04}$ (292)/$0.49^{\pm 0.05}$ (230)/$0.63^{\pm 0.06}$ (62) \\
$r_{F_{\rm p}{\rm AW}}$  & $0.43^{\pm 0.05}$ (160)/$0.43^{\pm 0.06}$ (132)/$0.44^{\pm 0.09}$ (28) & $0.22^{\pm 0.05}$ (105)/$0.29^{\pm 0.06}$ (75)/$0.09^{\pm u}$ (30)   & $0.35^{\pm 0.04}$ (292)/$0.36^{\pm 0.05}$ (230)/$0.28^{\pm 0.07}$ (62) \\
\hline
\multicolumn{4}{c}{{\bf Partial}} \\
$r_{F_{\rm p}I_{\rm X}.V_{\rm C}}$ & $0.19^{\pm 0.11}$ (131)/$0.06^{\pm 0.23}$ (35)/$0.31^{\pm 0.31}$ (26) & $0.28^{\pm 0.17}$ (51)/$0.33^{\pm 0.24}$ (30)/$0.17^{\pm 0.22}$ (21) & $0.22^{\pm 0.07}$ (234)/$0.23^{\pm 0.09}$ (154)/$0.22^{\pm 0.12}$ (80) \\
$r_{F_{\rm p}I_{\rm X}.{\rm AW}}$  & $0.17^{\pm 0.10}$ (131)/$0.13^{\pm 0.23}$ (35)/$0.33^{\pm 0.21}$ (26) & $0.32^{\pm 0.11}$ (51)/$0.34^{\pm 0.17}$ (30)/$0.05^{\pm u}$ (21) & $0.22^{\pm 0.06}$ (234)/$0.22^{\pm 0.08}$ (154)/$0.31^{\pm 0.09}$ (80) \\
$r_{F_{\rm p}F_{\rm X}.V_{\rm C}}$ & $0.23^{\pm 0.12}$ (130)/$0.15^{\pm 0.21}$ (35)/$0.26^{\pm 0.25}$ (25) & $0.23^{\pm 0.19}$ (50)/$0.28^{\pm 0.21}$ (29)/$0.04^{\pm 0.30}$ (21) & $0.29^{\pm 0.08}$ (233)/$0.28^{\pm 0.10}$ (153)/$0.31^{\pm 0.12}$ (80) \\
$r_{F_{\rm p}F_{\rm X}.{\rm AW}}$  & $0.27^{\pm 0.10}$ (130)/$0.19^{\pm 0.21}$ (35)/$0.46^{\pm 0.20}$ (25) & $0.49^{\pm 0.16}$ (50)/$0.46^{\pm 0.17}$ (29)/$-0.01^{\pm u}$ (21) & $0.34^{\pm 0.07}$ (233)/$0.33^{\pm 0.09}$ (153)/$0.42^{\pm 0.11}$ (80) \\
$r_{F_{\rm p}V_{\rm C}.I_{\rm X}}$ & $0.24^{\pm 0.06}$ (131)/$0.08^{\pm 0.11}$ (35)/$0.36^{\pm 0.14}$ (26) & $0.36^{\pm 0.08}$ (51)/$0.38^{\pm 0.12}$ (30)/$0.23^{\pm 0.11}$ (21) & $0.27^{\pm 0.05}$ (234)/$0.27^{\pm 0.06}$ (154)/$0.28^{\pm 0.08}$ (80) \\
$r_{F_{\rm p}V_{\rm C}.F_{\rm X}}$ & $0.28^{\pm 0.07}$ (130)/$0.19^{\pm 0.10}$ (35)/$0.37^{\pm 0.12}$ (25) & $0.37^{\pm 0.07}$ (50)/$0.41^{\pm 0.10}$ (29)/$0.09^{\pm 0.09}$ (21) & $0.35^{\pm 0.05}$ (233)/$0.32^{\pm 0.07}$ (153)/$0.42^{\pm 0.08}$ (80) \\
$r_{F_{\rm p}{\rm AW}.I_{\rm X}}$  & $0.19^{\pm 0.06}$ (131)/$0.16^{\pm 0.10}$ (35)/$0.25^{\pm 0.15}$ (26) & $0.24^{\pm 0.11}$ (51)/$0.26^{\pm 0.14}$ (30)/$0.02^{\pm u}$ (21) & $0.21^{\pm 0.06}$ (234)/$0.22^{\pm 0.07}$ (154)/$0.26^{\pm 0.08}$ (80) \\
$r_{F_{\rm p}{\rm AW}.F_{\rm X}}$  & $0.28^{\pm 0.06}$ (130)/$0.24^{\pm 0.11}$ (35)/$0.42^{\pm 0.15}$ (25) & $0.47^{\pm 0.12}$ (50)/$0.44^{\pm 0.14}$ (29)/$-0.01^{\pm u}$ (21) & $0.32^{\pm 0.06}$ (233)/$0.32^{\pm 0.08}$ (153)/$0.39^{\pm 0.08}$ (80) \\
\hline
\end{tabular}
\label{T-lowE-cc_Fp}
\end{sidewaystable}

\begin{sidewaystable}
\tiny
\caption{Table of log$-$log correlation coefficients between the high energy proton fluence $F_{\rm 50\,MeV}$ and flare/CME properties. The number of events used for specific calculation is given in brackets. u - uncertain.}
\begin{tabular}{clll}
\hline
Correlation             & SC23$^\prime$: 09/1996$-$08/2004           & SC24$^\prime$: 01/2009$-$12/2016           & Entire period: 01/1996$-$12/2016 \\
coefficient             & All events/W-origin sample/E-origin sample & All events/W-origin sample/E-origin sample & All events/W-origin sample/E-origin sample \\
\hline
\multicolumn{4}{c}{{\bf Linear}} \\
$r_{F_{\rm p}I_{\rm X}}$ & $0.31^{\pm 0.07}$ (138)/$0.31^{\pm 0.09}$ (102)/$0.30^{\pm 0.17}$ (32) & $0.41^{\pm 0.08}$ (72)/$0.39^{\pm 0.10}$ (55)/$0.49^{\pm 0.13}$ (17) & $0.38^{\pm 0.06}$ (237)/$0.36^{\pm 0.06}$ (178)/$0.44^{\pm 0.11}$ (55) \\
$r_{F_{\rm p}F_{\rm X}}$ & $0.38^{\pm 0.08}$ (137)/$0.36^{\pm 0.11}$ (102)/$0.47^{\pm 0.11}$ (32) & $0.37^{\pm 0.08}$ (72)/$0.43^{\pm 0.09}$ (55)/$0.15^{\pm 0.25}$ (17) & $0.42^{\pm 0.06}$ (236)/$0.42^{\pm 0.08}$ (178)/$0.44^{\pm 0.09}$ (55) \\
$r_{F_{\rm p}V_{\rm C}}$ & $0.54^{\pm 0.05}$ (145)/$0.53^{\pm 0.05}$ (117)/$0.68^{\pm 0.10}$ (28) & $0.46^{\pm 0.07}$ (96)/$0.45^{\pm 0.09}$ (73)/$0.47^{\pm 0.11}$ (23) & $0.51^{\pm 0.04}$ (268)/$0.50^{\pm 0.04}$ (213)/$0.61^{\pm 0.07}$ (55) \\
$r_{F_{\rm p}{\rm AW}}$  & $0.41^{\pm 0.06}$ (145)/$0.38^{\pm 0.07}$ (117)/$0.60^{\pm 0.08}$ (28) & $0.20^{\pm 0.05}$ (96)/$0.20^{\pm 0.07}$ (73)/$0.21^{\pm u}$ (23)   & $0.32^{\pm 0.05}$ (268)/$0.29^{\pm 0.06}$ (213)/$0.48^{\pm 0.07}$ (55) \\
\hline
\multicolumn{4}{c}{{\bf Partial}} \\
$r_{F_{\rm p}I_{\rm X}.V_{\rm C}}$ & $0.18^{\pm 0.10}$ (124)/$0.06^{\pm 0.24}$ (33)/$0.23^{\pm 0.30}$ (26) & $0.23^{\pm 0.17}$ (47)/$0.24^{\pm 0.27}$ (30)/$0.18^{\pm 0.18}$ (17) & $0.22^{\pm 0.07}$ (219)/$0.23^{\pm 0.09}$ (147)/$0.20^{\pm 0.12}$ (72) \\
$r_{F_{\rm p}I_{\rm X}.{\rm AW}}$  & $0.15^{\pm 0.09}$ (124)/$0.12^{\pm 0.24}$ (33)/$0.20^{\pm 0.27}$ (26) & $0.20^{\pm 0.15}$ (47)/$0.21^{\pm 0.23}$ (30)/$0.07^{\pm u}$ (17) & $0.18^{\pm 0.06}$ (219)/$0.21^{\pm 0.08}$ (147)/$0.19^{\pm 0.09}$ (72) \\
$r_{F_{\rm p}F_{\rm X}.V_{\rm C}}$ & $0.24^{\pm 0.12}$ (124)/$0.15^{\pm 0.22}$ (33)/$0.31^{\pm 0.24}$ (26) & $0.28^{\pm 0.17}$ (47)/$0.31^{\pm 0.20}$ (30)/$0.10^{\pm 0.31}$ (17) & $0.30^{\pm 0.08}$ (219)/$0.30^{\pm 0.10}$ (147)/$0.32^{\pm 0.12}$ (72) \\
$r_{F_{\rm p}F_{\rm X}.{\rm AW}}$  & $0.28^{\pm 0.11}$ (124)/$0.19^{\pm 0.21}$ (33)/$0.40^{\pm 0.23}$ (26) & $0.40^{\pm 0.16}$ (47)/$0.41^{\pm 0.19}$ (30)/$-0.05^{\pm u}$ (17) & $0.31^{\pm 0.07}$ (219)/$0.34^{\pm 0.10}$ (147)/$0.25^{\pm 0.09}$ (72) \\
$r_{F_{\rm p}V_{\rm C}.I_{\rm X}}$ & $0.23^{\pm 0.06}$ (124)/$0.09^{\pm 0.12}$ (33)/$0.35^{\pm 0.13}$ (26) & $0.31^{\pm 0.09}$ (47)/$0.37^{\pm 0.11}$ (30)/$0.18^{\pm 0.20}$ (17) & $0.26^{\pm 0.05}$ (219)/$0.27^{\pm 0.06}$ (147)/$0.25^{\pm 0.08}$ (72) \\
$r_{F_{\rm p}V_{\rm C}.F_{\rm X}}$ & $0.28^{\pm 0.06}$ (124)/$0.19^{\pm 0.13}$ (33)/$0.39^{\pm 0.15}$ (26) & $0.38^{\pm 0.09}$ (47)/$0.42^{\pm 0.12}$ (30)/$0.15^{\pm 0.16}$ (17) & $0.34^{\pm 0.05}$ (219)/$0.32^{\pm 0.07}$ (147)/$0.41^{\pm 0.09}$ (72) \\
$r_{F_{\rm p}{\rm AW}.I_{\rm X}}$  & $0.17^{\pm 0.07}$ (124)/$0.16^{\pm 0.10}$ (33)/$0.27^{\pm 0.11}$ (26) & $0.22^{\pm 0.11}$ (47)/$0.28^{\pm 0.11}$ (30)/$0.04^{\pm u}$ (17) & $0.17^{\pm 0.07}$ (219)/$0.20^{\pm 0.08}$ (147)/$0.17^{\pm 0.08}$ (72) \\
$r_{F_{\rm p}{\rm AW}.F_{\rm X}}$  & $0.28^{\pm 0.08}$ (124)/$0.23^{\pm 0.11}$ (33)/$0.45^{\pm 0.14}$ (26) & $0.44^{\pm 0.13}$ (47)/$0.47^{\pm 0.13}$ (30)/$-0.04^{\pm u}$ (17) & $0.28^{\pm 0.07}$ (219)/$0.31^{\pm 0.09}$ (147)/$0.23^{\pm 0.08}$ (72) \\
\hline
\end{tabular}
\label{T-highE-cc_Fp}
\end{sidewaystable}


\bibliographystyle{spr-mp-sola}
\bibliography{sola_bibliography}  

\begin{thebibliography}{52}
\ifx\bisbn     \undefined \def\bisbn  #1{ISBN #1}\fi
\ifx\binits    \undefined \def\binits#1{#1}\fi
\ifx\bauthor   \undefined \def\bauthor#1{#1}\fi
\ifx\batitle   \undefined \def\batitle#1{#1}\fi
\ifx\bjtitle   \undefined \def\bjtitle#1{\textit{#1}}\fi
\ifx\bvolume   \undefined \def\bvolume#1{\textbf{#1}}\fi
\ifx\byear     \undefined \def\byear#1{#1}\fi
\ifx\bissue    \undefined \def\bissue#1{#1}\fi
\ifx\bfpage    \undefined \def\bfpage#1{#1}\fi
\ifx\blpage    \undefined \def\blpage #1{#1}\fi
\ifx\burl      \undefined \def\burl#1{\textsf{#1}}\fi
\ifx\href      \undefined \def\href#1#2{\textsf{#2}}\fi
\ifx\betal     \undefined \def\betal{\textit{et al.}}\fi
\ifx\bctitle   \undefined \def\bctitle#1{#1}\fi
\ifx\beditor   \undefined \def\beditor#1{#1}\fi
\ifx\bbtitle   \undefined \def\bbtitle#1{\textit{#1}}\fi
\ifx\bedition  \undefined \def\bedition#1{#1}\fi
\ifx\bseriesno \undefined \def\bseriesno#1{\textbf{#1}}\fi
\ifx\blocation \undefined \def\blocation#1{#1}\fi
\ifx\bsertitle \undefined \def\bsertitle#1{\textit{#1}}\fi
\ifx\bsnm      \undefined \def\bsnm#1{#1}\fi
\ifx\bsuffix   \undefined \def\bsuffix#1{#1}\fi
\ifx\bparticle \undefined \def\bparticle#1{#1}\fi
\ifx\barticle  \undefined \def\barticle#1{}\fi
\ifx\binstitute  \undefined \def\binstitute#1{#1}\fi
\ifx\bpublisher  \undefined \def\bpublisher#1{#1}\fi
\ifx\doiurl    \undefined
  \def\doiurl#1{\href{http://dx.doi.org/#1}{\textsf{DOI}}}\fi
\ifx\arxivurl  \undefined
  \def\arxivurl#1{\href{http://arxiv.org/abs/#1}{\textsf{arXiv}}}\fi
\ifx\adsurl    \undefined
  \def\adsurl#1{\href{http://adsabs.harvard.edu/abs/#1}{\textsf{ADS}}}\fi
\ifx\botherref \undefined \def\botherref#1{}\fi
\ifx\url       \undefined \def\url#1{\textsf{#1}}\fi
\ifx\bchapter  \undefined \def\bchapter#1{}\fi
\ifx\bbook     \undefined \def\bbook#1{}\fi
\ifx\bcomment  \undefined \def\bcomment#1{#1}\fi
\ifx\oauthor   \undefined \def\oauthor#1{#1}\fi
\ifx\citeauthoryear \undefined\def \citeauthoryear#1{#1}\fi
\ifx\endbibitem\undefined \def\endbibitem{}\fi
\ifx\bconflocation  \undefined \def\bconflocation#1{#1} \fi

\bibitem[\protect\citeauthoryear{{Bazilevskaya}}{2017}]{2017JPhCS.798a2034B}
\begin{bchapter}
\bauthor{\bsnm{{Bazilevskaya}}, \binits{G.A.}}:
\byear{2017},
\bctitle{{Once again about origin of the solar cosmic rays}}.
In: \bbtitle{Journal of Physics Conference Series},
\bsertitle{Journal of Physics Conference Series}
\bseriesno{798},
\bfpage{012034}.
\doiurl{10.1088/1742-6596/798/1/012034}.
\adsurl{2017JPhCS.798a2034B}.
\end{bchapter}
\endbibitem

\bibitem[\protect\citeauthoryear{{Brueckner}
  \textit{et~al.}}{1995}]{1995SoPh..162..357B}
\begin{barticle}
\bauthor{\bsnm{{Brueckner}}, \binits{G.E.}},
\bauthor{\bsnm{{Howard}}, \binits{R.A.}},
\bauthor{\bsnm{{Koomen}}, \binits{M.J.}},
\bauthor{\bsnm{{Korendyke}}, \binits{C.M.}},
\bauthor{\bsnm{{Michels}}, \binits{D.J.}},
\bauthor{\bsnm{{Moses}}, \binits{J.D.}},
\bauthor{\bsnm{{Socker}}, \binits{D.G.}},
\bauthor{\bsnm{{Dere}}, \binits{K.P.}},
\bauthor{\bsnm{{Lamy}}, \binits{P.L.}},
\bauthor{\bsnm{{Llebaria}}, \binits{A.}},
\bauthor{\bsnm{{Bout}}, \binits{M.V.}},
\bauthor{\bsnm{{Schwenn}}, \binits{R.}},
\bauthor{\bsnm{{Simnett}}, \binits{G.M.}},
\bauthor{\bsnm{{Bedford}}, \binits{D.K.}},
\bauthor{\bsnm{{Eyles}}, \binits{C.J.}}:
\byear{1995},
\batitle{{The Large Angle Spectroscopic Coronagraph (LASCO)}}.
\bjtitle{\solphys}
\bvolume{162},
\bfpage{357}.
\doiurl{10.1007/BF00733434}.
\adsurl{1995SoPh..162..357B}.
\end{barticle}
\endbibitem

\bibitem[\protect\citeauthoryear{{Cane}, {McGuire}, and {von
  Rosenvinge}}{1986}]{1986ApJ...301..448C}
\begin{barticle}
\bauthor{\bsnm{{Cane}}, \binits{H.V.}},
\bauthor{\bsnm{{McGuire}}, \binits{R.E.}},
\bauthor{\bsnm{{von Rosenvinge}}, \binits{T.T.}}:
\byear{1986},
\batitle{{Two classes of solar energetic particle events associated with
  impulsive and long-duration soft X-ray flares}}.
\bjtitle{\apj}
\bvolume{301},
\bfpage{448}.
\doiurl{10.1086/163913}.
\adsurl{1986ApJ...301..448C}.
\end{barticle}
\endbibitem

\bibitem[\protect\citeauthoryear{{Cane}, {Richardson}, and {von
  Rosenvinge}}{2010}]{2010JGRA..11508101C}
\begin{barticle}
\bauthor{\bsnm{{Cane}}, \binits{H.V.}},
\bauthor{\bsnm{{Richardson}}, \binits{I.G.}},
\bauthor{\bsnm{{von Rosenvinge}}, \binits{T.T.}}:
\byear{2010},
\batitle{{A study of solar energetic particle events of 1997-2006: Their
  composition and associations}}.
\bjtitle{Journal of Geophysical Research (Space Physics)}
\bvolume{115},
\bfpage{A08101}.
\doiurl{10.1029/2009JA014848}.
\adsurl{2010JGRA..11508101C}.
\end{barticle}
\endbibitem

\bibitem[\protect\citeauthoryear{{Chandra}
  \textit{et~al.}}{2013}]{2013AdSpR..52.2102C}
\begin{barticle}
\bauthor{\bsnm{{Chandra}}, \binits{R.}},
\bauthor{\bsnm{{Gopalswamy}}, \binits{N.}},
\bauthor{\bsnm{{M{\"a}kel{\"a}}}, \binits{P.}},
\bauthor{\bsnm{{Xie}}, \binits{H.}},
\bauthor{\bsnm{{Yashiro}}, \binits{S.}},
\bauthor{\bsnm{{Akiyama}}, \binits{S.}},
\bauthor{\bsnm{{Uddin}}, \binits{W.}},
\bauthor{\bsnm{{Srivastava}}, \binits{A.K.}},
\bauthor{\bsnm{{Joshi}}, \binits{N.C.}},
\bauthor{\bsnm{{Jain}}, \binits{R.}},
\bauthor{\bsnm{{Awasthi}}, \binits{A.K.}},
\bauthor{\bsnm{{Manoharan}}, \binits{P.K.}},
\bauthor{\bsnm{{Mahalakshmi}}, \binits{K.}},
\bauthor{\bsnm{{Dwivedi}}, \binits{V.C.}},
\bauthor{\bsnm{{Choudhary}}, \binits{D.P.}},
\bauthor{\bsnm{{Nitta}}, \binits{N.V.}}:
\byear{2013},
\batitle{{Solar energetic particle events during the rise phases of solar
  cycles 23 and 24}}.
\bjtitle{Advances in Space Research}
\bvolume{52},
\bfpage{2102}.
\doiurl{10.1016/j.asr.2013.09.006}.
\adsurl{2013AdSpR..52.2102C}.
\end{barticle}
\endbibitem

\bibitem[\protect\citeauthoryear{{Cliver}}{2009}]{2009CEAB...33..253C}
\begin{barticle}
\bauthor{\bsnm{{Cliver}}, \binits{E.W.}}:
\byear{2009},
\batitle{{A Revised Classification Scheme for Solar Energetic Particle
  Events}}.
\bjtitle{Central European Astrophysical Bulletin}
\bvolume{33},
\bfpage{253}.
\adsurl{2009CEAB...33..253C}.
\end{barticle}
\endbibitem

\bibitem[\protect\citeauthoryear{{Cliver}}{2016}]{2016ApJ...832..128C}
\begin{barticle}
\bauthor{\bsnm{{Cliver}}, \binits{E.W.}}:
\byear{2016},
\batitle{{Flare vs. Shock Acceleration of High-energy Protons in Solar
  Energetic Particle Events}}.
\bjtitle{\apj}
\bvolume{832},
\bfpage{128}.
\doiurl{10.3847/0004-637X/832/2/128}.
\adsurl{2016ApJ...832..128C}.
\end{barticle}
\endbibitem

\bibitem[\protect\citeauthoryear{{Crosby}
  \textit{et~al.}}{2015}]{2015SpWea..13..406C}
\begin{barticle}
\bauthor{\bsnm{{Crosby}}, \binits{N.}},
\bauthor{\bsnm{{Heynderickx}}, \binits{D.}},
\bauthor{\bsnm{{Jiggens}}, \binits{P.}},
\bauthor{\bsnm{{Aran}}, \binits{A.}},
\bauthor{\bsnm{{Sanahuja}}, \binits{B.}},
\bauthor{\bsnm{{Truscott}}, \binits{P.}},
\bauthor{\bsnm{{Lei}}, \binits{F.}},
\bauthor{\bsnm{{Jacobs}}, \binits{C.}},
\bauthor{\bsnm{{Poedts}}, \binits{S.}},
\bauthor{\bsnm{{Gabriel}}, \binits{S.}},
\bauthor{\bsnm{{Sandberg}}, \binits{I.}},
\bauthor{\bsnm{{Glover}}, \binits{A.}},
\bauthor{\bsnm{{Hilgers}}, \binits{A.}}:
\byear{2015},
\batitle{{SEPEM: A tool for statistical modeling the solar energetic particle
  environment}}.
\bjtitle{Space Weather}
\bvolume{13},
\bfpage{406}.
\doiurl{10.1002/2013SW001008}.
\adsurl{2015SpWea..13..406C}.
\end{barticle}
\endbibitem

\bibitem[\protect\citeauthoryear{{Desai} and
  {Giacalone}}{2016}]{2016LRSP...13....3D}
\begin{barticle}
\bauthor{\bsnm{{Desai}}, \binits{M.}},
\bauthor{\bsnm{{Giacalone}}, \binits{J.}}:
\byear{2016},
\batitle{{Large gradual solar energetic particle events}}.
\bjtitle{Living Reviews in Solar Physics}
\bvolume{13},
\bfpage{3}.
\doiurl{10.1007/s41116-016-0002-5}.
\adsurl{2016LRSP...13....3D}.
\end{barticle}
\endbibitem

\bibitem[\protect\citeauthoryear{{Dierckxsens}
  \textit{et~al.}}{2015}]{2015SoPh..290..841D}
\begin{barticle}
\bauthor{\bsnm{{Dierckxsens}}, \binits{M.}},
\bauthor{\bsnm{{Tziotziou}}, \binits{K.}},
\bauthor{\bsnm{{Dalla}}, \binits{S.}},
\bauthor{\bsnm{{Patsou}}, \binits{I.}},
\bauthor{\bsnm{{Marsh}}, \binits{M.S.}},
\bauthor{\bsnm{{Crosby}}, \binits{N.B.}},
\bauthor{\bsnm{{Malandraki}}, \binits{O.}},
\bauthor{\bsnm{{Tsiropoula}}, \binits{G.}}:
\byear{2015},
\batitle{{Relationship between Solar Energetic Particles and Properties of
  Flares and CMEs: Statistical Analysis of Solar Cycle 23 Events}}.
\bjtitle{\solphys}
\bvolume{290},
\bfpage{841}.
\doiurl{10.1007/s11207-014-0641-4}.
\adsurl{2015SoPh..290..841D}.
\end{barticle}
\endbibitem

\bibitem[\protect\citeauthoryear{{Gopalswamy}}{2012}]{2012AIPC.1500...14G}
\begin{bchapter}
\bauthor{\bsnm{{Gopalswamy}}, \binits{N.}}:
\byear{2012},
\bctitle{{Energetic particle and other space weather events of solar cycle
  24}}.
In: \beditor{\bsnm{{Hu}}, \binits{Q.}},
\beditor{\bsnm{{Li}}, \binits{G.}},
\beditor{\bsnm{{Zank}}, \binits{G.P.}},
\beditor{\bsnm{{Ao}}, \binits{X.}},
\beditor{\bsnm{{Verkhoglyadova}}, \binits{O.}},
\beditor{\bsnm{{Adams}}, \binits{J.H.}} (eds.)
\bbtitle{American Institute of Physics Conference Series},
\bsertitle{American Institute of Physics Conference Series}
\bseriesno{1500},
\bfpage{14}.
\doiurl{10.1063/1.4768738}.
\adsurl{2012AIPC.1500...14G}.
\end{bchapter}
\endbibitem

\bibitem[\protect\citeauthoryear{{Gopalswamy}
  \textit{et~al.}}{2003}]{2003GeoRL..30lSEP3G}
\begin{barticle}
\bauthor{\bsnm{{Gopalswamy}}, \binits{N.}},
\bauthor{\bsnm{{Yashiro}}, \binits{S.}},
\bauthor{\bsnm{{Lara}}, \binits{A.}},
\bauthor{\bsnm{{Kaiser}}, \binits{M.L.}},
\bauthor{\bsnm{{Thompson}}, \binits{B.J.}},
\bauthor{\bsnm{{Gallagher}}, \binits{P.T.}},
\bauthor{\bsnm{{Howard}}, \binits{R.A.}}:
\byear{2003},
\batitle{{Large solar energetic particle events of cycle 23: A global view}}.
\bjtitle{\grl}
\bvolume{30}(\bissue{12}),
\bfpage{8015}.
\doiurl{10.1029/2002GL016435}.
\adsurl{2003GeoRL..30lSEP3G}.
\end{barticle}
\endbibitem

\bibitem[\protect\citeauthoryear{{Gopalswamy}
  \textit{et~al.}}{2008}]{2008AnGeo..26.3033G}
\begin{barticle}
\bauthor{\bsnm{{Gopalswamy}}, \binits{N.}},
\bauthor{\bsnm{{Yashiro}}, \binits{S.}},
\bauthor{\bsnm{{Akiyama}}, \binits{S.}},
\bauthor{\bsnm{{M{\"a}kel{\"a}}}, \binits{P.}},
\bauthor{\bsnm{{Xie}}, \binits{H.}},
\bauthor{\bsnm{{Kaiser}}, \binits{M.L.}},
\bauthor{\bsnm{{Howard}}, \binits{R.A.}},
\bauthor{\bsnm{{Bougeret}}, \binits{J.L.}}:
\byear{2008},
\batitle{{Coronal mass ejections, type II radio bursts, and solar energetic
  particle events in the SOHO era}}.
\bjtitle{Annales Geophysicae}
\bvolume{26},
\bfpage{3033}.
\doiurl{10.5194/angeo-26-3033-2008}.
\adsurl{2008AnGeo..26.3033G}.
\end{barticle}
\endbibitem

\bibitem[\protect\citeauthoryear{{Gopalswamy}
  \textit{et~al.}}{2009}]{2009EM&P..104..295G}
\begin{barticle}
\bauthor{\bsnm{{Gopalswamy}}, \binits{N.}},
\bauthor{\bsnm{{Yashiro}}, \binits{S.}},
\bauthor{\bsnm{{Michalek}}, \binits{G.}},
\bauthor{\bsnm{{Stenborg}}, \binits{G.}},
\bauthor{\bsnm{{Vourlidas}}, \binits{A.}},
\bauthor{\bsnm{{Freeland}}, \binits{S.}},
\bauthor{\bsnm{{Howard}}, \binits{R.}}:
\byear{2009},
\batitle{{The SOHO/LASCO CME Catalog}}.
\bjtitle{Earth Moon and Planets}
\bvolume{104},
\bfpage{295}.
\doiurl{10.1007/s11038-008-9282-7}.
\adsurl{2009EM\%26P..104..295G}.
\end{barticle}
\endbibitem

\bibitem[\protect\citeauthoryear{{Gopalswamy}
  \textit{et~al.}}{2014}]{2014EP&S...66..104G}
\begin{barticle}
\bauthor{\bsnm{{Gopalswamy}}, \binits{N.}},
\bauthor{\bsnm{{Xie}}, \binits{H.}},
\bauthor{\bsnm{{Akiyama}}, \binits{S.}},
\bauthor{\bsnm{{M{\"a}kel{\"a}}}, \binits{P.A.}},
\bauthor{\bsnm{{Yashiro}}, \binits{S.}}:
\byear{2014},
\batitle{{Major solar eruptions and high-energy particle events during solar
  cycle 24}}.
\bjtitle{Earth, Planets, and Space}
\bvolume{66},
\bfpage{104}.
\doiurl{10.1186/1880-5981-66-104}.
\adsurl{2014EP\%26S...66..104G}.
\end{barticle}
\endbibitem

\bibitem[\protect\citeauthoryear{{Grechnev}
  \textit{et~al.}}{2015}]{2015SoPh..290.2827G}
\begin{barticle}
\bauthor{\bsnm{{Grechnev}}, \binits{V.V.}},
\bauthor{\bsnm{{Kiselev}}, \binits{V.I.}},
\bauthor{\bsnm{{Meshalkina}}, \binits{N.S.}},
\bauthor{\bsnm{{Chertok}}, \binits{I.M.}}:
\byear{2015},
\batitle{{Relations Between Microwave Bursts and Near-Earth High-Energy Proton
  Enhancements and Their Origin}}.
\bjtitle{\solphys}
\bvolume{290},
\bfpage{2827}.
\doiurl{10.1007/s11207-015-0797-6}.
\adsurl{2015SoPh..290.2827G}.
\end{barticle}
\endbibitem

\bibitem[\protect\citeauthoryear{{Haggerty} and
  {Roelof}}{2002}]{2002ApJ...579..841H}
\begin{barticle}
\bauthor{\bsnm{{Haggerty}}, \binits{D.K.}},
\bauthor{\bsnm{{Roelof}}, \binits{E.C.}}:
\byear{2002},
\batitle{{Impulsive Near-relativistic Solar Electron Events: Delayed Injection
  with Respect to Solar Electromagnetic Emission}}.
\bjtitle{\apj}
\bvolume{579},
\bfpage{841}.
\doiurl{10.1086/342870}.
\adsurl{2002ApJ...579..841H}.
\end{barticle}
\endbibitem

\bibitem[\protect\citeauthoryear{{Kahler}}{1982}]{1982JGR....87.3439K}
\begin{barticle}
\bauthor{\bsnm{{Kahler}}, \binits{S.W.}}:
\byear{1982},
\batitle{{The role of the big flare syndrome in correlations of solar energetic
  proton fluxes and associated microwave burst parameters}}.
\bjtitle{\jgr}
\bvolume{87},
\bfpage{3439}.
\doiurl{10.1029/JA087iA05p03439}.
\adsurl{1982JGR....87.3439K}.
\end{barticle}
\endbibitem

\bibitem[\protect\citeauthoryear{{Kahler}}{2001}]{2001JGR...10620947K}
\begin{barticle}
\bauthor{\bsnm{{Kahler}}, \binits{S.W.}}:
\byear{2001},
\batitle{{The correlation between solar energetic particle peak intensities and
  speeds of coronal mass ejections: Effects of ambient particle intensities and
  energy spectra}}.
\bjtitle{\jgr}
\bvolume{106},
\bfpage{20947}.
\doiurl{10.1029/2000JA002231}.
\adsurl{2001JGR...10620947K}.
\end{barticle}
\endbibitem

\bibitem[\protect\citeauthoryear{{Kahler}}{2007}]{2007SSRv..129..359K}
\begin{barticle}
\bauthor{\bsnm{{Kahler}}, \binits{S.W.}}:
\byear{2007},
\batitle{{Solar Sources of Heliospheric Energetic Electron Events - Shocks or
  Flares?}}
\bjtitle{\ssr}
\bvolume{129},
\bfpage{359}.
\doiurl{10.1007/s11214-007-9143-0}.
\adsurl{2007SSRv..129..359K}.
\end{barticle}
\endbibitem

\bibitem[\protect\citeauthoryear{{Kaiser}
  \textit{et~al.}}{2008}]{2008SSRv..136....5K}
\begin{barticle}
\bauthor{\bsnm{{Kaiser}}, \binits{M.L.}},
\bauthor{\bsnm{{Kucera}}, \binits{T.A.}},
\bauthor{\bsnm{{Davila}}, \binits{J.M.}},
\bauthor{\bsnm{{St.~Cyr}}, \binits{O.C.}},
\bauthor{\bsnm{{Guhathakurta}}, \binits{M.}},
\bauthor{\bsnm{{Christian}}, \binits{E.}}:
\byear{2008},
\batitle{{The STEREO Mission: An Introduction}}.
\bjtitle{\ssr}
\bvolume{136},
\bfpage{5}.
\doiurl{10.1007/s11214-007-9277-0}.
\adsurl{2008SSRv..136....5K}.
\end{barticle}
\endbibitem

\bibitem[\protect\citeauthoryear{{Klecker}
  \textit{et~al.}}{2006}]{2006SSRv..123..217K}
\begin{barticle}
\bauthor{\bsnm{{Klecker}}, \binits{B.}},
\bauthor{\bsnm{{Kunow}}, \binits{H.}},
\bauthor{\bsnm{{Cane}}, \binits{H.V.}},
\bauthor{\bsnm{{Dalla}}, \binits{S.}},
\bauthor{\bsnm{{Heber}}, \binits{B.}},
\bauthor{\bsnm{{Kecskemety}}, \binits{K.}},
\bauthor{\bsnm{{Klein}}, \binits{K.-L.}},
\bauthor{\bsnm{{Kota}}, \binits{J.}},
\bauthor{\bsnm{{Kucharek}}, \binits{H.}},
\bauthor{\bsnm{{Lario}}, \binits{D.}},
\bauthor{\bsnm{{Lee}}, \binits{M.A.}},
\bauthor{\bsnm{{Popecki}}, \binits{M.A.}},
\bauthor{\bsnm{{Posner}}, \binits{A.}},
\bauthor{\bsnm{{Rodriguez-Pacheco}}, \binits{J.}},
\bauthor{\bsnm{{Sanderson}}, \binits{T.}},
\bauthor{\bsnm{{Simnett}}, \binits{G.M.}},
\bauthor{\bsnm{{Roelof}}, \binits{E.C.}}:
\byear{2006},
\batitle{{Energetic Particle Observations}}.
\bjtitle{\ssr}
\bvolume{123},
\bfpage{217}.
\doiurl{10.1007/s11214-006-9018-9}.
\adsurl{2006SSRv..123..217K}.
\end{barticle}
\endbibitem

\bibitem[\protect\citeauthoryear{{Klein} and
  {Trottet}}{2001}]{2001SSRv...95..215K}
\begin{barticle}
\bauthor{\bsnm{{Klein}}, \binits{K.-L.}},
\bauthor{\bsnm{{Trottet}}, \binits{G.}}:
\byear{2001},
\batitle{{The Origin of Solar Energetic Particle Events: Coronal Acceleration
  versus Shock Wave Acceleration}}.
\bjtitle{\ssr}
\bvolume{95},
\bfpage{215}.
\adsurl{2001SSRv...95..215K}.
\end{barticle}
\endbibitem

\bibitem[\protect\citeauthoryear{{Klein}, {Trottet}, and
  {Klassen}}{2010}]{2010SoPh..263..185K}
\begin{barticle}
\bauthor{\bsnm{{Klein}}, \binits{K.-L.}},
\bauthor{\bsnm{{Trottet}}, \binits{G.}},
\bauthor{\bsnm{{Klassen}}, \binits{A.}}:
\byear{2010},
\batitle{{Energetic Particle Acceleration and Propagation in Strong CME-Less
  Flares}}.
\bjtitle{\solphys}
\bvolume{263},
\bfpage{185}.
\doiurl{10.1007/s11207-010-9540-5}.
\adsurl{2010SoPh..263..185K}.
\end{barticle}
\endbibitem

\bibitem[\protect\citeauthoryear{{Klein}
  \textit{et~al.}}{2008}]{2008A&A...486..589K}
\begin{barticle}
\bauthor{\bsnm{{Klein}}, \binits{K.-L.}},
\bauthor{\bsnm{{Krucker}}, \binits{S.}},
\bauthor{\bsnm{{Lointier}}, \binits{G.}},
\bauthor{\bsnm{{Kerdraon}}, \binits{A.}}:
\byear{2008},
\batitle{{Open magnetic flux tubes in the corona and the transport of solar
  energetic particles}}.
\bjtitle{\aap}
\bvolume{486},
\bfpage{589}.
\doiurl{10.1051/0004-6361:20079228}.
\adsurl{2008A\%26A...486..589K}.
\end{barticle}
\endbibitem

\bibitem[\protect\citeauthoryear{{Kouloumvakos}
  \textit{et~al.}}{2015}]{2015A&A...580A..80K}
\begin{barticle}
\bauthor{\bsnm{{Kouloumvakos}}, \binits{A.}},
\bauthor{\bsnm{{Nindos}}, \binits{A.}},
\bauthor{\bsnm{{Valtonen}}, \binits{E.}},
\bauthor{\bsnm{{Alissandrakis}}, \binits{C.E.}},
\bauthor{\bsnm{{Malandraki}}, \binits{O.}},
\bauthor{\bsnm{{Tsitsipis}}, \binits{P.}},
\bauthor{\bsnm{{Kontogeorgos}}, \binits{A.}},
\bauthor{\bsnm{{Moussas}}, \binits{X.}},
\bauthor{\bsnm{{Hillaris}}, \binits{A.}}:
\byear{2015},
\batitle{{Properties of solar energetic particle events inferred from their
  associated radio emission}}.
\bjtitle{\aap}
\bvolume{580},
\bfpage{A80}.
\doiurl{10.1051/0004-6361/201424397}.
\adsurl{2015A\%26A...580A..80K}.
\end{barticle}
\endbibitem

\bibitem[\protect\citeauthoryear{{Krucker}
  \textit{et~al.}}{1999}]{1999ApJ...519..864K}
\begin{barticle}
\bauthor{\bsnm{{Krucker}}, \binits{S.}},
\bauthor{\bsnm{{Larson}}, \binits{D.E.}},
\bauthor{\bsnm{{Lin}}, \binits{R.P.}},
\bauthor{\bsnm{{Thompson}}, \binits{B.J.}}:
\byear{1999},
\batitle{{On the Origin of Impulsive Electron Events Observed at 1 AU}}.
\bjtitle{\apj}
\bvolume{519},
\bfpage{864}.
\doiurl{10.1086/307415}.
\adsurl{1999ApJ...519..864K}.
\end{barticle}
\endbibitem

\bibitem[\protect\citeauthoryear{{Krucker}
  \textit{et~al.}}{2007}]{2007ApJ...663L.109K}
\begin{barticle}
\bauthor{\bsnm{{Krucker}}, \binits{S.}},
\bauthor{\bsnm{{Kontar}}, \binits{E.P.}},
\bauthor{\bsnm{{Christe}}, \binits{S.}},
\bauthor{\bsnm{{Lin}}, \binits{R.P.}}:
\byear{2007},
\batitle{{Solar Flare Electron Spectra at the Sun and near the Earth}}.
\bjtitle{\apjl}
\bvolume{663},
\bfpage{L109}.
\doiurl{10.1086/519373}.
\adsurl{2007ApJ...663L.109K}.
\end{barticle}
\endbibitem

\bibitem[\protect\citeauthoryear{{K{\"u}hl}
  \textit{et~al.}}{2017}]{2017SoPh..292...10K}
\begin{barticle}
\bauthor{\bsnm{{K{\"u}hl}}, \binits{P.}},
\bauthor{\bsnm{{Dresing}}, \binits{N.}},
\bauthor{\bsnm{{Heber}}, \binits{B.}},
\bauthor{\bsnm{{Klassen}}, \binits{A.}}:
\byear{2017},
\batitle{{Solar Energetic Particle Events with Protons Above 500 MeV Between
  1995 and 2015 Measured with SOHO/EPHIN}}.
\bjtitle{\solphys}
\bvolume{292},
\bfpage{10}.
\doiurl{10.1007/s11207-016-1033-8}.
\adsurl{2017SoPh..292...10K}.
\end{barticle}
\endbibitem

\bibitem[\protect\citeauthoryear{{Laurenza}
  \textit{et~al.}}{2009}]{2009SpWea...7.4008L}
\begin{barticle}
\bauthor{\bsnm{{Laurenza}}, \binits{M.}},
\bauthor{\bsnm{{Cliver}}, \binits{E.W.}},
\bauthor{\bsnm{{Hewitt}}, \binits{J.}},
\bauthor{\bsnm{{Storini}}, \binits{M.}},
\bauthor{\bsnm{{Ling}}, \binits{A.G.}},
\bauthor{\bsnm{{Balch}}, \binits{C.C.}},
\bauthor{\bsnm{{Kaiser}}, \binits{M.L.}}:
\byear{2009},
\batitle{{A technique for short-term warning of solar energetic particle events
  based on flare location, flare size, and evidence of particle escape}}.
\bjtitle{Space Weather}
\bvolume{7},
\bfpage{S04008}.
\doiurl{10.1029/2007SW000379}.
\adsurl{2009SpWea...7.4008L}.
\end{barticle}
\endbibitem

\bibitem[\protect\citeauthoryear{{Mewaldt} \textit{et~al.}}{2015}]{2015ICRC..M}
\begin{bchapter}
\bauthor{\bsnm{{Mewaldt}}, \binits{R.A.}},
\bauthor{\bsnm{{Cohen}}, \binits{C.M.S.}},
\bauthor{\bsnm{{Mason}}, \binits{G.M.}},
\bauthor{\bsnm{{von Rosenvinge}}, \binits{T.}},
\bauthor{\bsnm{{Li}}, \binits{G.}},
\bauthor{\bsnm{{Smith}}, \binits{S.W.}},
\bauthor{\bsnm{{Vourlidas}}, \binits{A.}}:
\byear{2015},
\bctitle{{Investigating the Causes of Solar-Cycle Variations in Solar Energetic
  Particle Fluences and Composition}}.
In: \bbtitle{Proceedings of the The 34th International Cosmic Ray Conference 30
  July - 6 August 2015 The Hague, The Netherlands}.
\end{bchapter}
\endbibitem

\bibitem[\protect\citeauthoryear{{Miteva}, {Samwel}, and
  {Costa-Duarte}}{2017}]{2017JASTP...M}
\begin{botherref}
\oauthor{\bsnm{{Miteva}}, \binits{R.}},
\oauthor{\bsnm{{Samwel}}, \binits{S.W.}},
\oauthor{\bsnm{{Costa-Duarte}}, \binits{M.V.}}:
2017,
{Solar energetic particle catalogs: assumptions, uncertainties and validity of
  reports}.
\textit{JASTP (in press)}.
\end{botherref}
\endbibitem

\bibitem[\protect\citeauthoryear{{Miteva}
  \textit{et~al.}}{2013}]{2013SoPh..282..579M}
\begin{barticle}
\bauthor{\bsnm{{Miteva}}, \binits{R.}},
\bauthor{\bsnm{{Klein}}, \binits{K.-L.}},
\bauthor{\bsnm{{Malandraki}}, \binits{O.}},
\bauthor{\bsnm{{Dorrian}}, \binits{G.}}:
\byear{2013},
\batitle{{Solar Energetic Particle Events in the 23rd Solar Cycle:
  Interplanetary Magnetic Field Configuration and Statistical Relationship with
  Flares and CMEs}}.
\bjtitle{\solphys}
\bvolume{282},
\bfpage{579}.
\doiurl{10.1007/s11207-012-0195-2}.
\adsurl{2013SoPh..282..579M}.
\end{barticle}
\endbibitem

\bibitem[\protect\citeauthoryear{{Miteva}
  \textit{et~al.}}{2016}]{2016simi.conf...27M}
\begin{bchapter}
\bauthor{\bsnm{{Miteva}}, \binits{R.}},
\bauthor{\bsnm{{Samwel}}, \binits{S.W.}},
\bauthor{\bsnm{{Costa-Duarte}}, \binits{M.V.}},
\bauthor{\bsnm{{Danov}}, \binits{D.}}:
\byear{2016},
\bctitle{{The online catalog of Wind/EPACT proton events}}.
In: \beditor{\bsnm{{Georgieva}}, \binits{K.}},
\beditor{\bsnm{{Kirov}}, \binits{B.}},
\beditor{\bsnm{{Danov}}, \binits{D.}} (eds.)
\bbtitle{Proceedings of the Eighth Workshop ''Solar Influences on the
  Magnetosphere, Ionosphere and Atmosphere'', 30 May-3 June 2016 Sunny Beach,
  Bulgaria. ISSN: 2367-7570},
\bfpage{27}.
\adsurl{2016simi.conf...27M}.
\end{bchapter}
\endbibitem

\bibitem[\protect\citeauthoryear{{Miteva}
  \textit{et~al.}}{2017}]{2017SunGe..12...11M}
\begin{botherref}
\oauthor{\bsnm{{Miteva}}, \binits{R.}},
\oauthor{\bsnm{{Samwel}}, \binits{S.W.}},
\oauthor{\bsnm{{Costa-Duarte}}, \binits{M.V.}},
\oauthor{\bsnm{{Malandraki}}, \binits{O.E.}}:
2017,
{Solar cycle dependence of Wind/EPACT protons, solar flares and coronal mass
  ejections}.
\textit{Sun and Geosphere}.
\end{botherref}
\endbibitem

\bibitem[\protect\citeauthoryear{{M{\"u}ller-Mellin}
  \textit{et~al.}}{1995}]{1995SoPh..162..483M}
\begin{barticle}
\bauthor{\bsnm{{M{\"u}ller-Mellin}}, \binits{R.}},
\bauthor{\bsnm{{Kunow}}, \binits{H.}},
\bauthor{\bsnm{{Flei{\ss}ner}}, \binits{V.}},
\bauthor{\bsnm{{Pehlke}}, \binits{E.}},
\bauthor{\bsnm{{Rode}}, \binits{E.}},
\bauthor{\bsnm{{R{\"o}schmann}}, \binits{N.}},
\bauthor{\bsnm{{Scharmberg}}, \binits{C.}},
\bauthor{\bsnm{{Sierks}}, \binits{H.}},
\bauthor{\bsnm{{Rusznyak}}, \binits{P.}},
\bauthor{\bsnm{{McKenna-Lawlor}}, \binits{S.}},
\bauthor{\bsnm{{Elendt}}, \binits{I.}},
\bauthor{\bsnm{{Sequeiros}}, \binits{J.}},
\bauthor{\bsnm{{Meziat}}, \binits{D.}},
\bauthor{\bsnm{{Sanchez}}, \binits{S.}},
\bauthor{\bsnm{{Medina}}, \binits{J.}},
\bauthor{\bsnm{{Del Peral}}, \binits{L.}},
\bauthor{\bsnm{{Witte}}, \binits{M.}},
\bauthor{\bsnm{{Marsden}}, \binits{R.}},
\bauthor{\bsnm{{Henrion}}, \binits{J.}}:
\byear{1995},
\batitle{{COSTEP - Comprehensive Suprathermal and Energetic Particle
  Analyser}}.
\bjtitle{\solphys}
\bvolume{162},
\bfpage{483}.
\doiurl{10.1007/BF00733437}.
\adsurl{1995SoPh..162..483M}.
\end{barticle}
\endbibitem

\bibitem[\protect\citeauthoryear{{Onsager}
  \textit{et~al.}}{1996}]{1996SPIE.2812..281O}
\begin{bchapter}
\bauthor{\bsnm{{Onsager}}, \binits{T.}},
\bauthor{\bsnm{{Grubb}}, \binits{R.}},
\bauthor{\bsnm{{Kunches}}, \binits{J.}},
\bauthor{\bsnm{{Matheson}}, \binits{L.}},
\bauthor{\bsnm{{Speich}}, \binits{D.}},
\bauthor{\bsnm{{Zwickl}}, \binits{R.W.}},
\bauthor{\bsnm{{Sauer}}, \binits{H.}}:
\byear{1996},
\bctitle{{Operational uses of the GOES energetic particle detectors}}.
In: \beditor{\bsnm{{Washwell}}, \binits{E.R.}} (ed.)
\bbtitle{GOES-8 and Beyond},
\bsertitle{SPIE}
\bseriesno{2812},
\bfpage{281}.
\doiurl{10.1117/12.254075}.
\adsurl{1996SPIE.2812..281O}.
\end{bchapter}
\endbibitem

\bibitem[\protect\citeauthoryear{{Paassilta}
  \textit{et~al.}}{2017}]{2017JSWSC...7A..14P}
\begin{barticle}
\bauthor{\bsnm{{Paassilta}}, \binits{M.}},
\bauthor{\bsnm{{Raukunen}}, \binits{O.}},
\bauthor{\bsnm{{Vainio}}, \binits{R.}},
\bauthor{\bsnm{{Valtonen}}, \binits{E.}},
\bauthor{\bsnm{{Papaioannou}}, \binits{A.}},
\bauthor{\bsnm{{Siipola}}, \binits{R.}},
\bauthor{\bsnm{{Riihonen}}, \binits{E.}},
\bauthor{\bsnm{{Dierckxsens}}, \binits{M.}},
\bauthor{\bsnm{{Crosby}}, \binits{N.}},
\bauthor{\bsnm{{Malandraki}}, \binits{O.}},
\bauthor{\bsnm{{Heber}}, \binits{B.}},
\bauthor{\bsnm{{Klein}}, \binits{K.-L.}}:
\byear{2017},
\batitle{{Catalogue of 55-80 MeV solar proton events extending through solar
  cycles 23 and 24}}.
\bjtitle{Journal of Space Weather and Space Climate}
\bvolume{7}(\bissue{27}),
\bfpage{A14}.
\doiurl{10.1051/swsc/2017013}.
\adsurl{2017JSWSC...7A..14P}.
\end{barticle}
\endbibitem

\bibitem[\protect\citeauthoryear{{Papaioannou}
  \textit{et~al.}}{2016}]{2016JSWSC...6A..42P}
\begin{barticle}
\bauthor{\bsnm{{Papaioannou}}, \binits{A.}},
\bauthor{\bsnm{{Sandberg}}, \binits{I.}},
\bauthor{\bsnm{{Anastasiadis}}, \binits{A.}},
\bauthor{\bsnm{{Kouloumvakos}}, \binits{A.}},
\bauthor{\bsnm{{Georgoulis}}, \binits{M.K.}},
\bauthor{\bsnm{{Tziotziou}}, \binits{K.}},
\bauthor{\bsnm{{Tsiropoula}}, \binits{G.}},
\bauthor{\bsnm{{Jiggens}}, \binits{P.}},
\bauthor{\bsnm{{Hilgers}}, \binits{A.}}:
\byear{2016},
\batitle{{Solar flares, coronal mass ejections and solar energetic particle
  event characteristics}}.
\bjtitle{Journal of Space Weather and Space Climate}
\bvolume{6}(\bissue{27}),
\bfpage{A42}.
\doiurl{10.1051/swsc/2016035}.
\adsurl{2016JSWSC...6A..42P}.
\end{barticle}
\endbibitem

\bibitem[\protect\citeauthoryear{Posner}{2007}]{SWE:SWE185}
\begin{barticle}
\bauthor{\bsnm{Posner}, \binits{A.}}:
\byear{2007},
\batitle{Up to 1-hour forecasting of radiation hazards from solar energetic ion
  events with relativistic electrons}.
\bjtitle{Space Weather}
\bvolume{5}(\bissue{5}),
\bfpage{n/a}.
\bcomment{S05001}.
\doiurl{10.1029/2006SW000268}.
\burl{http://dx.doi.org/10.1029/2006SW000268}.
\end{barticle}
\endbibitem

\bibitem[\protect\citeauthoryear{{Pulkkinen}}{2007}]{2007LRSP....4....1P}
\begin{barticle}
\bauthor{\bsnm{{Pulkkinen}}, \binits{T.}}:
\byear{2007},
\batitle{{Space Weather: Terrestrial Perspective}}.
\bjtitle{Living Reviews in Solar Physics}
\bvolume{4},
\bfpage{1}.
\doiurl{10.12942/lrsp-2007-1}.
\adsurl{2007LRSP....4....1P}.
\end{barticle}
\endbibitem

\bibitem[\protect\citeauthoryear{{Reames}}{1999}]{1999SSRv...90..413R}
\begin{barticle}
\bauthor{\bsnm{{Reames}}, \binits{D.V.}}:
\byear{1999},
\batitle{{Particle acceleration at the Sun and in the heliosphere}}.
\bjtitle{\ssr}
\bvolume{90},
\bfpage{413}.
\doiurl{10.1023/A:1005105831781}.
\adsurl{1999SSRv...90..413R}.
\end{barticle}
\endbibitem

\bibitem[\protect\citeauthoryear{{Reames}}{2013}]{2013SSRv..175...53R}
\begin{barticle}
\bauthor{\bsnm{{Reames}}, \binits{D.V.}}:
\byear{2013},
\batitle{{The Two Sources of Solar Energetic Particles}}.
\bjtitle{\ssr}
\bvolume{175},
\bfpage{53}.
\doiurl{10.1007/s11214-013-9958-9}.
\adsurl{2013SSRv..175...53R}.
\end{barticle}
\endbibitem

\bibitem[\protect\citeauthoryear{{Richardson}, {von Rosenvinge}, and
  {Cane}}{2015}]{2015SoPh..290.1741R}
\begin{barticle}
\bauthor{\bsnm{{Richardson}}, \binits{I.G.}},
\bauthor{\bsnm{{von Rosenvinge}}, \binits{T.T.}},
\bauthor{\bsnm{{Cane}}, \binits{H.V.}}:
\byear{2015},
\batitle{{The Properties of Solar Energetic Particle Event-Associated Coronal
  Mass Ejections Reported in Different CME Catalogs}}.
\bjtitle{\solphys}
\bvolume{290},
\bfpage{1741}.
\doiurl{10.1007/s11207-015-0701-4}.
\adsurl{2015SoPh..290.1741R}.
\end{barticle}
\endbibitem

\bibitem[\protect\citeauthoryear{{Schwenn}}{2006}]{2006LRSP....3....2S}
\begin{barticle}
\bauthor{\bsnm{{Schwenn}}, \binits{R.}}:
\byear{2006},
\batitle{{Space Weather: The Solar Perspective}}.
\bjtitle{Living Reviews in Solar Physics}
\bvolume{3}.
\doiurl{10.12942/lrsp-2006-2}.
\adsurl{2006LRSP....3....2S}.
\end{barticle}
\endbibitem

\bibitem[\protect\citeauthoryear{{Torsti}
  \textit{et~al.}}{1995}]{1995SoPh..162..505T}
\begin{barticle}
\bauthor{\bsnm{{Torsti}}, \binits{J.}},
\bauthor{\bsnm{{Valtonen}}, \binits{E.}},
\bauthor{\bsnm{{Lumme}}, \binits{M.}},
\bauthor{\bsnm{{Peltonen}}, \binits{P.}},
\bauthor{\bsnm{{Eronen}}, \binits{T.}},
\bauthor{\bsnm{{Louhola}}, \binits{M.}},
\bauthor{\bsnm{{Riihonen}}, \binits{E.}},
\bauthor{\bsnm{{Schultz}}, \binits{G.}},
\bauthor{\bsnm{{Teittinen}}, \binits{M.}},
\bauthor{\bsnm{{Ahola}}, \binits{K.}},
\bauthor{\bsnm{{Holmlund}}, \binits{C.}},
\bauthor{\bsnm{{Kelh{\"a}}}, \binits{V.}},
\bauthor{\bsnm{{Lepp{\"a}l{\"a}}}, \binits{K.}},
\bauthor{\bsnm{{Ruuska}}, \binits{P.}},
\bauthor{\bsnm{{Str{\"o}mmer}}, \binits{E.}}:
\byear{1995},
\batitle{{Energetic Particle Experiment ERNE}}.
\bjtitle{\solphys}
\bvolume{162},
\bfpage{505}.
\doiurl{10.1007/BF00733438}.
\adsurl{1995SoPh..162..505T}.
\end{barticle}
\endbibitem

\bibitem[\protect\citeauthoryear{{Trottet}
  \textit{et~al.}}{2015}]{2015SoPh..290..819T}
\begin{barticle}
\bauthor{\bsnm{{Trottet}}, \binits{G.}},
\bauthor{\bsnm{{Samwel}}, \binits{S.}},
\bauthor{\bsnm{{Klein}}, \binits{K.-L.}},
\bauthor{\bsnm{{Dudok de Wit}}, \binits{T.}},
\bauthor{\bsnm{{Miteva}}, \binits{R.}}:
\byear{2015},
\batitle{{Statistical Evidence for Contributions of Flares and Coronal Mass
  Ejections to Major Solar Energetic Particle Events}}.
\bjtitle{\solphys}
\bvolume{290},
\bfpage{819}.
\doiurl{10.1007/s11207-014-0628-1}.
\adsurl{2015SoPh..290..819T}.
\end{barticle}
\endbibitem

\bibitem[\protect\citeauthoryear{{Tylka}
  \textit{et~al.}}{2003}]{2003ICRC....6.3305T}
\begin{barticle}
\bauthor{\bsnm{{Tylka}}, \binits{A.J.}},
\bauthor{\bsnm{{Cohen}}, \binits{C.M.S.}},
\bauthor{\bsnm{{Dietrich}}, \binits{W.F.}},
\bauthor{\bsnm{{Krucker}}, \binits{S.}},
\bauthor{\bsnm{{McGuire}}, \binits{R.E.}},
\bauthor{\bsnm{{Mewaldt}}, \binits{R.A.}},
\bauthor{\bsnm{{Ng}}, \binits{C.K.}},
\bauthor{\bsnm{{Reames}}, \binits{D.V.}},
\bauthor{\bsnm{{Share}}, \binits{G.H.}}:
\byear{2003},
\batitle{{Onsets and Release Times in Solar Particle Events}}.
\bjtitle{International Cosmic Ray Conference}
\bvolume{6},
\bfpage{3305}.
\adsurl{2003ICRC....6.3305T}.
\end{barticle}
\endbibitem

\bibitem[\protect\citeauthoryear{{Vainio}
  \textit{et~al.}}{2013}]{2013JSWSC...3A..12V}
\begin{barticle}
\bauthor{\bsnm{{Vainio}}, \binits{R.}},
\bauthor{\bsnm{{Valtonen}}, \binits{E.}},
\bauthor{\bsnm{{Heber}}, \binits{B.}},
\bauthor{\bsnm{{Malandraki}}, \binits{O.E.}},
\bauthor{\bsnm{{Papaioannou}}, \binits{A.}},
\bauthor{\bsnm{{Klein}}, \binits{K.-L.}},
\bauthor{\bsnm{{Afanasiev}}, \binits{A.}},
\bauthor{\bsnm{{Agueda}}, \binits{N.}},
\bauthor{\bsnm{{Aurass}}, \binits{H.}},
\bauthor{\bsnm{{Battarbee}}, \binits{M.}},
\bauthor{\bsnm{{Braune}}, \binits{S.}},
\bauthor{\bsnm{{Dr{\"o}ge}}, \binits{W.}},
\bauthor{\bsnm{{Ganse}}, \binits{U.}},
\bauthor{\bsnm{{Hamadache}}, \binits{C.}},
\bauthor{\bsnm{{Heynderickx}}, \binits{D.}},
\bauthor{\bsnm{{Huttunen-Heikinmaa}}, \binits{K.}},
\bauthor{\bsnm{{Kiener}}, \binits{J.}},
\bauthor{\bsnm{{Kilian}}, \binits{P.}},
\bauthor{\bsnm{{Kopp}}, \binits{A.}},
\bauthor{\bsnm{{Kouloumvakos}}, \binits{A.}},
\bauthor{\bsnm{{Maisala}}, \binits{S.}},
\bauthor{\bsnm{{Mishev}}, \binits{A.}},
\bauthor{\bsnm{{Miteva}}, \binits{R.}},
\bauthor{\bsnm{{Nindos}}, \binits{A.}},
\bauthor{\bsnm{{Oittinen}}, \binits{T.}},
\bauthor{\bsnm{{Raukunen}}, \binits{O.}},
\bauthor{\bsnm{{Riihonen}}, \binits{E.}},
\bauthor{\bsnm{{Rodr{\'{\i}}guez-Gas{\'e}n}}, \binits{R.}},
\bauthor{\bsnm{{Saloniemi}}, \binits{O.}},
\bauthor{\bsnm{{Sanahuja}}, \binits{B.}},
\bauthor{\bsnm{{Scherer}}, \binits{R.}},
\bauthor{\bsnm{{Spanier}}, \binits{F.}},
\bauthor{\bsnm{{Tatischeff}}, \binits{V.}},
\bauthor{\bsnm{{Tziotziou}}, \binits{K.}},
\bauthor{\bsnm{{Usoskin}}, \binits{I.G.}},
\bauthor{\bsnm{{Vilmer}}, \binits{N.}}:
\byear{2013},
\batitle{{The first SEPServer event catalogue \~{}68-MeV solar proton events
  observed at 1 AU in 1996-2010}}.
\bjtitle{Journal of Space Weather and Space Climate}
\bvolume{3}(\bissue{27}),
\bfpage{A12}.
\doiurl{10.1051/swsc/2013030}.
\adsurl{2013JSWSC...3A..12V}.
\end{barticle}
\endbibitem

\bibitem[\protect\citeauthoryear{{von Rosenvinge}
  \textit{et~al.}}{1995}]{1995SSRv...71..155V}
\begin{barticle}
\bauthor{\bsnm{{von Rosenvinge}}, \binits{T.T.}},
\bauthor{\bsnm{{Barbier}}, \binits{L.M.}},
\bauthor{\bsnm{{Karsch}}, \binits{J.}},
\bauthor{\bsnm{{Liberman}}, \binits{R.}},
\bauthor{\bsnm{{Madden}}, \binits{M.P.}},
\bauthor{\bsnm{{Nolan}}, \binits{T.}},
\bauthor{\bsnm{{Reames}}, \binits{D.V.}},
\bauthor{\bsnm{{Ryan}}, \binits{L.}},
\bauthor{\bsnm{{Singh}}, \binits{S.}},
\bauthor{\bsnm{{Trexel}}, \binits{H.}},
\bauthor{\bsnm{{Winkert}}, \binits{G.}},
\bauthor{\bsnm{{Mason}}, \binits{G.M.}},
\bauthor{\bsnm{{Hamilton}}, \binits{D.C.}},
\bauthor{\bsnm{{Walpole}}, \binits{P.}}:
\byear{1995},
\batitle{{The Energetic Particles: Acceleration, Composition, and Transport
  (EPACT) investigation on the WIND spacecraft}}.
\bjtitle{\ssr}
\bvolume{71},
\bfpage{155}.
\doiurl{10.1007/BF00751329}.
\adsurl{1995SSRv...71..155V}.
\end{barticle}
\endbibitem

\bibitem[\protect\citeauthoryear{{Wall} and
  {Jenkins}}{2003}]{2003psa..book.....W}
\begin{bbook}
\bauthor{\bsnm{{Wall}}, \binits{J.V.}},
\bauthor{\bsnm{{Jenkins}}, \binits{C.R.}}:
\byear{2003},
\bbtitle{{Practical Statistics for Astronomers. Cambridge observing handbooks
  for research astronomers, vol. 3. Cambridge, UK: Cambridge University Press,
  2003}}.
\adsurl{http://cdsads.u-strasbg.fr/abs/2003psa..book.....W}.
\end{bbook}
\endbibitem

\bibitem[\protect\citeauthoryear{{Yashiro}
  \textit{et~al.}}{2004}]{2004JGRA..10907105Y}
\begin{barticle}
\bauthor{\bsnm{{Yashiro}}, \binits{S.}},
\bauthor{\bsnm{{Gopalswamy}}, \binits{N.}},
\bauthor{\bsnm{{Michalek}}, \binits{G.}},
\bauthor{\bsnm{{St.~Cyr}}, \binits{O.C.}},
\bauthor{\bsnm{{Plunkett}}, \binits{S.P.}},
\bauthor{\bsnm{{Rich}}, \binits{N.B.}},
\bauthor{\bsnm{{Howard}}, \binits{R.A.}}:
\byear{2004},
\batitle{{A catalog of white light coronal mass ejections observed by the SOHO
  spacecraft}}.
\bjtitle{Journal of Geophysical Research (Space Physics)}
\bvolume{109},
\bfpage{A07105}.
\doiurl{10.1029/2003JA010282}.
\adsurl{2004JGRA..10907105Y}.
\end{barticle}
\endbibitem

\end{thebibliography}

\IfFileExists{\jobname.bbl}{} {\typeout{}
\typeout{****************************************************}
\typeout{****************************************************}
\typeout{** Please run "bibtex \jobname" to obtain} \typeout{**
the bibliography and then re-run LaTeX} \typeout{** twice to fix
the references !}
\typeout{****************************************************}
\typeout{****************************************************}
\typeout{}}

\end{article} 

\end{document}